\documentclass[twocolumn]{aastex631}

\graphicspath{{./}{figures/}}
\usepackage{amsmath}
\usepackage{array}   
\usepackage{enumitem}
\usepackage{CJK}
\usepackage{hyperref}
\received{June 22, 2022}
\revised{September 1, 2022}
\accepted{September 5, 2022}
\submitjournal{\apj}

\shorttitle{Radio emission in local LIRGs}
\shortauthors{Song et al.}

\begin{document}

\title{Characterizing Compact 15 - 33\,GHz Radio Continuum Sources in Local U/LIRGs}

\correspondingauthor{Yiqing Song}
\email{ys7jf@virginia.edu}

\author[0000-0002-3139-3041]{Y. Song}
\affil{Department of Astronomy, University of Virginia, 530 McCormick Road, Charlottesville, VA 22903, U.S.A.}
\affiliation{National Radio Astronomy Observatory, 520 Edgemont Road, Charlottesville, VA 22903, USA}

\author[0000-0002-1000-6081]{S. T. Linden}
\affiliation{Department of Astronomy, University of Massachusetts, LGRT-B 618, 710 North Pleasant Street, Amherst, U.S.A.}

\author[0000-0003-2638-1334]{A. S. Evans}
\affiliation{Department of Astronomy, University of Virginia, 530 McCormick Road, Charlottesville, VA 22903, U.S.A.}
\affiliation{National Radio Astronomy Observatory, 520 Edgemont Road, Charlottesville, VA 22903, USA}

\author[0000-0003-0057-8892]{L. Barcos-Mu\~{n}oz}
\affiliation{National Radio Astronomy Observatory, 520 Edgemont Road, Charlottesville, VA 22903, USA}

\author[0000-0001-7089-7325]{E. J. Murphy}
\affiliation{National Radio Astronomy Observatory, 520 Edgemont Road, Charlottesville, VA 22903, USA}

\author[0000-0003-3168-5922]{E. Momjian}
\affiliation{National Radio Astronomy Observatory, P.O. Box O, Socorro, NM 87801, USA}

\author[0000-0003-0699-6083]{T. D\'{i}az-Santos}
\affiliation{Institute of Astrophysics, Foundation for Research and Technology-Hellas (FORTH), Heraklion, 70013, Greece}
\affiliation{School of Sciences, European University Cyprus, Diogenes street, Engomi, 1516 Nicosia, Cyprus}

\author[0000-0003-3917-6460]{K. L. Larson}
\affiliation{AURA for the European Space Agency (ESA), Space Telescope Science Institute, 3700 San Martin Drive, Baltimore, MD 21218, USA}

\author[0000-0003-3474-1125]{G. C. Privon}
\affiliation{National Radio Astronomy Observatory, 520 Edgemont Road, Charlottesville, VA 22903, USA}
\affiliation{Department of Astronomy, University of Florida, P.O. Box 112055, Gainesville, FL 32611, USA}

\author[0000-0003-2868-489X]{X. Huang}
\affiliation{Department of Astronomy, University of Virginia, 530 McCormick Road, Charlottesville, VA 22903, U.S.A.}

\author[0000-0003-3498-2973]{L. Armus}
\affiliation{IPAC, California Institute of Technology, Pasadena, CA 91125, USA}

\author[0000-0002-8204-8619]{J. M. Mazzarella}
\affiliation{IPAC, California Institute of Technology, Pasadena, CA 91125, USA}

\author[0000-0002-1912-0024]{V. U}
\affiliation{Department of Physics and Astronomy, 4129 Frederick Reines Hall, University of California, Irvine, CA 92697, USA}

\author[0000-0003-4268-0393]{H. Inami}
\affiliation{Hiroshima Astrophysical Science Center, Hiroshima University, 1-3-1 Kagamiyama, Higashi-Hiroshima, Hiroshima 739-8526, Japan}

\author[0000-0002-2688-1956]{V. Charmandaris}
\affiliation{Institute of Astrophysics, Foundation for Research and Technology-Hellas (FORTH), Heraklion, 70013, Greece}
\affiliation{Department of Physics, University of Crete, Heraklion, 71003, Greece}
\affiliation{School of Sciences, European University Cyprus, Diogenes street, Engomi, 1516 Nicosia, Cyprus}

\author[0000-0001-5231-2645]{C. Ricci}
\affiliation{N\'{u}cleo de Astronom\'{i}a de la Facultad de Ingenier\'{i}a y Ciencias, Universidad Diego Portales, Av. Ej\'{e}rcito Libertador 441, Santiago, 8320000, Chile}

\author[0000-0001-6527-6954]{K. L. Emig}
\altaffiliation{NRAO Jansky Fellow}
\affiliation{National Radio Astronomy Observatory, 520 Edgemont Road, Charlottesville, VA 22903, USA}

\author[0000-0002-6149-8178]{J. McKinney}
\affiliation{Department of Astronomy, University of Massachusetts, LGRT-B 618, 710 North Pleasant Street, Amherst, U.S.A.}


\author[0000-0001-9163-0064]{I. Yoon}
\affiliation{National Radio Astronomy Observatory, 520 Edgemont Road, Charlottesville, VA 22903, USA}


\author[0000-0002-1568-579X]{D. Kunneriath}
\affiliation{National Radio Astronomy Observatory, 520 Edgemont Road, Charlottesville, VA 22903, USA}

\author[0000-0001-8490-6632]{T. S.-Y. Lai}
\affiliation{IPAC, California Institute of Technology, Pasadena, CA 91125, USA}

\author[0000-0001-6956-0987]{E. E. Rodas-Quito}
\affiliation{Departamento de Arqueoastronom\'{i}a y Astronom\'{i}a Cultural, Universidad Nacional Aut\'{o}noma de Honduras, Edificio K1 y K2, Ciudad Universitaria, Tegucigalpa, M. D. C. Honduras}

\author[0000-0003-4546-3810]{A. Saravia}
\affiliation{Department of Astronomy, University of Virginia, 530 McCormick Road, Charlottesville, VA 22903, U.S.A.}

\author[0000-0002-1158-6372]{T. Gao}
\affiliation{Department of Astronomy, Beijing Normal University, Beijing 100875, China}

\author{W. Meynardie}
\affil{Department of Astronomy, University of Virginia, 530 McCormick Road, Charlottesville, VA 22903, U.S.A.}

\author[0000-0002-1233-9998]{D. B. Sanders}
\affiliation{Institute for Astronomy, University of Hawaii, 2680 Woodlawn Drive, Honolulu, HI 96822, USA}





\begin{abstract}
We present the analysis of $\sim 100$\,pc-scale compact radio continuum sources detected in 63 local (Ultra) Luminous Infrared Galaxies (U/LIRGs; $L_{\rm IR} \ge 10^{11}\,L_\odot$), using FWHM$\,\lesssim 0\farcs1 - 0\farcs2$ resolution 15 and 33 GHz observations with the Karl G. Jansky Very Large Array. We identify a total of 133 compact radio sources with effective radii of 8 -- 170\,pc, which are classified into four main categories -- ``AGN" (AGN), ``AGN/SBnuc" (AGN-starburst composite nucleus), ``SBnuc" (starburst nucleus) and ``SF" (star-forming clumps) -- based on ancillary datasets and the literature. We find that ``AGN" and ``AGN/SBnuc" more frequently occur in late-stage mergers and have up to 3\,dex higher 33\,GHz luminosities and surface densities compared with ``SBnuc" and ``SF", which may be attributed to extreme nuclear starburst and/or AGN activity in the former. Star formation rates (SFRs) and surface densities ($\Sigma_{\rm SFR}$) are measured for ``SF" and ``SBnuc" using both the total 33\,GHz continuum emission (SFR $\sim 0.14 - 13$\,M$_\odot$ yr$^{-1}$, $\Sigma_{\rm SFR} \sim 13 - 1600$\,M$_\odot$ yr$^{-1}$ kpc$^{-2}$) and the thermal free-free emission from HII regions (median SFR$_{\rm th} \sim 0.4$\,M$_\odot$ yr$^{-1}$, $\Sigma_{\rm SFR_{th}} \sim 44$\,M$_\odot$ yr$^{-1}$ kpc$^{-2}$). These values are 1 -- 2\,dex higher than those measured for similar-sized clumps in nearby normal (non-U/LIRGs). The latter also have much flatter median 15 -- 33 GHz spectral index ($\sim -0.08$) compared with ``SBnuc" and ``SF" ($\sim -0.46$), which may reflect higher non-thermal contribution from supernovae and/or ISM densities in local U/LIRGs that directly result from and/or lead to their extreme star-forming activities on 100\,pc scales.
\end{abstract}



\section{Introduction}\label{sec:intro}
Luminous Infrared Galaxies (LIRGs; $ 10^{11} \leq
L_\mathrm{IR}$[8--1000$\mu$m]$\,< 10^{12} L_\odot$) and Ultra-luminous Infrared
Galaxies (ULIRG; $L_\mathrm{IR}$[8--1000$\mu$m]$ \geq 10^{12} L_\odot$) are an
important class of objects for understanding massive galaxy evolution. Despite
their rarity in the local Universe, U/LIRGs are the dominant contributors
to the co-moving infrared (IR) luminosity density and star formation rate (SFR)
density at $z \gtrsim 1$
\citep{chary01,lefloch05,magnelli11,magnelli13,gruppioni13}, and ULIRGs are
about a thousand times more common at $z \sim 2$ compared to $z \sim 0$ \citep[e.g.][]{chapman05,magnelli13}. \\
\indent Observations of U/LIRGs in the local Universe revealed that a
significant fraction of local LIRGs and nearly all local ULIRGs are interacting
or merging gas-rich spirals \cite[e.g.][]{lonsdale84, armus87,
sanders96}. Simulations of galaxy interactions have been used to infer that such a process
typically drives large fractions of interstellar materials into the central kpc
of each galaxy \citep[e.g.][]{barnes92}, triggering intense nuclear
starbursts \citep[e.g.][]{mihos96, moreno20} and/or fueling of powerful Active
Galactic Nuclei \citep[AGN; e.g.][]{matteo05}. This nuclear activity is thought
to play a key role in the transformation of gas-rich systems into massive
elliptical galaxies, the formation of quasars and the co-evolution of
supermassive black holes (SMBH) and stellar bulges \citep[e.g.][]{sanders88,
hopkins06}. While the discovery of heavily-obscured luminous AGN
in local U/LIRGs \citep[][]{iwasawa11,treister12,ricci17, koss18, torres-alba18,ricci21} has provided strong supporting evidence for this evolutionary scenario \citep[see also review by][]{u2022review}, details regarding the interplay between
star formation and AGN activity, as well as how they together (or separately)
act upon the transformation of these extreme systems, still remain ambiguous. However, the extraordinary
star-forming properties of local U/LIRGs relative to nearby normal galaxies
\citep[i.e. galaxies with $L_{\rm IR} < 10^{11} L_\odot$; e.g.][]{condon91, lonsdale84, howell10, stierwalt14,
piqueras16, diaz-santos17, linden19, larson20, song21, linden21}, and the prevalence of outflows observed in
starburst-dominated local U/LIRGs \citep[e.g.][]{rupke05,cazzoli16,barcos18,fluetsch19,u19,fluetsch20}
highlight the pivotal role of star formation-driven feedback in regulating their
evolution. To better quantify the physical processes governing the evolution of
local U/LIRGs, a robust characterization of the most energetic regions in these
systems is necessary.\\
\indent Due to heavy dust obscuration, especially in the central kpc of U/LIRGs, observations of the optically-thin radio emission provide the best tools for probing into the most obscured but energetically-dominant regions in these systems. \cite{condon91} conducted the first sub-kpc scale radio continuum study of a sample of 40 local U/LIRGs at 8.4\,GHz using the Very Large Array, and concluded that most of their dust-obscured nuclei are powered by starbursts, with many as compact as 100\,pc in radius. With the upgraded bandwidth of the Karl G. Jansky Very Large Array (VLA), observations at higher frequencies are now possible, allowing access to the faint, thermal free-free emission directly arising from ionizing photons from HII regions \citep[][]{condon92,murphy11} at sub-arcsecond resolutions. Using 33 GHz continuum VLA observations, \cite{barcos15,barcos17} constrained the sizes and star formation rates (SFR) for the nuclei of the 22 most luminous local U/LIRGs. In the western nucleus of ULIRG Arp\,220, the authors derived a SFR surface density of $10^{4.1}$ M$_\odot$\,yr$^{-1}$\,kpc$^{-2}$, the highest value ever measured, and far exceeding the theoretical limits for starbursts supported by supernovae feedback and dust-reprocessed radiation \citep[][]{thompson05, kim15}. What drives these extreme SFR surface densities, and are such conditions also observed in LIRGs at lower IR luminosity?\\
\indent This paper aims to investigate the above questions. We present results from the high-resolution ($0\farcs1 - 0\farcs2$) component of a new multi-frequency multi-resolution radio continuum snapshot survey of 68 local U/LIRGs from the Great Observatories All-sky LIRG Survey \citep[GOALS;][]{armus09}. In contrast to the previous radio surveys \citep[e.g.][]{condon91,barcos17} that focused on the most luminous objects, these 68 U/LIRGs span the entire IR luminosity range of the full GOALS sample of 201 U/LIRGs in the local Universe (i.e. $10^{11} - 10^{12.5} \rm L_\odot$), as demonstrated in Figure \ref{fig:sample} (see Section \ref{sec:sampledata} for details), and therefore represents a more diverse range of physical environments, including ones that more closely resemble nearby normal galaxies. These new observations also serve as an excellent companion to the Star Formation in Radio Survey \citep[SFRS;][]{murphy12, murphy18,linden20}, a study of 56 nearby normal galaxies observed at the same frequencies and physical scales as the U/LIRG sample presented in this paper.\\
\indent \cite{linden19} presented the first results from our new U/LIRG survey based on $\sim\,1\arcsec$ resolution observations. Despite finding extra-nuclear star formation enhancement relative to SFRS galaxies on kpc scales , they concluded that nuclear star formation must drive GOALS systems above the Star Formation Main Sequence \citep[SFMS; e.g.][]{elbaz11,speagle14} occupied by the SFRS galaxies. Subsequently in \cite{song21}, four nuclear rings detected in our survey were examined at $\sim$\,100\,pc ($0\farcs1$) scales and compared to five nuclear rings detected in the SFRS galaxies. The nuclear ring star formation was found to contribute more significantly to the total star formation of the host galaxies for LIRGs compared with normal galaxies. In this paper, we extend the methodology adopted in \cite{song21} to study the $\sim 100\,$pc-scale compact radio continuum sources detected in 63 U/LIRGs in the survey, with the aim of constraining the nature and physical properties of these energetic regions at frequencies unimpeded by dust extinction.\\
\indent This paper is divided into six sections. We describe our sample, data, observation information and reduction procedures in \S \ref{sec:sampledata}. In \S \ref{sec:analysis}, we describe the methods used to identify and characterize individual regions of compact radio continuum emission. We further classify individual regions in each U/LIRG system into different types using ancillary multi-wavelength datasets and information from the literature in \S \ref{sec:results}, and present the derived region quantities for different region types. In \S \ref{sec:discuss}, we discuss the limitations and implications of our results, complemented by results derived from observations of other U/LIRGs and nearby normal galaxies. Finally, \S \ref{sec:summary} summarizes major results and conclusions. \\
\indent Throughout this work we adopt $H_0 = 70$\,km/s/Mpc, $\Omega_\mathrm{matter} = 0.28$ and $\Omega_\mathrm{vacuum} = 0.72$ based on the five-year WMAP result \citep{hinshaw09}. These parameters are used with the 3-attractor model \citep{mould00} to calculate the luminosity distances of the
sample.\\
 
\section{Survey Description \& Data Reduction}\label{sec:sampledata}
\subsection{The GOALS ``Equatorial" Survey}
\begin{figure}[htb]
    \centering
    \includegraphics[scale=0.6]{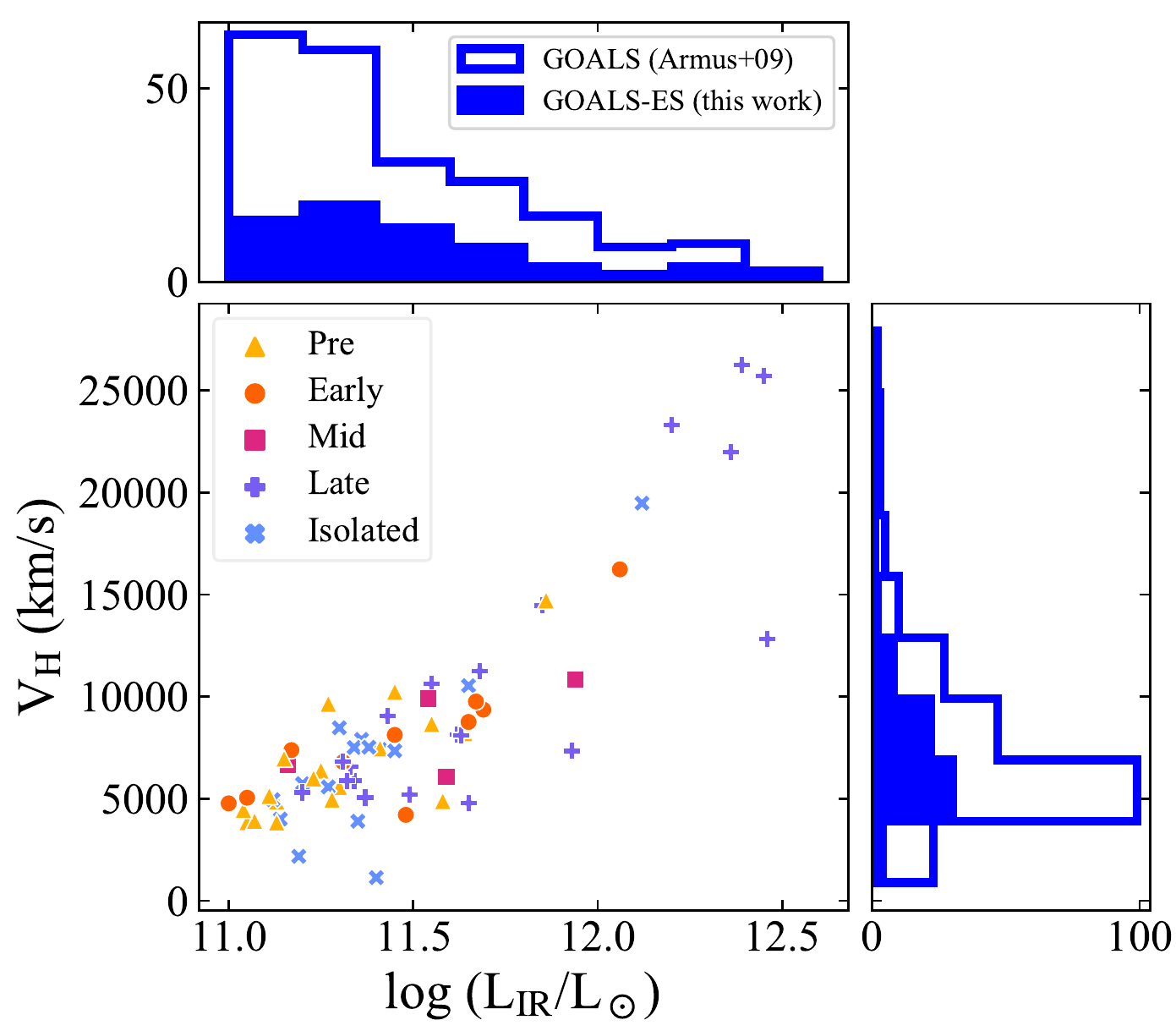}
    \caption{Basic properties (Heliocentric velocity and 8-1000$\mu$m IR luminosity) of the GOALS-ES sample. Each system is color-coded by its merger stage, using visual classification by \cite{stierwalt13}. The sample covers the full range of IR luminosities, distances and merger stages represented by the GOALS sample of all 201 local U/LIRGs.\label{fig:sample}}
\end{figure}

GOALS \citep[][]{armus09} is a multi-wavelength imaging and spectroscopic campaign dedicated to studying the complete subset of over 200 local ($z < 0.088$) U/LIRGs from the IRAS Revised Bright Galaxy Sample of 629 extragalactic objects \citep[RBGS; $S_{60 \mu \rm{m}} > 5.24$\,Jy, $l > 5^\circ$;][]{sanders03}.
The GOALS ``equatorial" radio survey \citep[hereafter GOALS-ES; see also][]{linden19} is a multi-frequency, multi-resolution snapshot
VLA survey designed to map the brightest radio continuum emission in all 68
U/LIRGs from GOALS that have declination of $|\delta| < 20^\circ$. This
equatorial selection allows detailed follow-up studies using
ground-based facilities from both Hemispheres. The sample covers the entire range of $L_\mathrm{IR}$ ($10^{11} - 10^{12.5} \rm L_\odot$),
distances ($V_{\rm H}$ = 1137 - 26249 km/s), and merger stages spanned by the full GOALS sample of 201 systems, as shown in Figure \ref{fig:sample}. A two-sample Kolmogorov-Smirnov (K-S) test on the $L_\mathrm{IR}$ and $V_{\rm H}$ distributions of GOALS and the equatorial sample yields p-values of 0.86 and 0.74, respectively. Hence, this equatorial sample serves as a statistically robust representation of the local
U/LIRG population. Table \ref{tab:sample} lists the basic properties of the GOALS-ES sample. In total,  18 systems are in (a) ``pre-mergers", 10 in (b) ``early-stage" mergers, 4 in (c) ``mid-stage" mergers, 21 in (d) ``late-stage" mergers, and 15 are (N) ``non-mergers", based on visual classification by \cite{stierwalt13} using $\textit{Spitzer}$ imaging.


\centerwidetable
\startlongtable
\begin{deluxetable*}{clllRrrrrc}
\tablecaption{Basic Properties of the GOALS-ES Sample \label{tab:sample}}
\tablenum{1}
\tablecolumns{10}
\tablehead{
\colhead{ID} &
\colhead{IRAS} & 
\colhead{Galaxy Name} & 
\colhead{RA (J2000)} & 
\colhead{DEC(J2000)} & 
\colhead{log$(\frac{L_\mathrm{IR}}{L_\odot}$)} &
\colhead{V$_\mathrm{H}$(km/s)} & 
\colhead{D$_L$(Mpc)} & 
\colhead{Scale (pc/\arcsec)} &
\colhead{Stage} \\
\colhead{(1)} &
\colhead{(2)} & 
\colhead{(3)} & 
\colhead{(4)} & 
\colhead{(5)} & 
\colhead{(6)} & 
\colhead{(7)} & 
\colhead{(8)} & 
\colhead{(9)} & 
\colhead{(10)}
}
\startdata
1  & F00085-1223 & NGC 0034             & 00h11m06.56s & -12\degr06\arcmin28\farcs2 & 11.34 & 5881  & 84  & 393         & d     \\
2  & F00163-1039 & MCG -02-01-052       & 00h18m50.90s & -10\degr22\arcmin36\farcs7 & 11.45 & 8125  & 117 & 540         & b     \\
3  & F01053-1746 & IC 1623 (VV 114)     & 01h07m47.59s & -17\degr30\arcmin24\farcs2 & 11.59 & 6087  & 87  & 400         & c     \\
4  & F01076-1707 & MCG -03-04-014       & 01h10m08.93s & -16\degr51\arcmin09\farcs9 & 11.65 & 10536 & 152 & 689         & N     \\
5  & F01173+1405 & CGCG 436-030         & 01h20m02.63s & +14\degr21\arcmin42\farcs3 & 11.69 & 9362  & 134 & 612         & b     \\
6  & F01364-1042 &                      & 01h38m52.79s & -10\degr27\arcmin12\farcs1 & 11.85 & 14464 & 211 & 931         & d     \\
7  & F01417+1651 & III Zw 035            & 01h44m30.56s & +17\degr06\arcmin09\farcs0 & 11.64 & 8214  & 117 & 539         & a     \\
8  & F02071-1023 & NGC 0838             & 02h09m38.66s & -10\degr08\arcmin47\farcs2 & 11.05 & 3851  & 54  & 257         & a     \\
9  & F02114+0456 & IC 0214              & 02h14m05.56s & +05\degr10\arcmin23\farcs7 & 11.43 & 9061  & 130 & 592         & d     \\
10 & F02152+1418 & NGC 0877             & 02h17m53.26s & +14\degr31\arcmin18\farcs4 & 11.07 & 3913  & 55  & 261         & a     \\
11 & F02281-0309 & NGC 0958             & 02h30m42.84s & -02\degr56\arcmin20\farcs5 & 11.20  & 5738  & 81  & 379         & N     \\
12 & F02401-0013 & NGC 1068             & 02h42m40.72s & -00\degr00\arcmin47\farcs9 & 11.40  & 1137  & 17  & 80          & N     \\
13 & F02435+1253 & UGC 02238            & 02h46m17.46s & +13\degr05\arcmin44\farcs6 & 11.33 & 6560  & 93  & 433         & d     \\
14 & F02512+1446 & UGC 02369            & 02h54m01.75s & +14\degr58\arcmin36\farcs4 & 11.67 & 9761  & 140 & 640         & b     \\
15 & F03359+1523 &                      & 03h38m47.07s & +15\degr32\arcmin54\farcs1 & 11.55 & 10613 & 153 & 693         & d     \\
16 & F03514+1546 & CGCG 465-012         & 03h54m15.95s & +15\degr55\arcmin43\farcs4 & 11.16 & 6662  & 95  & 442         & c     \\
17 & F04097+0525 & UGC 02982            & 04h12m22.68s & +05\degr32\arcmin49\farcs1 & 11.20  & 5305  & 76  & 355         & d     \\
18 & F04191-1855 & ESO 550-IG02         & 04h21m20.02s & -18\degr48\arcmin39\farcs6 & 11.27 & 9652  & 140 & 637         & a     \\
19 & F04315-0840 & NGC 1614             & 04h33m59.95s & -08\degr34\arcmin46\farcs6 & 11.65 & 4778  & 69  & 323         & d     \\
20 & F04326+1904 & UGC 03094            & 04h35m33.81s & +19\degr10\arcmin18\farcs0 & 11.41 & 7408  & 107 & 493         & N     \\
21 & F05053-0805 & NGC 1797             & 05h07m44.84s & -08\degr01\arcmin08\farcs7 & 11.04 & 4457  & 65  & 304         & a     \\
22 & F05054+1718 & CGCG 468-002         & 05h08m21.21s & +17\degr22\arcmin08\farcs0 & 11.05 & 5049  & 73  & 340         & b     \\
23 & F05187-1017 &                      & 05h21m06.53s & -10\degr14\arcmin46\farcs2 & 11.30  & 8474  & 123 & 566         & N     \\
24 & 05442+1732  &                      & 05h47m11.2s  & +17\degr33\arcmin46\farcs4 & 11.30  & 5582  & 81  & 381         & a     \\
25 & F06295-1735 & ESO 557-G002         & 06h31m47.2s  & -17\degr37\arcmin16\farcs6 & 11.25 & 6385  & 94  & 439         & a     \\
26 & 07251-0248  &                      & 07h27m37.62s & -02\degr54\arcmin54\farcs8 & 12.39 & 26249 & 401 & 1643        & d     \\
27 & F07329+1149 & MCG +02-20-003       & 07h35m43.44s & +11\degr42\arcmin34\farcs8 & 11.13 & 4873  & 74  & 345         & a     \\
28 & F09111-1007 &                      & 09h13m37.69s & -10\degr19\arcmin24\farcs6 & 12.06 & 16231 & 246 & 1073        & b     \\
29 & F09437+0317 & Arp 303 (IC 0563/4)  & 09h46m20.70s & +03\degr03\arcmin30\farcs4 & 11.23 & 6002  & 93  & 430         & a     \\
30 & F10015-0614 & NGC 3110             & 10h04m02.11s & -06\degr28\arcmin29\farcs5 & 11.37 & 5054  & 80  & 372         & d     \\
31 & F10173+0828 &                      & 10h20m00.24s & +08\degr13\arcmin32\farcs8 & 11.86 & 14716 & 224 & 986         & a     \\
32 & F11186-0242 & CGCG 011-076         & 11h21m12.24s & -02\degr59\arcmin02\farcs5 & 11.41 & 7464  & 117 & 538         & a     \\
33 & F11231+1456 & IC 2810              & 11h25m45.07s & +14\degr40\arcmin36\farcs0 & 11.45 & 10243 & 158 & 714         & a     \\
34 & F12112+0305 &                      & 12h13m46.02s & +02\degr48\arcmin42\farcs2 & 12.36 & 21980 & 340 & 1427        & d     \\
35 & F12224-0624 &                      & 12h25m03.9s  & -06\degr40\arcmin52\farcs1 & 11.36 & 7902  & 124 & 570         & N     \\
36 & F12243-0036 & NGC 4418             & 12h26m54.6s  & -00\degr52\arcmin39\farcs6 & 11.19 & 2179  & 36  & 170         & N     \\
37 & F12592+0436 & CGCG 043-099         & 13h01m50.28s & +04\degr20\arcmin00\farcs8 & 11.68 & 11237 & 174 & 782         & d     \\
38 & F12596-1529 & MCG -02-33-098       & 13h02m19.66s & -15\degr46\arcmin04\farcs2 & 11.17 & 4773  & 77  & 359         & b     \\
39 & F13188+0036 & NGC 5104             & 13h21m23.09s & +00\degr20\arcmin33\farcs2 & 11.27 & 5578  & 90  & 419         & N     \\
40 & F13197-1627 & MCG -03-34-064       & 13h22m24.45s & -16\degr43\arcmin42\farcs4 & 11.28 & 4959  & 80  & 375         & a     \\
41 & F13373+0105 & Arp 240 (NGC 5257/8) & 13h39m55.34s & +00\degr50\arcmin09\farcs5 & 11.62 & 6798  & 108 & 500         & b     \\
42 & F13497+0220 & NGC 5331             & 13h52m16.32s & +02\degr06\arcmin18\farcs0 & 11.66 & 9906  & 154 & 699         & c     \\
43 & F14348-1447 &                      & 14h37m38.28s & -15\degr00\arcmin24\farcs2 & 12.39 & 24883 & 387 & 1596        & d     \\
44 & F15107+0724 & CGCG 049-057         & 15h13m13.07s & +07\degr13\arcmin32\farcs1 & 11.35 & 3897  & 64  & 302         & N     \\
45 & F15276+1309 & NGC 5936             & 15h30m00.85s & +12\degr59\arcmin22\farcs1 & 11.14 & 4004  & 66  & 310         & N     \\
46 & F15437+0234 & NGC 5990             & 15h46m16.41s & +02\degr24\arcmin55\farcs6 & 11.13 & 3839  & 63  & 297         & a     \\
47 & F16164-0746 &                      & 16h19m11.75s & -07\degr54\arcmin03\farcs0 & 11.62 & 8140  & 127 & 583         & d     \\
48 & F16284+0411 & CGCG 052-037         & 16h30m56.53s & +04\degr04\arcmin58\farcs7 & 11.45 & 7342  & 115 & 531         & N     \\
49 & F16399-0937 &                      & 16h42m40.11s & -09\degr43\arcmin13\farcs7 & 11.63 & 8098  & 126 & 579         & d     \\
50 & F16504+0228 & NGC 6240             & 16h52m58.9s  & +02\degr24\arcmin03\farcs3 & 11.93 & 7339  & 115 & 529         & d     \\
51 & F16516-0948 &                      & 16h54m23.72s & -09\degr53\arcmin20\farcs9 & 11.31 & 6807  & 107 & 495         & d     \\
52 & F17138-1017 &                      & 17h16m35.68s & -10\degr20\arcmin40\farcs5 & 11.49 & 5197  & 83  & 386         & d     \\
53 & 17208-0014  &                      & 17h23m21.97s & -00\degr17\arcmin00\farcs7 & 12.46 & 12834 & 197 & 874         & d     \\
54 & 17578-0400  &                      & 18h00m31.86s & -04\degr00\arcmin53\farcs4 & 11.48 & 4210  & 67  & 317         & b     \\
55 & 18090+0130  &                      & 18h11m33.41s & +01\degr31\arcmin42\farcs4 & 11.65 & 8759  & 134 & 610         & b     \\
56 & F19297-0406 &                      & 19h32m22.30s & -04\degr00\arcmin01\farcs1 & 12.45 & 25701 & 394 & 1619        & d     \\
57 & 19542+1110  &                      & 19h56m35.78s & +11\degr19\arcmin04\farcs9 & 12.12 & 19473 & 294 & 1257        & N     \\
58 & F20304-0211 & NGC 6926             & 20h33m06.13s & -02\degr01\arcmin38\farcs9 & 11.32 & 5880  & 88  & 412         & d     \\
59 & F20550+1655 & II Zw 096             & 20h57m24.38s & +17\degr07\arcmin39\farcs2 & 11.94 & 10822 & 160 & 724         & c     \\
60 & F22287-1917 & ESO 602-G025         & 22h31m25.48s & -19\degr02\arcmin04\farcs0 & 11.34 & 7507  & 110 & 506         & N     \\
61 & F22491-1808 &                      & 22h51m49.35s & -17\degr52\arcmin24\farcs9 & 12.20  & 23312 & 351 & 1466        & d     \\
62 & F23007+0836 & NGC 7469             & 23h03m15.64s & +08\degr52\arcmin25\farcs5 & 11.58 & 4892  & 71  & 332         & a     \\
63 & F23024+1916 & CGCG 453-062         & 23h04m56.55s & +19\degr33\arcmin07\farcs1 & 11.38 & 7524  & 109 & 502         & N     \\
64 & F23157+0618 & NGC 7591             & 23h18m16.25s & +06\degr35\arcmin09\farcs1 & 11.12 & 4956  & 71  & 335         & N     \\
65 & F23157-0441 & NGC 7592             & 23h18m22.19s & -04\degr24\arcmin57\farcs4 & 11.40  & 7380  & 107 & 490         & b     \\
66 & F23254+0830 & NGC 7674             & 23h27m56.71s & +08\degr46\arcmin44\farcs3 & 11.55 & 8671  & 125 & 573         & a     \\
67 & 23262+0314  & NGC 7679             & 23h28m46.62s & +03\degr30\arcmin41\farcs4 & 11.11 & 5138  & 74  & 346         & a     \\
68 & F23394-0353 & MCG -01-60-022       & 23h42m00.91s & -03\degr36\arcmin54\farcs4 & 11.15 & 6966  & 100 & 464         & \ a     
\enddata
\tablecomments{(1): Unique identifier for each IRAS system. (2): IRAS system name;
(3): Commonly used galaxy name; (4) \& (5): J2000 coordinates for galaxy based on
\textit{Spitzer} IRAC 8$\mu$m imaging \citep[Mazzarella, in prep.;][]{Chu17}.
(6): 8 - 1000 $\mu$m infrared luminosity in solar units. (7) Heliocentric
velocity from the NASA/IPAC Extragalactic Database (NED). (8) \& (9): Luminosity distance
of the system and physical scale corresponding to 1\arcsec at the distance of the system, calculated using the 3-attractor model \citep{mould00} and Ned Wright's
Cosmology Calculator \citep[][]{wright06}, based on values from (4), (5) and (7). (10): Merger
stage based on visual classification, from \cite{stierwalt13}: a - pre-merger; b
- early-merger; c - mid-merger; d - late-merger; N - isolated galaxy; see
\cite{stierwalt13} for more details.}
\end{deluxetable*}
\subsection{Observations \& Data Reduction}
\begin{figure*}[htb]
    \centering
    \includegraphics[scale=0.8]{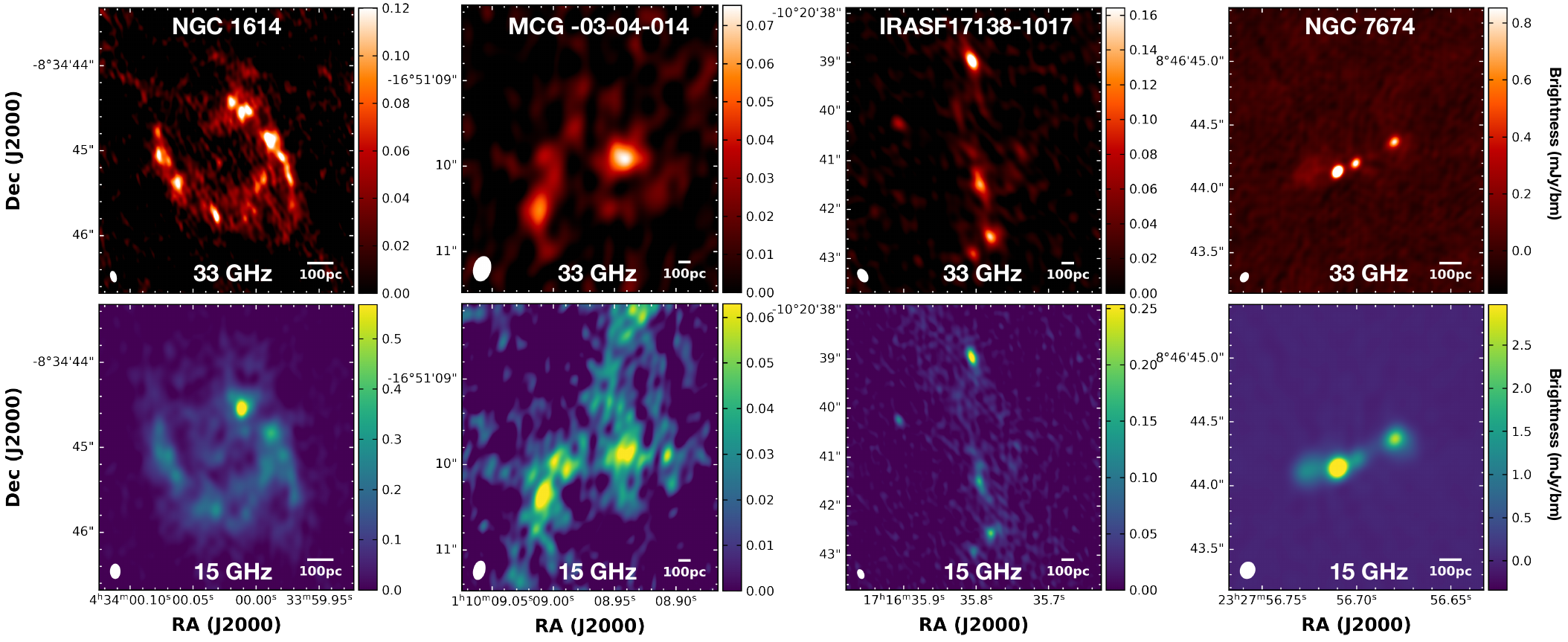}
    \caption{Examples of native resolution 33 (top) and 15 GHz (bottom) images used in this work. Each image is displayed in linear stretch with bilinear interpolation, and the colorbars show the brightness of the radio continuum emission, in the range of 0 to 80\% of the peak pixel value, in units of mJy\,beam$^{-1}$. Synthesized beams (lower left) and scale bars of 100 pc (lower right) are shown.\label{fig:example}}
\end{figure*}

The VLA observations for the GOALS-ES utilizes three receiver
bands: \textit{S}-band (2--4 GHz), \textit{Ku}-band (12--18 GHz) and
\textit{Ka}-band (26.5--40 GHz), which has enabled us to sample a 
wide frequency range for
characterizing the radio spectral energy distribution (SED). Each target was observed at each band in both \textit{A}-configuration (synthesized beam FWHM $\sim0\farcs06 - 0\farcs6$) and \textit{C}-configuration (synthesized beam FWHM  $\sim 0\farcs6 - 7\farcs0$) to detect bright compact regions as well as large-scale diffuse structures. Ten systems in the sample were additionally observed with
\textit{Ka}-band in \textit{B}-configuration (beam FWHM $\sim 0\farcs2$) due to
poor \textit{A}-configuration detections. In this work, we focus on analyzing
observations taken at \textit{Ku}-band (15 GHz) and \textit{Ka}-band (33 GHz) in
\textit{A}- and/or \textit{B}-configuration where the nuclear star-forming
structures are resolved at sub-kpc scales at the distances of these U/LIRGs (i.e. $D_L\sim 100\,$Mpc, $1\arcsec \sim 500$\,pc).
For simplicity, here we only provide descriptions on these relevant observations
and datasets. Information on \textit{C}-configuration observations and datasets
can be found in \cite{linden19}. A comprehensive description and atlas for
all VLA images from the GOALS-ES will be presented in a
forthcoming paper.\\
\indent \textit{A}-configuration observations at \textit{Ku}-band and
\textit{Ka}-band were initially carried out in two separate \textit{A}-configuration
cycles, in 2014 March 06 - May 10 (14A-471, PI: A. Evans), and 2016 October 7 - 12 (16A-204, PI: S.
Linden). For these observations, each galaxy was observed with 5-minute on-source time at \textit{Ku}-band, and 10 minutes at \textit{Ka}-band. Additional \textit{B}-configuration observations at \textit{Ka}-band were carried out in 2020 (20A-401, PI: Y. Song). These observations focused on ten systems with extended emission that were clearly detected at \textit{Ku}-band during the 14A-471 campaign, but had poor detections at \textit{Ka}-band due to limited sensitivity of the snapshots. Therefore, these ten systems were observed with longer on-source time ($\gtrsim$
30 minutes) to ensure good detections for comparison with \textit{Ku}-band
observations. Additionally, \textit{Ku}-band observations for six systems from 14A-471 were unsuccessful due to temporary malfunction of the requantizer, and were re-observed on December 10 (20B-313, PI: Y. Song) with 5 minutes on-source time. The project codes for the observations of each system used in this work are provided in Table \ref{tab:vlaimaging} in the Appendix.\\ 
\indent All raw datasets from project 14A-471 and 16A-204 were first reduced and calibrated into Measurement Sets (MS) using the Common Astronomy Software
Applications \citep[CASA;][]{mcmullin07} VLA data calibration pipeline (v4.7.0).
For observations from 20A-401 and 20B-313, we acquired the calibrated Measurement Sets
directly from the NRAO Science Ready Data Products (SRDP) data archive (CASA
v5.6.2 for 20A-401, v5.4.2 for 20B-313).\\
\indent We then visually inspected the calibrated MS, flagged bad
data related to RFI and specific antennae or channels, and then re-ran the appropriate versions of VLA pipelines on the flagged
MS without Hanning smoothing. We repeated this procedure until all bad data were
removed from the Measurement Sets.\\
\indent We proceeded to image each science observation using \texttt{tclean} in
CASA, utilizing the same versions that calibrations were performed with. In general,
we adopted Briggs weighting with a robust parameter of 0.5, using the Multi-Term (Multi-Scale)
Multi-Frequency Synthesis deconvolving algorithm \citep{rau11} with \texttt{scales} = [0, 10, 30] pixels and \texttt{nterm} = 2. In cases where sensitivity was
poor (peak S/N $<$ 10), Natural weighting or a robust parameter of 1.0 was
adopted instead to enhance sensitivity at the expense of the angular resolution. Cleaning masks were determined visually using the CASA \texttt{viewer}. Self-calibrations were not performed.\\
\indent We detected emission at SNR$ \gtrsim 5$ in 63 of the 68 GOALS-ES systems, at a resolution of
$\sim$ 0\farcs1 -- 0\farcs2 at 15 (\textit{Ku}) and/or 33 GHz (\textit{Ka}),
corresponding to $\sim$ 10 -- 160\,pc at the distances of these systems. To our
knowledge, this is the largest sample of local U/LIRGs that have been
observed at high ($> 10$ GHz) radio frequencies on $\sim$ 100\,pc scales. The characteristics of the native resolution images used in this paper are
listed in Table \ref{tab:vlaimaging}. In Figure \ref{fig:example}, we show several examples of the native resolution images used for our analysis. The full image atlas will be presented in the upcoming survey paper.
\subsection{Ancillary VLA Data}\label{sec:ancillary}
\indent To expand our study, in Section \ref{sec:discuss} we include comparisons between properties of compact radio continuum sources detected in the GOALS-ES systems and of those detected in other local U/LIRGs and nearby normal galaxies. To do this, we utilize VLA continuum observations of 22 of the most luminous local U/LIRGs presented in \cite{barcos17} (hereafter BM17) and of 56 nearby normal galaxies from the SFRS \citep{murphy18,linden20}. \\
\indent Observations for BM17 were taken with all four VLA configurations at both 6 and 33\,GHz, but we only utilize the high-resolution (VLA \textit{A-} or \textit{B-}configuration) 33 GHz observations here to complement our the \textit{A-} or \textit{B-}configuration GOALS-ES observations. Observations for the SFRS were taken with the VLA in \textit{D-}configuration at 33\,GHz, \textit{C-}configuration at 15\,GHz and \textit{B-}configuration at 3\,GHz. We only use the 15 and 33\,GHz observations in this work for comparing with our GOALS-ES observations at the same frequencies. These ancillary VLA observations were reduced using CASA by BM17 and the SFRS team, and relevant details are provided in the original publications. The synthesized beams have FWHM $\sim 0\farcs06 - 0\farcs2$ for the BM17 images, and  FWHM $\sim 2\arcsec$ for the SFRS images. At the distances of the BM17 ($D_L \sim 170\,$Mpc) and SFRS galaxies ($D_L \sim 11\,$Mpc), these values corresponding to spatial resolutions of 20 -- 200\,pc and 30--290\,pc, respectively, which are similar to the 10 -- 160\,pc resolutions reached by the GOALS-ES observations.\\
\indent To ensure consistent comparisons, we re-analyzed these ancillary VLA images from BM17 and the SFRS using the same methods adopted here for the GOALS-ES images (see \S \ref{sec:analysis} and \S \ref{sec:results}). We present the results of these ancillary analysis in Appendix \ref{ap:ancillary}, and compare them with the GOALS-ES results (see \S \ref{sec:results}) in \S \ref{sec:discuss}.

\section{Analysis}\label{sec:analysis} 
\subsection{Regions identification \& measurements}
\begin{figure}
    \centering
    \includegraphics[scale=0.70]{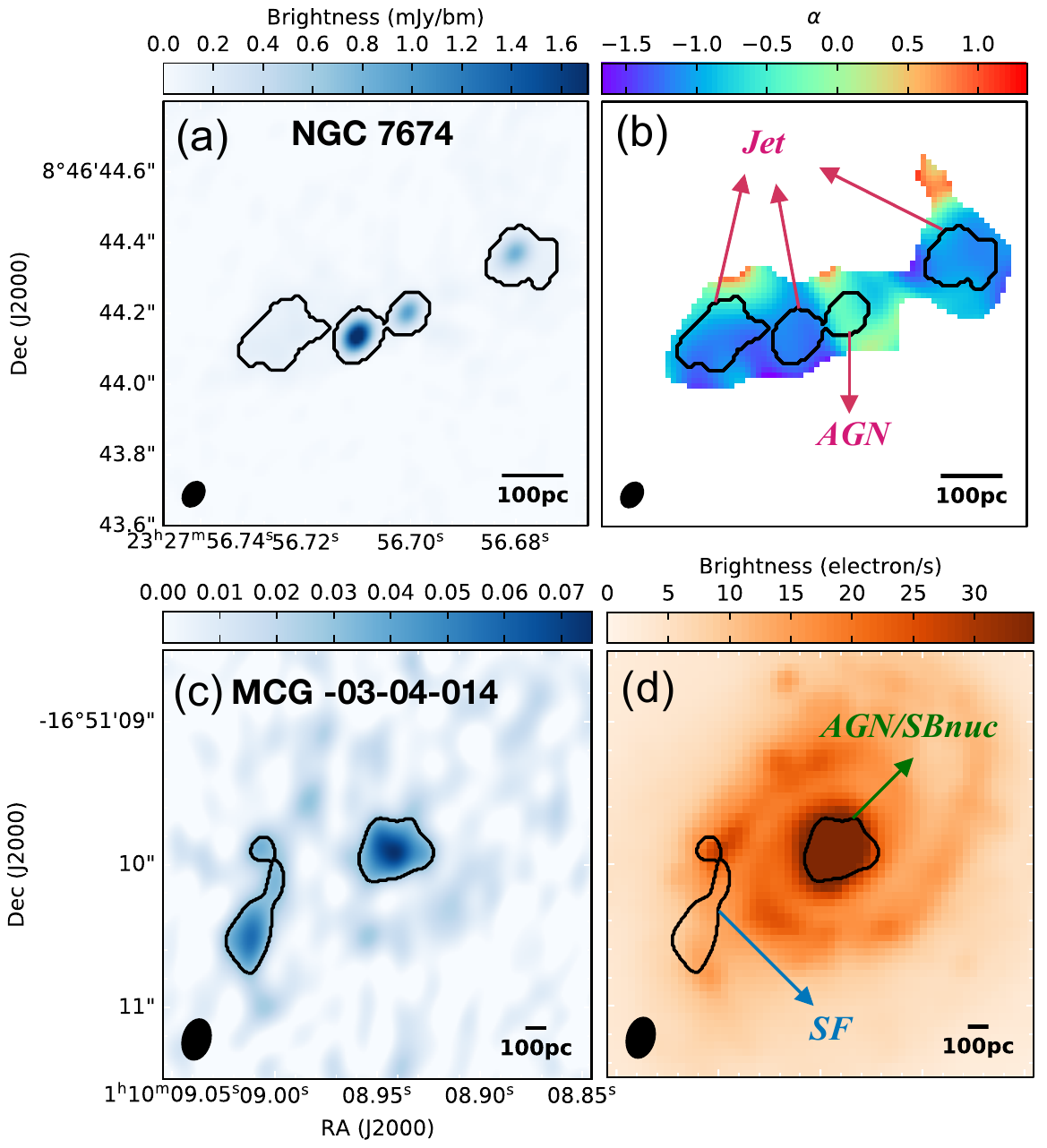}
    \caption{Images of NGC 7674 (a,b) and MCG-03-04-014 (c,d). These example images illustrate region identification, measurement and classification procedures described in \S \ref{sec:analysis} and \ref{sec:reg_type}. (a) \& (c): native-resolution 33 GHz images, as displayed in Figure \ref{fig:example}, with black contours outlining the areas of individual radio regions identified as \texttt{leaf} in a single run of \textit{Astrodendro}. (b) 15 - 33\,GHz spectral index map for NGC 7674 allows for the identification of synchrotron-dominated radio jets with steep spectra ($\alpha \sim -1$) and AGN with flat-spectrum ($\alpha \sim 0$). Black contours are the same as in (a). (d) Archival \textit{HST/NICMOS} F160W is used to locate the nucleus and off-nuclear star-forming region identified in MCG-03-04-014, as shown in (c). In (b) \& (d), region types classified following the methodology outlined in \S \ref{sec:reg_type} are labeled. In each panel, the black ellipse represents the native synthesized beam. \label{fig:dendro}}
\end{figure}

\indent To characterize the properties of compact radio sources detected in our VLA observations, we use the Python package \textit{Astrodendro} \citep{astrodendro} for region identification and measurements. \textit{Astrodendro} identifies and categorizes structures in an
image into \texttt{trunk}, \texttt{branch} and \texttt{leaf}, based on three
input parameters: the minimum brightness required for a structure to be
physically meaningful (\texttt{min\_value}), the minimum number of pixels in a
structure (\texttt{min\_npix}), and the minimum brightness relative to the local
background required for a structure to be considered independent
(\texttt{min\_delta}). Structures identified as \texttt{leaf} are of the highest
hierarchical order and are the independent regions of compact radio emission that we are interested in,
while \texttt{branch} and \texttt{trunk} are the surrounding relatively diffuse emission. \\
\indent To ensure that we only identify physically meaningful structures, we ran \textit{Astrodendro} on both the 15 and 33 GHz images of each system with \texttt{min\_value}$=5\sigma_\mathrm{rms}$ and
\texttt{min\_delta}=$1\sigma_\mathrm{rms}$ where $\sigma_\mathrm{rms}$ is the rms noise measured
in an emission-free region of the image before primary beam correction. We follow \cite{song21} and set
\texttt{min\_npix} to be a quarter of the area of the synthesized beam, to avoid identifying noise spikes yet allowing detection of small unresolved regions. Despite that extended diffuse emission is largely filtered out in these observations, complex structures encompassing \texttt{trunk}, \texttt{branch}, and \texttt{leaf} are identified in several systems. For our purpose of characterizing the most compact radio sources, we only focus on the identified \texttt{leaf} structures in subsequent analysis.  
\\
\indent Because the 33\,GHz radio continuum more directly traces thermal free-free emission from star formation \citep[e.g.][]{condon92}, in general we use \textit{Astrodendro} results derived at 33\,GHz for region identification and characterization. This also allows more robust constraints on the region sizes and surface brightness, given that 33\,GHz observations either have higher native resolutions than 15\,GHz observations, or better sensitivity (i.e. observations from 20A-401). In ten systems and NGC 5258 in Arp 240, only the 15\,GHz emission is bright enough to be identified via \textit{Astrodendro} at native resolutions, and hence 15\,GHz results were used instead. We also visually inspected all images and \textit{Astrodendro} results to ensure that any identified structures associated with image artifacts are excluded from further analysis.
\\
\indent To account for the image noise and its influence on size and
flux measurements of the identified regions, we
re-ran \textit{Astrodendro} 1000 times, randomly adjusting the brightness of each
pixel sampling from a Gaussian distribution defined by the rms noise
$\sigma_\mathrm{rms}$ and a VLA flux calibration error (10\%)\footnote{While the fundamental accuracy of flux density scale calibration is 3-5\%, here we conservatively assume an accuracy of 10\% instead since flux density calibrators and complex gain calibrators were not observed at similar elevations given the nature of our snapshot observations. See \href{https://science.nrao.edu/facilities/vla/docs/manuals/oss/performance/fdscale}{https://science.nrao.edu/facilities/vla/docs/manuals/oss\\/performance/fdscale}.}.
The standard deviations of the results from the 1000 runs are used to quantify
the uncertainties in measured flux densities and sizes. Figure \ref{fig:dendro} shows two examples of \textit{Astrodendro} output for a single run. Given that we only focus on the most compact and distinct clumps in these systems, the uncertainties introduced by the image noise are estimated to be low in general, on the order of a few percents. Additionally, large-scale diffuse emission that could more significantly influence the source identification and measurements is largely filtered out in the A-/B-configuration VLA observations used here.
\subsection{15 -- 33 GHz Spectral Index}
To better understand the nature of the identified regions of compact radio continuum emission, we measured the 15 -- 33 GHz radio spectral index
associated with each region, which can be used to estimate the relative contribution of thermal free-free emission to the total radio continuum emission at 33 GHz. To accomplish this, we smoothed and re-gridded the 15 and 33 GHz image of each galaxy to have a common
resolution and pixel scale. Next, we again performed \textit{Astrodendro} analysis on these beam-matched images following the same procedures described in the previous Section. For most regions, boundaries identified using the 33 GHz images were
used to measure their 15 and 33 GHz flux densities, which are then used to calculate the 15 - 33 GHz spectral index, $\alpha$, given by
the slope between 15 and 33 GHz flux measurements with respect to frequency:
\begin{equation}
    \alpha=\frac{\log S_{\nu_1} - \log S_{\nu_2}}{\log \nu_1 - \log \nu_2}
\end{equation} where $\nu_1 = 33$ GHz and $\nu_2 = 15$ GHz in our case. The uncertainty in $\alpha$ is calculated via error propagation, accounting for uncertainties in flux density measurements. For regions that were only identified at 15\,GHz, either due to limited sensitivity or intrinsic faintness at 33 GHz, boundaries identified from the 15 GHz images were used instead and hence the derived spectral indices are upper-limits. For each unresolved region that 
has an area smaller than the matched-beam, we use the flux density measured within a beam-sized aperture centered on the region to estimate its spectral index.

\section{Results}\label{sec:results}
For each of the 63 systems with detections, at least one region was identified using \textit{Astrodendro}. In total, we identified and characterized 133 regions at native resolutions at 33 and/or 15\,GHz, 19 of which are unresolved by the native beams. Because the 15 -- 33\,GHz matched-beams are 2-5 times larger than the native beams at 33\,GHz, distinct compact regions at native resolutions are blended together into larger, more extended regions at matched resolutions. Therefore, at matched resolutions, only 115 regions were identified, including 12 regions unresolved by the matched beams.\\
\indent To better distinguish regions identified at native and matched resolutions, for the rest of this paper, we use ``native regions" to refer to regions characterized at native resolutions, and refer to those characterized at matched resolutions as ``matched regions". All matched regions encompass at least one native region. \\
\indent In the following sections, we present the derived properties of the native and matched regions. First, we classify regions into different types on the basis of their AGN activity (\S \ref{sec:reg_type}). In \S \ref{sec:T_b} and \ref{sec:lum} we use measurements made for the native regions to constrain the brightness temperatures, physical sizes and luminosity surface densities of various region types. In \S \ref{sec:fth} and \ref{sec:sfr} we use measurements for the matched regions that are not associated with AGN activity to estimate their total and thermal-only SFR and surface densities. Measured and derived quantities for the native regions are presented in Table \ref{tab:native_reg}, and those for the matched regions are in Table \ref{tab:kaku_reg}.
\subsection{Region Classification}\label{sec:reg_type}
Before deriving the physical quantities associated with each region, it is crucial that we first identify the potential source powering the 33 and/or 15\,GHz radio continuum emission. Radio continuum emission at frequencies $> 30$\,GHz is widely used as a tracer of SFR \citep[e.g.][]{murphy12,murphy18}. Yet emission from AGN, if present, can completely dominate the observed radio emission at a physical scale of $\sim$ 100\,pc \citep{lonsdale03}, in which case the radio-derived SFR would be over-estimated. Further, separating AGN and SF-dominated regions first will allow us to more clearly examine and better understand the radio properties of each population. \\
\indent Although high brightness temperature ($T_b > 10^5$\,K) is typically used to identify radio AGN \citep[e.g.][]{condon91}, beam-dilution may reduce the brightness temperatures observed on 100\,pc scales to the level that is characteristic of starbursts (see \S \ref{sec:T_b}). Therefore, here we adopt a multi-wavelength approach to classify the native and matched regions characterized in \S \ref{sec:analysis} into different categories based on whether or not they may contain energetically-dominant AGN. We describe this two-step procedure below, which is illustrated in Figure \ref{fig:dendro} and summarized in Figure \ref{fig:flowchart}, and provide more details on individual sources in Appendix A.
\subsubsection{Region Location}
\indent As a first step, we separate regions into three initial categories -- ``nuclear", ``off-nuclear" and ``extra-nuclear" -- based on their relative
location in their host galaxies. These locations are determined visually by first overlaying the 33 and/or 15 GHz radio images on top of optical $y$-band images of the host galaxy from PanSTARRS1 \citep{chambers16, flewelling20} as well as \textit{Spitzer} IRAC channel 1 and channel 4 maps \citep[J. Mazzarella in prep.;][]{goalsdoi}. Afterwards, we overlay an ellipse representing the size of the unresolved Mid-IR (MIR, $\lambda=13.2\,\mu$m) galaxy ``core" reported in \cite{diaz-santos10}, which is the FWHM of the Gaussian fit to the \textit{Spitzer} IRS spectra of the galaxy that have spatial resolutions of $\sim 3\farcs6$. The MIR traces warm dust emission ($\sim 300\,K$) from obscured starburst and/or AGN activity, and hence provides useful constraints on the spatial extent of the most energetic component of the galaxy. The ellipse is then projected using galaxy position angles provided in the HyperLeda database \citep{hyperleda} and the Two Micron All Sky Survey (2MASS) Extended Source Catalog \citep{2mass,2massdoi}, along with galaxy inclination derived from galaxy axis ratio reported in \cite{kim13} and \cite{jin19} using the recipe given by \cite{dale97}.\\
\indent In general, we found agreement between the astrometry of the multi-wavelength images within a few arcseconds. Regions that spatially coincide with the optical and MIR galaxy peak are considered to be the galactic nuclei and hence are classified as ``nuclear". Regions that are not ``nuclear" but also lie within the MIR galaxy core are ``off-nuclear", and regions lying completely outside of the MIR galaxy core are ``extra-nuclear". In II Zw 096, the identified region is co-spatial with the brightest MIR component that has previously been identified as a powerful starburst region triggered on the outer edge of the merging galaxy pair \citep{inami10, inami22}, therefore we classify it as an ``extra-nuclear" region. Regions residing within the MIR galaxy core (i.e. ``nuclear" and ``off-nuclear") are labelled with ``n" in Table \ref{tab:native_reg} and \ref{tab:kaku_reg} (column 2), and ``extra-nuclear" regions are labelled with ``e".\\
\indent Due to the comparatively low spatial resolution of the Pan-STARRS1 and IRAC images, determining whether a given region is ``nuclear" or ``off-nuclear" can be challenging when there are multiple regions within the MIR galaxy core. For seven galaxies, we were able to rely on direct comparisons with high-resolution \textit{HST} and/or ALMA datasets publicly available from the archives to pinpoint the location of the galactic nuclei (often the kinematic center) and hence separate ``nuclear" and ``off-nuclear" regions. For 17 native regions residing in eight U/LIRGs without sufficient ancillary information from high-resolution imaging and/or gas kinematics, we assign them a final type ``Ud" (undetermined) in Table \ref{tab:native_reg} and \ref{tab:kaku_reg} (column 3). Images of these eight systems are shown in the Appendix. We carry out further classification for the remaining 116 native regions (57 ``nuclear", 49 ``off-nuclear", 10 ``extra-nuclear") in the following section.
\subsubsection{Host AGN Classification}
\begin{figure}[htb]
    \centering
    \includegraphics[scale=0.4]{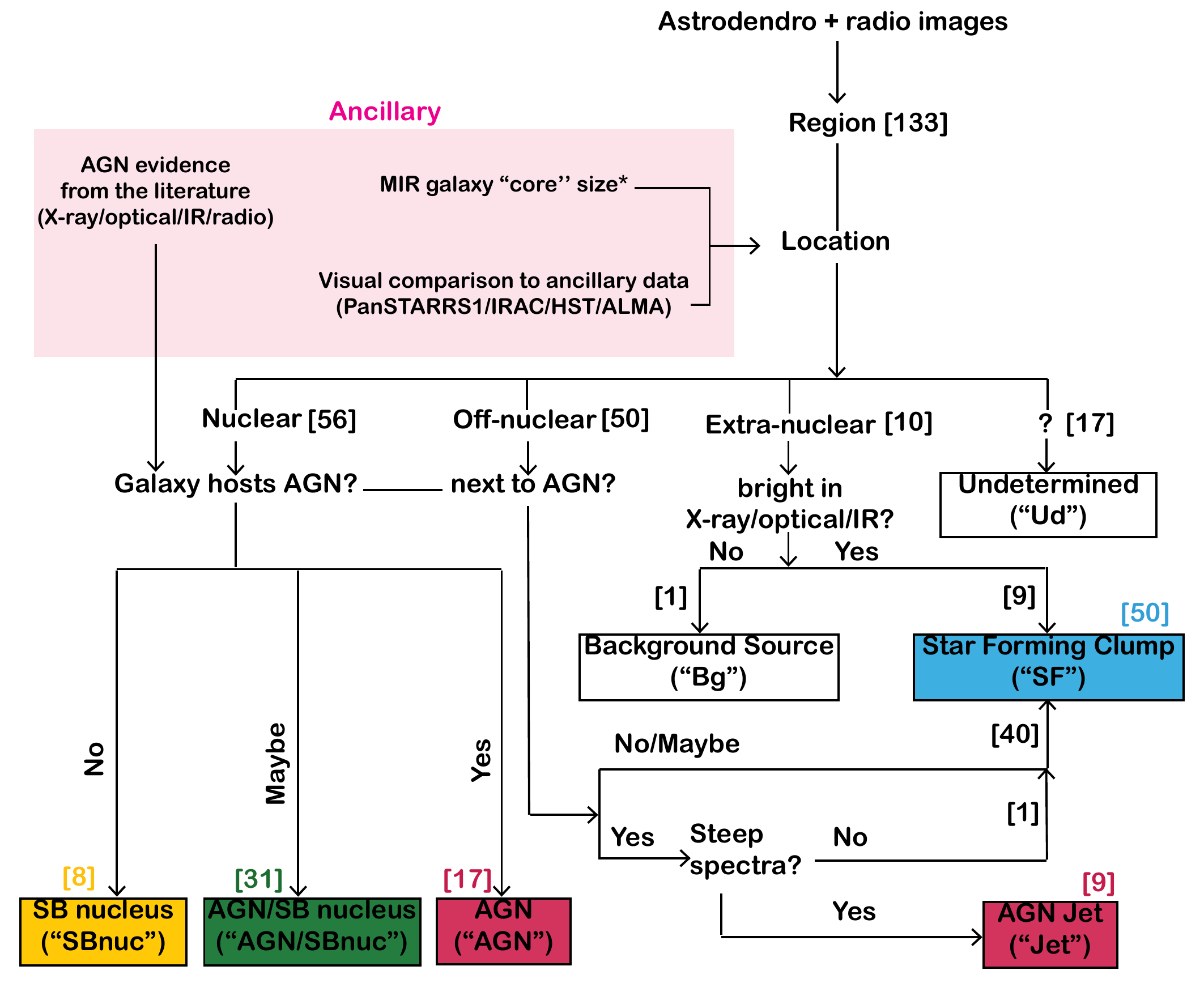}
    \caption{Region classification scheme described in \S \ref{sec:reg_type} and demonstrated in Figure \ref{fig:dendro}. Descriptions of individual galaxies are provided in Appendix A, including their multi-wavelength AGN classifications and and the ancillary datasets used. Numbers of native regions classified at each step are shown in brackets. $^*$MIR galaxy ``core" sizes are measured by \cite{diaz-santos10}. \label{fig:flowchart}}
\end{figure}

\indent For the next step, we search in the literature for multi-wavelength (i.e. X-ray, optical, MIR, radio/sub-mm) evidence for AGN presence in each of the 63 U/LIRGs with detections, summarized in Table \ref{tab:AGN} in the Appendix. Mainly, we build upon optical classifications by \cite{veilleux95} and \cite{yuan10}, as well as results from previous surveys of local U/LIRGs with \textit{NuSTAR} \citep[e.g.][]{ricci17,ricci21}, \textit{Chandra} \citep[][]{iwasawa11,torres-alba18}, \textit{Spitzer} \citep[e.g.][]{petric11,stierwalt13}, \textit{AKARI} \citep[e.g.][]{inami18}, \textit{VLA} \citep[e.g. at 1.4 and 8.4\,GHz;][]{condon95,vardoulaki15} and \textit{VLBA} \citep[e.g.][]{smith98}. The compiled information is used in combination with the initial location classifications to further narrow down whether a region may contain an AGN that could dominate the radio emission: \\
\indent \textendash \ For ``nucler'' regions: if the host galaxies have been identified as hosting AGN in the literature at more than one wavelength range, we classify them as ``AGN''. For example, we classify the nucleus of NGC 0034 as “AGN”, given that the host galaxy
is classified as a Seyfert 2 galaxy based on optical line ratios \citep[][]{veilleux95, yuan10} and as an obscured AGN based on Chandra X-ray analysis revealing excess hard-band X-ray emission and an absorbing column density $N_{\rm H} \sim 10^{23}\rm {cm}^{-2}$ \citep{torres-alba18}. If the host galaxy has only been identified as AGN at one wavelength range but lacks identification at other wavelengths, or if evidence for AGN is ambiguous or inconsistent across all wavelengths, we classify the ``nuclear" regions as ``AGN/SBnuc''. For example, the nucleus of IRAS F17138-1017 is classified as ``AGN/SBnuc", because the X-ray spectral shape of the host galaxy is consistent with either star formation or an obscured AGN \citep{ricci17, torres-alba18}, and the galaxy is classified as LINER in the optical, which may be powered by low-luminosity AGN, evolved stars, or both \citep[][]{singh13}. Another example is the nucleus in MCG-03-04-014, which we classify as ``AGN/SBnuc" given that the observed nuclear optical line ratios indicates emission from both AGN and starburst \citep{yuan10}, despite that the 3.3 and 6.2\,$\mu$m PAH feature have large equivalent widths consistent with starburst-dominated emission \citep{stierwalt13,inami18}. Lastly, if no AGN evidence has been found at any wavelength range for the host galaxy, we then classify the region as starburst-dominated nucleus (``SBnuc''). An example of this is NGC 5257 (Arp 240N). \\
\indent \textendash \ For ``off-nuclear'' regions: if they form a linear structure with an identified ``AGN'' and show steep 15 - 33 GHz spectral indices ($\alpha \lesssim -0.8$) indicative of synchrotron-dominated emission \citep[e.g.][]{condon91}, we classify them as radio jets (``Jets'') associated with the AGN. An example of this is NGC 7674 (see Figure \ref{fig:dendro}). While ``off-nuclear" regions next to ``AGN/SBnuc" may be jets from unconfirmed AGN or star-forming clumps, we classify them as ``SF" given that AGN with jets tend to dominate the nuclear emission and likely would have been identified as AGN at multiple wavelengths. This reasoning has been adopted to classify the ``SF" regions in IC 1623B, MCG-03-04-014, CGCG 436-030, III Zw 035 and IRAS F17138-1017, and we note that these regions also all have optical/IR counterparts. While highly-energetic optical/IR synchrotron jets have been observed in powerful quasars \citep[e.g.][]{floyd06a,floyd06b}, we argue that this scenario is unlikely given the lack of clear AGN evidence reported for the above U/LIRGs in our sample.\\
\indent \textendash \ For ``extra-nuclear'' regions: if they are detected in the X-rays or have visible optical/IR counterparts, we classify them as star-forming regions (``SF''). Only one ``extra-nuclear" region, in IC 0214, does not show any X-ray, optical or IR counterpart. Hence, it is likely a background radio source (``Bg") that is not associated with the galaxy and therefore eliminated from further analysis. \\
\indent In summary, out of the 116 native regions with identified locations (i.e. not ``Ud"),  17 ``AGN", 9 ``Jet", 8 ``SBnuc", 31 ``AGN/SBnuc", 50 ``SF" (41 ``off-nuclear" and 9 ``extra-nuclear") regions are classified, excluding one ``Bg" region detected near IC 0214. At matched resolutions, many ``off-nuclear" native regions are blended with the ``nuclear" native regions. In these cases, the larger blended matched regions are designated with the ``nuclear" classifications (i.e. ``AGN", ``AGN/SBnuc", ``SBnuc"). As a result, 17 ``AGN", 6 ``Jet", 30 ``AGN/SBnuc", 8 ``SBnuc" and 40 ``SF" (32 ``off-nuclear" and 8 ``extra-nuclear") matched regions are classified. Figure \ref{fig:flowchart} summarizes our region classification scheme, and all region classifications are reported in Table \ref{tab:native_reg} and \ref{tab:kaku_reg}, and described in Appendix A in more detail. We note that only two systems in the sample (MCG-03-34-064, NGC 7674) have previously been classified as ``radio-loud" AGN based on the excess radio over FIR emission on galaxy scales \citep[][]{condon91q,condon95}, which emphasizes the necessity of the above two-step approach in constraining the sources of radio emission at resolved scales. In the upcoming survey paper we will further investigate the kpc-scale radio-IR correlations in the GOALS-ES systems for the different region types classified here.

\subsection{Brightness Temperature}\label{sec:T_b}
\begin{figure*}[!tbhp]
    \centering
    \includegraphics[scale=0.5]{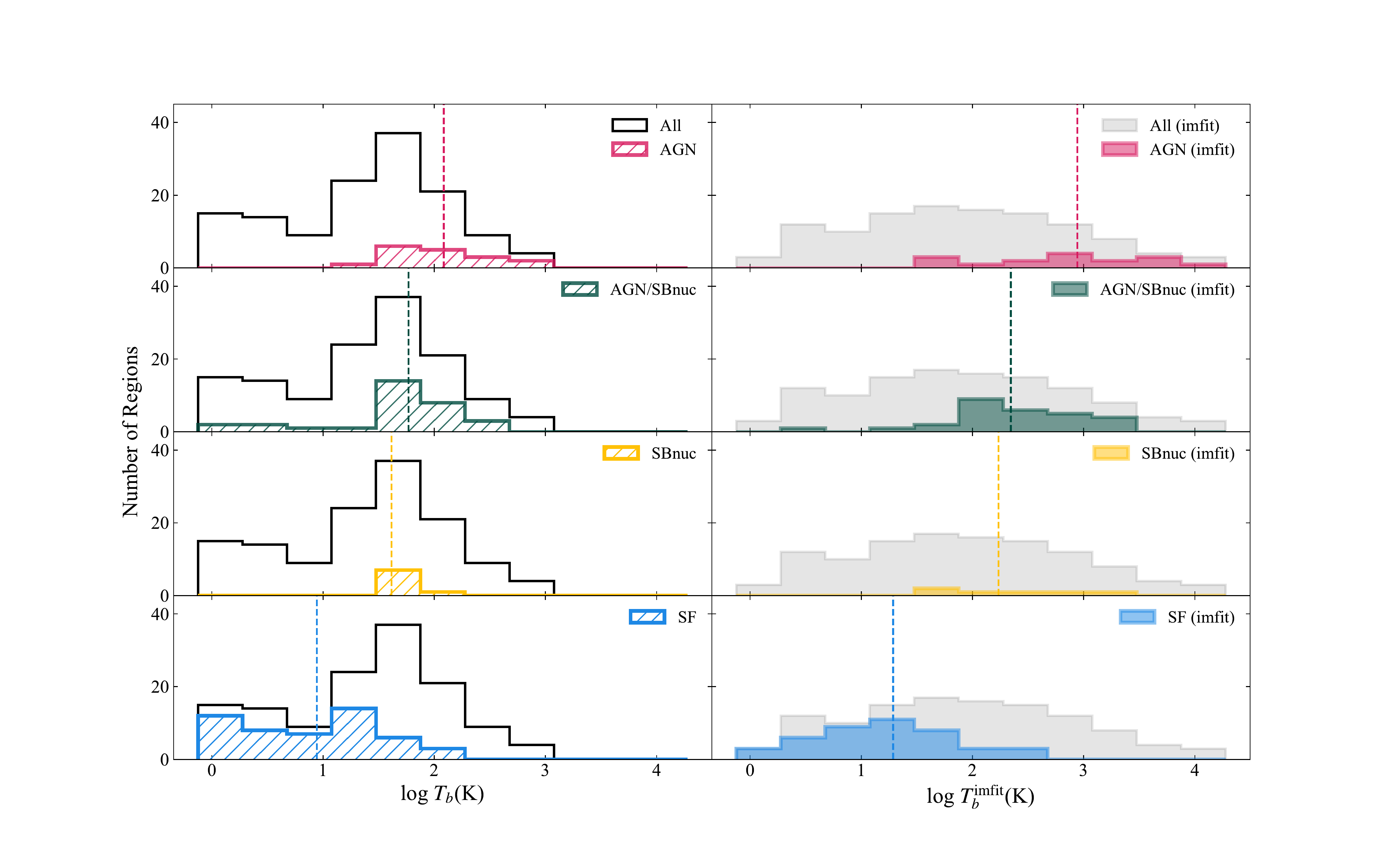}
    \caption{The distribution of brightness temperatures of regions identified at native resolutions at 15 or 33\,GHz. \textit{Left:} values derived from \textit{Astrodendro} measurements of region areas and flux densities. The distribution for all 133 native regions are shown in black un-filled histogram. \textit{Right:} values derived from Gaussian-fitting results using CASA \texttt{imfit} task. The distribution for 116 regions with successful fits are shown in black filled histogram. In both panels, distributions for ``AGN" (magenta), ``AGN/SBnuc" (green), ``SBnuc" (yellow) and ``SF" (blue) are shown separately for comparison, with dashed lines marking the median values. Overall, ``AGN" have the highest brightness temperatures and Gaussian-fitting yields higher values, but only one ``AGN" (in NGC 1068) exceeds the maximum starburst brightness temperature ($\sim 10^4$\,K).
    The measured low brightness temperature for AGN is likely due to beam dilution. \label{fig:Tb}}
\end{figure*}

\indent \cite{condon91} derived the maximum brightness temperature $T_b$ for an optically-thick starburst radio source to be:
\begin{equation}\label{eq_Tb0}
    T_b \leq T_e \left({1 + 10 (\frac{\nu}{\rm GHz})^{0.1+\alpha_{\rm NT}}} \right)
\end{equation}
where $T_e \sim 10^4$\,K is the thermal electron temperature characteristic of massive star-formation, $\nu$ is the radio frequency at which measurements are made, and $\alpha_{\rm NT}$ is the non-thermal spectral index characteristic of synchrotron emission generated by electrons accelerated by Type II supernovae. This limit allowed \cite{condon91} to confirm the AGN nature of the compact radio source they detected in UGC 08058 (Mrk 231), which has $T_b \gtrsim 10^{6}$\,K at 8.4\,GHz. Here we assume $\alpha_{\rm NT} \sim -0.85$, based on resolved measurements of star-forming regions in the nearby disk galaxy NGC 6946 by \citep[][]{murphy11}, which gives a maximum starburst $T_b$ of $10^{4.2}$\,K at 33 GHz, and $10^{4.4}$\,K at 15 GHz. \\
\indent Using the Rayleigh-Jeans approximation, the brightness temperature $T_b$ of each native region can be calculated via \citep{condon92, perez-torres21}:
\begin{align}\label{eq_Tb}
    T_b & = \left( {\frac{S_\nu}{\Omega}} \right)\frac{c^2}{2k\nu^2} \nonumber\\
        & \simeq 1.6 \times 10^3 \left( {\frac{S_\nu}{\rm mJy}} \right)\left( {\frac{\nu}{\rm GHz} } \right)^{-2}\left( { \frac{\theta_M\theta_m}{\rm arcsec^{2}}} \right)\ \rm K
\end{align}
where $c$ is the speed of light, $k$ is the Boltzmann constant, $S_\nu$ is the region flux density measured at frequency $\nu$, $\Omega=\pi\theta_M\theta_m/(4\ln{2})$ is the region area assuming a Gaussian morphology, with $\theta_M$ and $\theta_m$ corresponding to the FWHM of the major and minor axis of the Gaussian. \\
\indent Because the identified regions have irregular morphology with unknown sub-beam structures, here we calculate the brightness temperature of each native region using two different methods. First, we use region flux density and area measured with \textit{Astrodendro} for $S_\nu$ and $\Omega$ in Equation \ref{eq_Tb}. Second, we perform Gaussian fitting and deconvolution on all native-resolution images using CASA task \texttt{imfit} and \textit{Astrodendro} results as initial guess inputs, assuming uniform background noise level as characterized by $\sigma_{\rm rms}$ (see Table \ref{tab:vlaimaging}), and calculate the deconvolved brightness temperature $T^{\rm imfit}_b$ of each region using the flux density of the fitted Gaussian model and the deconvolved $\theta_M$ and $\theta_m$, following \cite{condon91}. We note that by assuming a simple Gaussian morphology, the latter method allows tighter constraints on the intrinsic sizes of marginally-resolved regions, but does not reflect the observed diverse region morphology or the varying degree of surrounding diffuse emission present in each system, which leads to poor flux recovery especially for extended regions. Gaussian-fitting was also unsuccessful for 17 regions in 9 systems. Therefore we use the latter method only in this Section to illustrate the possible effect of beam dilution, but continue to use results derived with \textit{Astrodendro} (Section \ref{sec:analysis}) throughout the rest of the paper. Values of $T_b$ and $T^{\rm imfit}_b$ are reported in Table \ref{tab:native_reg} and compared in Figure \ref{fig:Tb}. \\
\indent Figure \ref{fig:Tb} shows that, regardless of the method used,
``AGN" and ``AGN/SBnuc" have higher brightness temperatures than ``SF". The
ranges of $T_b$ for ``AGN", ``AGN/SBnuc" and ``SF" are 19 -- 950\,K, 1 -- 360\,K
and 1 -- 160\,K, respectively. However, all regions, including ``AGN", have $T_b
< 10^{4}\,$K. For ``SF", the overall low $T_b$ is expected from optically-thin
emission associated with star formation. For ``AGN" and ``AGN/SBnuc", the
observed emission may come from a combination of AGN emission and nuclear star
formation, which may be further diluted by the beam. The effect of such beam
dilution is also demonstrated in Figure \ref{fig:Tb}, where we see that $T^{\rm
imfit}_b$ for ``AGN" (50 -- 11900\,K), ``AGN/SBnuc" (3 -- 2500\,K) and ``SBnuc"
(70 -- 1710\,K) are higher than $T_b$ by up to $\sim 1$\,dex. Nevertheless, no
``AGN" has $T^{\rm imfit}_b$ greater than $10^6$\,K, and the only ``AGN" with
$T^{\rm imfit}_b > 10^{4}$\,K is the one in NGC 1068, which is the most nearby
Seyfert in GOALS. Additionally, 19 regions (including the ``AGN/SBnuc" in IRAS
17208-0014) are unresolved by the beam, 7 of which, along with 12 other regions (including 3 ``AGN" and 2
``AGN/SBnuc") are separately determined as point-sources by CASA \texttt{imfit}.
For these regions, the calculated $T_b$ and $T^{\rm imfit}_b$ are lower-limits
and indicated in Table \ref{tab:native_reg}. \\ 
\indent Our results are similar to those found by \cite{barcos17}, who
measured an overall low $T_b$ ($\sim 100 - 1000$\,K) in the nuclei of the most
luminous local U/LIRGs at 100\,pc scales at 33 GHz. These results also
demonstrate the limitation of the current VLA observations for directly
identifying AGN using brightness temperatures. Future VLBI follow-up of the
``AGN/SBnuc" regions would significantly improve our ability to identify AGN in
many more local U/LIRGs, as well as isolate AGN emission from the compact
circumnuclear star formation prevalent in these systems
\citep[e.g.][]{condon91}.
\subsection{Size, Luminosity and Luminosity Surface Density}\label{sec:lum}
\begin{figure*}[tbhp]
    \centering
    \includegraphics[scale=0.6]{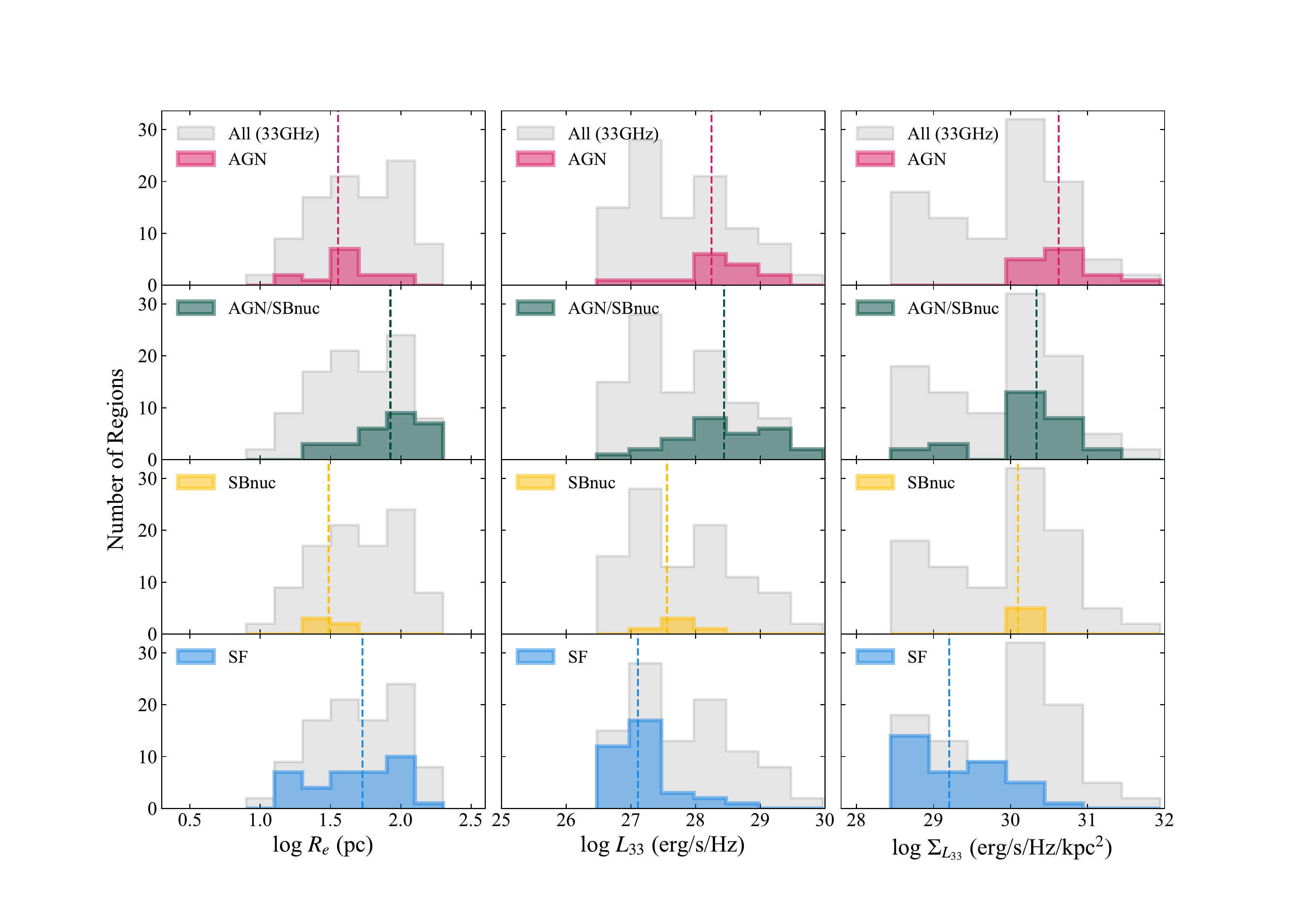}
    \caption{The distribution of derived properties of regions identified and characterized at native resolutions using \textit{Astrodendro}. For direct comparison, we only show results derived from 33\,GHz measurements, available for 99 native regions, excluding the ``Bg" source in IC 0214 and 17 unresolved regions. \textit{Left:} Effective radius $R_e$. \textit{Middle:} 33 GHz luminosity $L_{33}$. \textit{Right:} 33 GHz luminosity surface density $\Sigma_{L_{33}}$. In all three columns, we also show distributions for ``AGN" (magenta), ``AGN/SBnuc" (green), ``SBnuc" (yellow) and ``SF" regions (blue), and the corresponding median values (dashed lines), overlaid on the distributions of all 99 native regions (grey). Results for ``Jet" and ``Ud" regions are not separately shown for simplicity. While all regions types are distributed over similar size ranges, ``AGN" and ``AGN/SBnuc" have higher $L_{33}$ and $\Sigma_{L_{33}}$ than ``SF" by an order of magnitude. \label{fig:native_dist}}
\end{figure*}

For each native region, \textit{Astrodendro} can be used to measure its angular area $A$ and flux density $S_\nu$ at frequency $\nu$ using the region boundary identified by the algorithm (i.e. black contours in Figure \ref{fig:example}).  We use the mean values of these measurements from 1000 runs of \textit{Astrodendro} (see \S \ref{sec:analysis}) to calculate the spectral luminosity ($L_\nu = 4\pi S_\nu D^2_L$), effective radius ($R_e = \sqrt{(A/\pi)}$), and spectral luminosity surface density $\Sigma_{L_{\nu}}$ of each region, using the luminosity distance ($D_L$) and angular-to-physical conversion factor derived for each system, as listed in Table \ref{tab:sample}. For the 19 native regions with areas smaller than the synthesized beams even after accounting for uncertainties, we use the beam areas as upper-limit estimates for the region sizes, and thus the corresponding $\Sigma_{L_{\nu}}$ are lower-limits. The derived properties of a total of 133 native regions in 63 systems are reported in Table \ref{tab:native_reg}, of which 16 regions in 10 systems were measured at 15 GHz due to poor or non-detections at 33 GHz. In Figure \ref{fig:native_dist}, we show the distributions of the derived properties of 99 native regions with 33\,GHz measurements, excluding the ``Bg" in IC 0214 and 17 unresolved regions. \\
\indent The effective radii ($R_e$) of these 99 native regions span from 8 to 170\,pc, with no significant size differences among the region types, except for ``AGN/SBnuc" regions, which have the largest sizes at a median value of 80\,pc compared with $\sim 40\,$pc for ``AGN", ``SBnuc" and ``SF". As shown in Figure \ref{fig:native_dist}, the 33\,GHz luminosity ($L_{\rm 33}$) span three orders of magnitude, ranging from $3.0\times10^{26}$ to $3.4\times10^{29}$\,erg s$^{-1}$ Hz$^{-1}$. Unsurprisingly, ``AGN" regions are overall more luminous, with a $L_{\rm 33} = 8.0\times10^{26} - 1.7\times10^{29}$\,erg s$^{-1}$ Hz$^{-1}$ and a median of $1.7\times10^{28}$\,erg s$^{-1}$ Hz$^{-1}$, compared with ``SF" regions which have $L_{\rm 33} = 2.0\times10^{26} - 3.4\times10^{28}$\,erg s$^{-1}$ Hz$^{-1}$ and a median of $1.1\times10^{27}$\,erg s$^{-1}$ Hz$^{-1}$, about an order of magnitude lower. This difference is also evident in distribution of spectral luminosity surface density $\Sigma_{L_{\rm 33}}$: ``AGN" regions have $\Sigma_{L_{\rm 33}}$ ranging from $1.1\times10^{30}$ to $3.0\times10^{31}$\,erg s$^{-1}$ Hz$^{-1}$ kpc$^{-2}$ with a median of $4.2\times10^{30}$\,erg s$^{-1}$ Hz$^{-1}$ kpc$^{-2}$, which is also an order of magnitude higher than $2.3\times10^{29}$\,erg s$^{-1}$ Hz$^{-1}$ kpc$^{-2}$ for the ``SF" regions. When considering all 99 native regions, including 15 ``AGN", 9 ``Jet", 28``AGN/SBnuc", 5 ``SBnuc", 36 ``SF" (31 ``off-nuclear" and 5 ``extra-nuclear") and 9 ``Ud" regions, the median for $\Sigma_{L_{\rm 33}}$ is around  $1.1\times10^{30}$\,erg s$^{-1}$ Hz$^{-1}$ kpc$^{-2}$, below which the distribution is almost completely dominated by ``SF" regions. In \S \ref{dis:lum-size} we further discuss the implication of the differences we observe between the ``AGN" and ``SF" native regions, in the theoretical context of radiation feedback-regulated star formation in the dusty environments of U/LIRGs.
\subsection{Thermal Fraction at 33 GHz}\label{sec:fth}
Assuming a typical radio continuum SED for star-forming galaxies \citep[e.g.][]{condon92}, the 33 GHz radio continuum emission can be decomposed into thermal free-free emission with a flat spectrum ($S_\nu \propto \nu^{-0.1}$) and non-thermal synchrotron
emission with a steep spectrum ($S_\nu \propto \nu^{\alpha_\mathrm{NT}}$), where a non-thermal spectral index of $\alpha_\mathrm{NT} \sim -0.85$ has been found to be widely applicable in resolved star-forming regions detected in nearby disk galaxies \citep{murphy11,murphy12}. For each matched region, we derive the 33 GHz thermal fraction $f_\mathrm{th}$, which measures the fractional contribution of thermal free-free emission generated from plasma around massive young stars  (i.e. HII regions) using the measured 15 - 33 GHz spectral index $\alpha_{\rm 15 - 33}$ (see Section \ref{sec:analysis}), and Equation (11) from \cite{murphy12}:
\begin{equation}\label{eq_fth}
    f_{\rm th}=\dfrac{\Big(\frac{\nu_2}{\nu_1}\Big)^{\alpha} - \Big(\frac{\nu_2}{\nu_1}\Big)^{\alpha_{\mathrm{NT}}}}{\Big(\frac{\nu_2}{\nu_1}\Big)^{-0.1} - \Big(\frac{\nu_2}{\nu_1}\Big)^{\alpha_{\mathrm{NT}}}}
\end{equation}
where we set the spectral index $\alpha$ between $\nu_1$ and $\nu_2$ (33 and 15 GHz) to be our measured $\alpha_{\rm 15-33}$, and use error propagation to derive the uncertainties associated with flux calibration and image noise levels. We note that 17 matched regions were not identified with \textit{Astrodendro} at 33 GHz due to insufficient sensitivity, so the measured $\alpha_{\rm 15 - 33}$ for these regions are likely steeper than the intrinsic values. We label these values in Table \ref{tab:kaku_reg} as upper-limits, and mark the host systems with ``*".\\
\indent Of the 97 matched regions that were identified at both 15 and 33 GHz (excluding ``Bg" in IC 0214), 10 regions have steep spectra with $\alpha_{\rm 15-33} \lesssim -0.85$ after accounting for the estimated uncertainties. These regions include four ``Jet", three ``SF" and three ``AGN/SBnuc". The observed 33\,GHz emission in these regions are likely dominated by non-thermal synchrotron emission produced by relativistic electrons accelerated in AGN jets or supernovae. The ``AGN/SBnuc" in UGC 02238 and NGC 5104 have the steepest spectra, with $\alpha_{\rm 15-33} \sim -1.6\pm0.3$. In these cases we follow \cite{linden20} and set $\alpha_{\mathrm{NT}}=\alpha_{\rm 15-33}$, which gives $f_{\rm th} \sim 0$\%, on the basis that negative $f_{\rm th}$ are not physically meaningful. For three ``SF" regions, IC1623B\_n4, NGC5257\_e1 and IC2810\_e1, $\alpha_{\rm 15-33} \gtrsim 0$ after accounting for uncertainties, which is unexpected from optically-thin thermal free-free emission. Given that all ``SF" regions have brightness temperatures much lower than the optically-thick starburst temperature of $\sim$10$^4$\,K (see \S \ref{sec:T_b}), a potential cause for the higher than expected 33 GHz continuum flux may be anomalous microwave emission from spinning dust particles in heavily-obscured young starburst \citep{murphy20}. This possible explanation will require more high-resolution observations above and below 33 GHz to confirm. We note that the extra-nuclear region in NGC 5257 also shows the flattest 3 - 33 GHz spectrum among 48 extra-nuclear regions hosted in 25 U/LIRGs in the equatorial sample when measured on kpc scale, consistent with \cite{linden19}. For regions with $\alpha_{\rm 15-33} \gtrsim -0.1$ we adopt $f_{\rm th} \sim 100$\%.
\begin{figure}[htb!]
    \centering
    \includegraphics[scale=0.6]{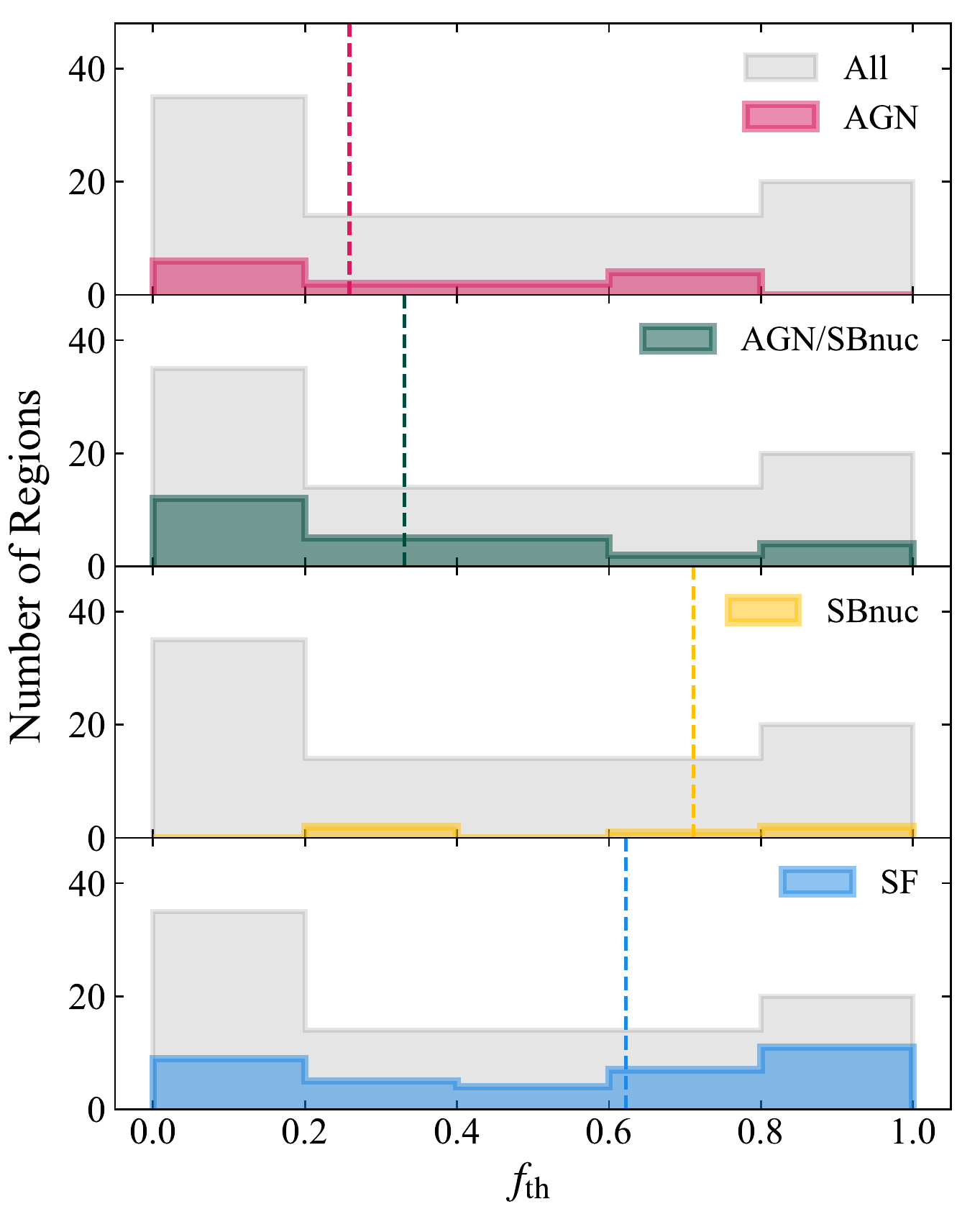}
    \caption{The distribution of the derived 33\,GHz thermal fraction, $f_{\rm th}$, for 97 matched regions identified at both 15 and 33\,GHz (in gray) excluding ``Bg", and for ``AGN" (magenta), ``AGN/SBnuc" (green), ``SBnuc" (yellow) and ``SF" regions (blue). The median values are $\sim$ 25\%, 33\%, 71\% and 62\%, respectively, and shown in dashed lines. Results for ``Jet" and ``Ud" regions are not separately shown. While $f_{\rm th}$ spans a wide range for all region types compared, ``AGN" and ``AGN/SBnuc" have lower  median $f_{\rm th}$ than ``SBnuc" and ``SF" regions. \label{fig:fth_dist}}
\end{figure}

\indent Figure \ref{fig:fth_dist} shows the distribution of $f_\mathrm{th}$ for all 97 matched regions as well as for different region types, which all span a wide range from $\sim$ 0\% (dominated by non-thermal emission) to 100\% (dominated by thermal emission). However, the median values for ``AGN" and ``AGN/SBnuc", at $f_{\rm th}\sim 30$\%, are noticeably lower than those for ``SBnuc" and ``SF", at $f_{\rm th}\sim 65\%$. This result is consistent with kpc-scale measurements of extra-nuclear star-forming regions by \cite{linden19} using GOALS-ES \textit{C-}configuration observations. For ``AGN" and ``AGN/SBnuc", mechanisms other than star formation may be producing excess non-thermal emission at 33\,GHz \citep[e.g.][]{panessa19}. Overall, the wide range of $f_\mathrm{th}$ spanned by different region types demonstrates that spectral shape and the derived $f_\mathrm{th}$ alone are insufficient for inferring the nature of radio emission in a given region at 100\,pc scales. In \S \ref{dis:fth} we further discuss the potential mechanisms that may be contributing to the 15 -- 33\,GHz radio continuum emission in these local U/LIRGs at 100\,pc scales.

\subsection{Star Formation Rates and Surface Densities}\label{sec:sfr}
For all matched-resolution ``SF" and ``SBnuc" regions, we use Equation (10) in \cite{murphy12} to convert the measured 33 or 15 GHz continuum flux density to a total star formation rate (SFR), accounting for both thermal free-free emission from HII regions ($< 10\,$Myr) and non-thermal synchrotron emission from supernovae ($\sim$\,10--100\,Myr):
\begin{align}\label{eq_sfr}
     \Big(\frac{\mathrm{SFR}}{\textup{M}_\odot \mathrm{yr}^{-1}}\Big)  & = 10^{-27}\Big[2.18\Big(\frac{T_e}{10^4 \mathrm{K}}\Big)^{0.45}\Big(\frac{\nu}{\textup{GHz}}\Big)^{-0.1} + \notag\\ 
    & 15.1\Big(\frac{\nu}{\textup{GHz}}\Big)^{\alpha^{\textup{NT}}}\Big]^{-1}\Big(\frac{L_\nu}{\textup{erg} \textup{s}^{-1} \textup{Hz}^{-1}}\Big),
\end{align} 
where a Kroupa Initial Mass Function (IMF) and continuous and constant star-forming history over 100\,Myr is assumed. In Equation \ref{eq_sfr}, $L_\nu$
is the spectral luminosity at the observed frequency $\nu$, given by $L_\nu =
4\pi D_L^2 S_\nu$, where $S_\nu$ is the measured flux density. Here we again adopt an electron temperature $T_e=10^4$\,K and a non-thermal spectral index $\alpha^{\mathrm{NT}}=-0.85$, as done in the previous Sections. If we only consider the thermal free-free emission from young massive stars, Equation \ref{eq_sfr} becomes \citep[Equation 6 in ][]{murphy12}:
\begin{align}\label{eq_sfr_th}
     \Big(\frac{\mathrm{SFR_{\rm th}}}{\textup{M}_\odot \mathrm{yr}^{-1}}\Big)  & = 4.6\times10^{-28}\Big(\frac{T_e}{10^4 \mathrm{K}}\Big)^{-0.45}\Big(\frac{\nu}{\textup{GHz}}\Big)^{0.1} \notag\\ 
    & \times \Big(\frac{L^{\rm T}_\nu}{\rm erg s^{-1} \rm Hz^{-1}}\Big),
\end{align}
where $L^{\rm T}_\nu = f_{\rm th} L_\nu$ is the thermal-only spectral luminosity. For regions with $f_{\rm th} \simeq 100\%$, thermal emission from young massive stars completely dominates the radio continuum, and $L^{\rm T}_\nu \simeq L_\nu$. For $f_{\rm th} \simeq 0\%$, SFR$_{\rm th} \simeq 0$\,$M_\odot$yr$^{-1}$.  \\
\indent For the 40 matched ``SF" regions, SFR ranges from 0.14 to 12\,$M_\odot$yr$^{-1}$, with a median of $\sim 0.7$\,$M_\odot$yr$^{-1}$. The SFR$_{\rm th}$ spans from $\sim 0$\,$M_\odot$yr$^{-1}$ to 12\,$M_\odot$yr$^{-1}$, corresponding to $f_{\rm th} \simeq 0\%$ to $f_{\rm th} \simeq 100\%$. The median $SFR_{\rm th}$ is 0.4\,$M_\odot$yr$^{-1}$. For the 8 ``SBnuc", the ranges of SFR and SFR$_{\rm th}$ are 0.2 -- 13\,$M_\odot$yr$^{-1}$ and 0 -- 11\,$M_\odot$yr$^{-1}$, similar to the ``SF" regions, but with higher median values, at 3.5 and 2\,$M_\odot$yr$^{-1}$, respectively. When taking account of the physical sizes of these matched regions, as calculated from the region boundaries defined by \textit{Astrodendro} with which flux density and spectral index of each region was measured, the SFR and SFR$_{\rm th}$ surface densities, $\Sigma_{\rm SFR}$ and $\Sigma_{\rm SFR_{\rm th}}$, range from 13 -- 1.6$\times$10$^3$\,$M_\odot$yr$^{-1}$kpc$^{-2}$ and 0 -- 1.7$\times$10$^3$\,$M_\odot$yr$^{-1}$kpc$^{-2}$ for the ``SF" regions including 8 unresolved regions. For ``SBnuc", $\Sigma_{\rm SFR}$ and $\Sigma_{\rm SFR_{\rm th}}$ have ranges of 22 -- 540\,$M_\odot$yr$^{-1}$kpc$^{-2}$ and 0 -- 400\,$M_\odot$yr$^{-1}$kpc$^{-2}$, including 1 unresolved region. The median values for the ``SBnuc" regions are higher than those for the ``SF" regions by about a factor of five. However, this result may not be representative given the limited numbers of ``SBnuc" identified in the sample. We report the above derived values in Table \ref{tab:kaku_reg}. For all other region types, given the unknown contribution of star formation to the observed radio continuum, we do not report values of SFR and SFR$_{\rm th}$. In \S \ref{dis:sfr} we compare these results to those derived for star-forming regions in nearby normal galaxies observed with the SFRS at $\sim$\,100\,pc scales.
\begin{figure*}[htb]
    \centering
    \includegraphics[scale=0.55]{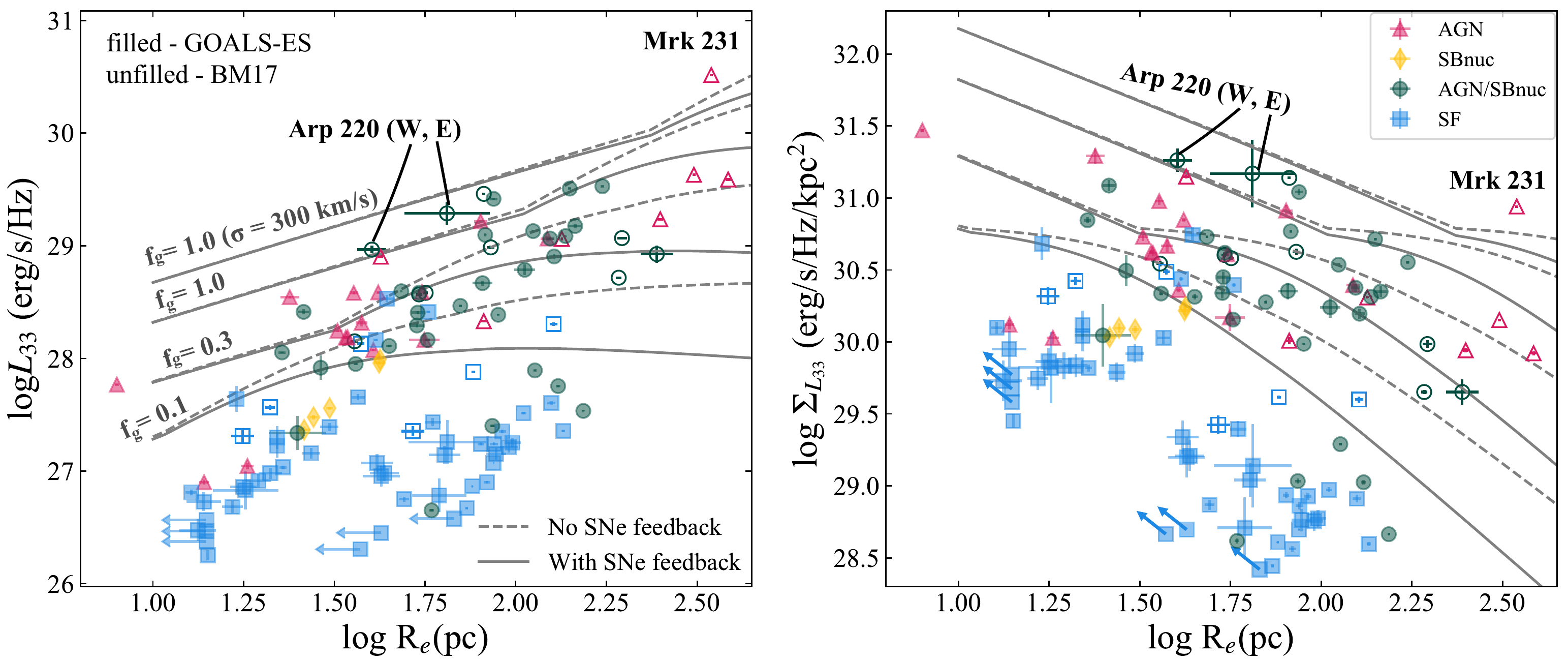}
    \caption{33\,GHz luminosity $L_{33}$ (left) and surface density $\Sigma_{L_{33}}$ (right) vs. effective radius for 95 native regions from GOALS-ES (filled symbols) and of additional 23 native regions characterized using observations of the most luminous U/LIRGs from \cite{barcos17} (BM17; unfilled symbols). In both panels, we show ``AGN" (magenta triangle), ``AGN/SBnuc" (green circles), ``SBnuc" (yellow diamonds) and ``SF" regions (blue squares). Regions classified as ``Jet" or ``Ud" are not shown for simplicity. Upper-limits in size measurements are indicated with arrows. Dashed grey curves are models for a steady-state radiation feedback-supported maximal starburst disk from \cite{thompson05} with assumed molecular gas fraction $f_g = 0.1, 0.3, 1.0$ and stellar velocity dispersion $\sigma = 200$\,km s$^{-1}$, as well as for $f_g = 1.0$ and $\sigma = 300$\,km s$^{-1}$. Solid grey curves show the same models with a modified term to include supernovae feedback, as adopted in \cite{barcos17}. Compared to ``SF" and ``SBnuc" regions, ``AGN" and ``AGN/SBnuc" regions have higher $L_{33}$ and $\Sigma_{L_{33}}$ by up to 3\,dex across the entire size range probed, but only the ``AGN" in Mrk 231 clearly exceeds the predicted $L_{33}$ and $\Sigma_{L_{33}}$ for a maximally star-forming nuclear disk. \label{fig:lum-size}}
\end{figure*}
\section{Discussion}\label{sec:discuss}
\subsection{What powers the compact 33 GHz continuum emission in local U/LIRGs?}\label{dis:lum-size}
As demonstrated in \S \ref{sec:T_b}, the radio data at hand does not allow for direct AGN identification using brightness temperatures, and multi-frequency VLBI observations at milli-arcsecond resolutions are needed to pinpoint the location of AGN and isolate their emission from the circumnuclear star formation in ``AGN" and ``AGN/SBnuc" regions. Nevertheless, it is evident from Figures \ref{fig:Tb} and \ref{fig:native_dist} that ``AGN" and ``SF" respectively dominate the upper and lower end of the distributions in brightness temperature and 33 GHz luminosity surface density. In Figure \ref{fig:lum-size}, we further illustrate this difference by showing the luminosity and luminosity surface density with respect to the effective radius characterized with \textit{Astrodendro} at 33 GHz (filled symbols), including an additional 23 regions from BM17 (see \S \ref{sec:ancillary} and Appendix \ref{ap:ancillary}). \\
\indent As illustrated in Figure \ref{fig:lum-size}, ``AGN" and ``AGN/SBnuc" almost always have higher  $L_{33}$ and $\Sigma_{L_{33}}$ relative to ``SF" across the entire size range probed; the difference can
be as much as $\sim 3$\,dex. This result suggests that more extreme mechanisms may be driving the observed radio emission in the ``AGN" and ``AGN/SBnuc" regions compared with the ``SF" and ``SBnuc" regions. In the following sections we discuss two mechanisms that may be simultaneously contributing to the elevated 33\,GHz emission observed in these ``AGN" and ``AGN/SBnuc" regions.
\subsubsection{Radiation pressure-supported nuclear starburst}
\indent Using analytical models, \cite{thompson05} (hereafter TQM05) predicted that intense starbursts triggered in the dust-obscured gas-rich nuclear environments of local U/LIRGs can potentially radiate at the Eddington-limit (for dust). In this scenario, IR radiation from dust-reprocessed UV or optical emission from massive young stars provides the dominant vertical support against gravitational collapse in an optically-thick starburst disk. The authors estimated the IR luminosity surface densities of 40 local U/LIRGs using radio observations by \cite{condon91} and found the values to agree with those predicted by their models, which have also been invoked to interpret compact radio/sub-mm sources observed in the most luminous local U/LIRGs on $\sim$ 100\,pc scales \citep[e.g.][]{barcos15, barcos17, ps21}. In Figure \ref{fig:lum-size} we compare our 33\,GHz measurements to a simplified version of the radiation pressure-supported starburst disk models presented in TQM05 to investigate the possibility that the observed compact regions of radio emission is driven by such radiation pressure-supported optically-thick starbursts. \\
\indent Following BM17, we also present additional solutions incorporating vertical support from supernovae feedback that can be approximated as  $10\rm{n_{\rm mol}}^{-1/7}$\citep{faucher13,kim15}, where $\rm{n_{\rm mol}}$ is the volume number density of the molecular gas of the modelled marginally-stable one-zone disk (Equation 1 and 7 from TQM05). The predicted IR luminosities are then converted into 33\,GHz luminosities by assuming both come from star formation, using Equations 10 and 15 from \cite{murphy12}. With this assumption, we expect that excess 33\,GHz emission from AGN activity would bring the ``AGN" and ``AGN/SBnuc" regions above the predicted values for maximal starbursts. \\
\indent However, as shown in Figure \ref{fig:lum-size}, only the nucleus in Mrk 231 has $L_{33}$ and $\Sigma L_{33}$ exceeding the model prediction for the highest molecular gas fraction and velocity dispersion assumed, suggesting dominant AGN contribution to the 33\,GHz emission at 100\,pc scales. While this result is unsurprising given that Mrk 231 hosts the closest quasar \citep[][]{adams72}, the fact that all other ``AGN" and ``AGN/SBnuc" do not exceed the model predictions points to the possibility that their 33\,GHz emission could be attributed to star formation. Most of the ``AGN/SBnuc" regions cluster around the solutions for a constant molecular gas fraction of 0.3, which is also the average value for local U/LIRGs derived by \cite{larson16} based on results from galaxy SED fitting by \cite{u12} (for stellar mass) and global molecular gas mass estimates from the literature. Therefore, in the context of this model comparison, the higher $L_{33}$ and $\Sigma_{L_{33}}$ of the ``AGN" and ``AGN/SBnuc" regions relative to the ``SF" regions appear to reflect a more extreme mode of star formation that maintains a radiation pressure-supported nuclear starburst disk, compared with star formation in relatively isolated Giant Molecular Clouds in the outskirts of the systems.  \\
\indent We note that many of these compact nuclei may have gas fractions higher than 0.3 at 100\,pc scales as molecular gas likely dominates the nuclear environments of local U/LIRGs \citep{downes93,larson20}. Additionally, radiation pressure may exceed the Eddington limit and drive outflows \citep[e.g.][]{murray05}, in which case measurements will also lie above the model predictions \citep[e.g.][]{ps21}. A notable example is the two nuclei in Arp 220 \citep[$f_g \sim 0.5, \sigma \sim 165 $\,km/s;][]{genzel01,solomon98}, around which outflows have been detected in different tracers \citep[e.g.][]{sakamoto09, tunnard15, sakamoto17, barcos18, perna20}. Although these outflows have collimated morphology that indicates an AGN origin \citep{sakamoto17, barcos18}, VLBI observation does not show evidence for a bright AGN radio core \citep[e.g.][]{smith98arp220, lonsdale06, parra07, varenius19}, which suggests that the elevated radio continuum emission of these two nuclei are likely powered by nuclear starbursts. Future follow-up high-resolution extinction-free measurements of the stellar and molecular gas distribution and kinematics in these nuclei are required to provide better constraints on the gas fraction and stellar velocity dispersion. 
\subsubsection{(Obscured) AGN activity} \label{dis:agn}
\begin{figure}[h]
    \centering
    \includegraphics[scale=0.55]{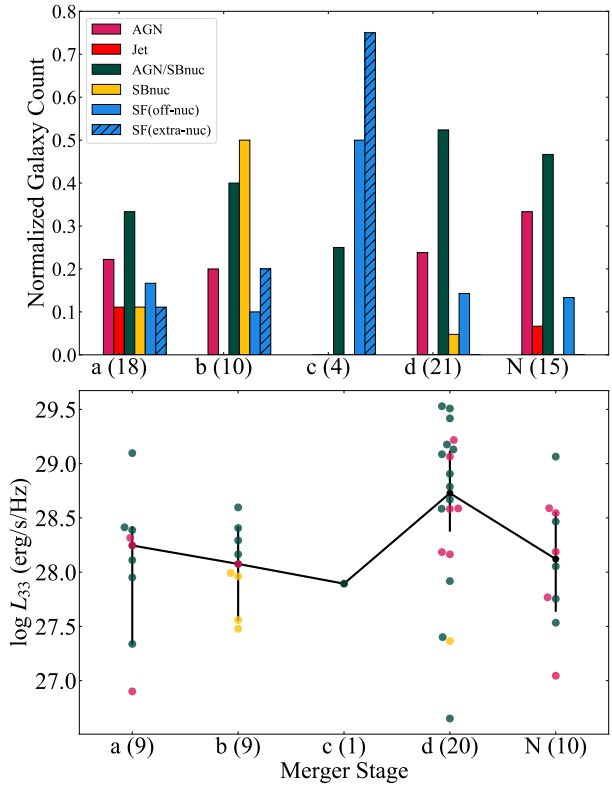}
    \caption{The fraction of systems hosting different region types (upper) and nuclear luminosity distribution (lower) vs. merger stage classification of the host system \citep[a (pre-merging), b (early-merger), c (mid-merger), d (late-merger), N (non-merger);][]{stierwalt13}. Upper: For each merger stage, the number of galaxies hosting each region type is normalized by the total number of galaxies with the specific merger classification, shown in parentheses on the horizontal axis. Galaxies often host more than one type of region, therefore the normalized galaxy counts at each merger stage do not add to 1. All native regions identified at 15 or 33\,GHz from GOALS-ES are accounted for, excluding ``Ud" and ``Bg". Lower: Individual values are color-coded by the nuclear region types (``AGN" - magenta, ``AGN/SBnuc" - green, ``SBnuc" - yellow), and median values and uncertainties at each merger stages are represented in solid black lines. The number of systems included at each merger stage are shown in parentheses on the horizontal axis. Only nuclei identified at 33\,GHz from GOALS-ES are included. Overall, ``AGN" and ``AGN/SBnuc" are more frequently found and are more luminous at 33\,GHz in the final merger stages. \label{fig:merger_hist}}
\end{figure}

\begin{figure*}[tbh]
    \centering
    \includegraphics[scale=0.85]{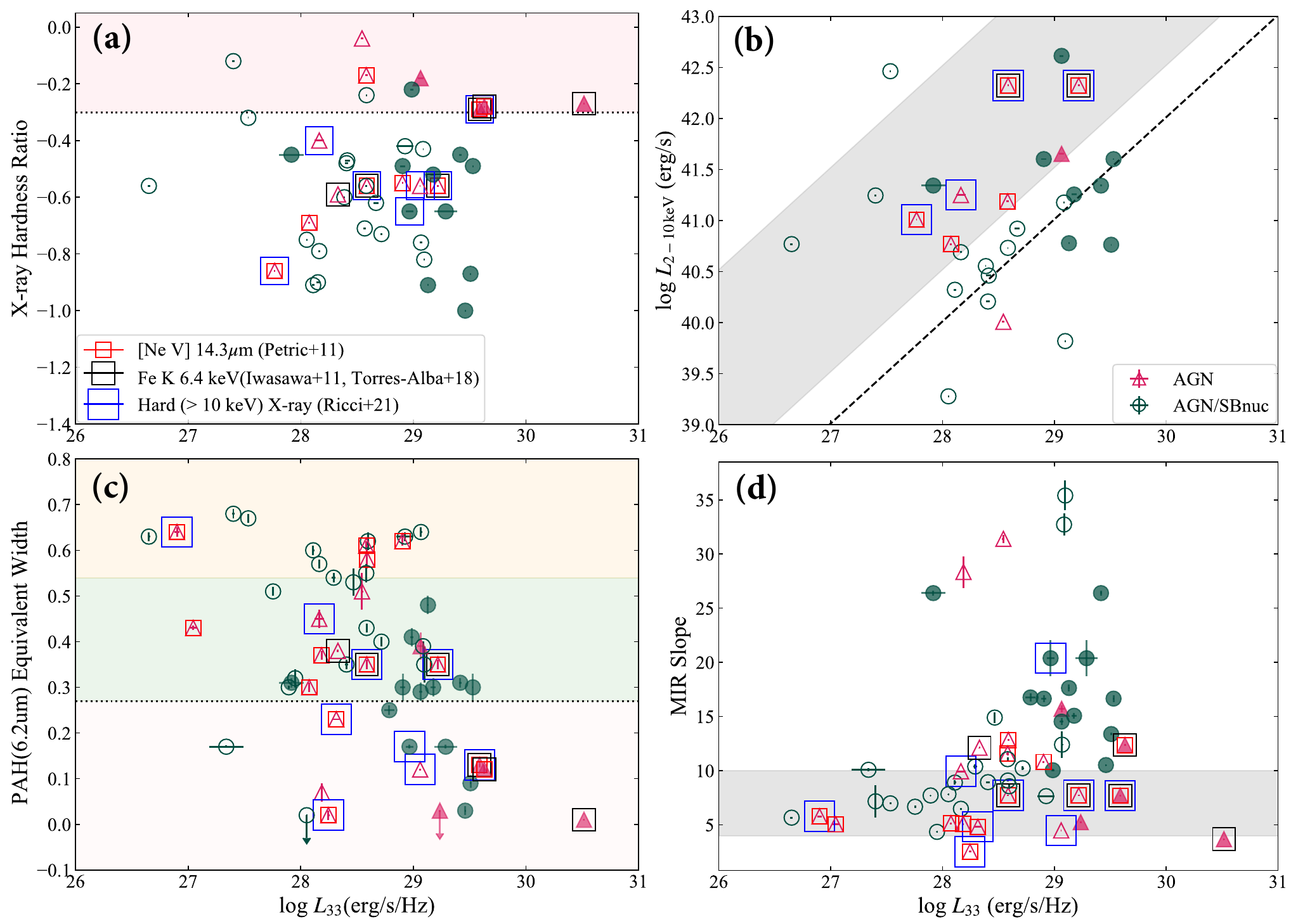}
    \caption{The 33 GHz continuum luminosity of nuclei in GOALS-ES and BM17 vs. AGN diagnostics in the X-rays (a,b) and Mid-IR (c,d). \textit{(a):} X-ray hardness ratios (HR) measured by \cite{iwasawa11} and \cite{torres-alba18} using \textit{Chandra} observations, defined as HR=(H-S)/(H+S), where H is the hard-band (2-7\,keV) flux and S is the soft-band (0.5 - 2\,keV) flux. The dotted horizontal line marks the empirical threshold above which the nucleus is considered to host an AGN due to excess hard X-ray emission (pink shaded area). \textit{(b):} X-ray luminosity at 2-10\,keV from \cite{iwasawa11} and \cite{torres-alba18}, corrected for galactic extinction. The grey shaded area represents the radio/X-ray luminosity ratio range for radio-quiet AGN from \cite{panessa19} (i.e. $\nu L_{\nu}$/$L_{\rm 2-10\,keV}$ = 10$^{-2}$ -- 10$^{-4}$, $\nu =$ 90 -- 100\,GHz), assuming the luminosities at 33 and 100\,GHz are similar. The black dashed line represents the 1.4 GHz radio/X-ray luminosity ratio for star-forming galaxies from \cite{ranalli03} (i.e. $\log L_{\rm 2-10\,keV} = \log L_{\rm 1.4} + 11.12$), assuming $\alpha_{1.4 - 33} \sim \alpha_{15 - 33} \sim -0.65$, as measured among ``AGN/SBnuc" in this work. \textit{(c):} The 6.2\,$\mu$m PAH equivalent widths (EW) measured by \cite{stierwalt13} using \textit{Spitzer} observations. The horizontal dotted line marks the empirical threshold, 0.27\,$\mu$m, below which the MIR nuclear emission is considered to be dominated by AGN (pink shaded area). The green shaded area represents the empirical range (0.27\,$\mu$m - 0.54\,$\mu$m) where the nuclear emission is considered to have some but non-dominant AGN contribution, and nuclei in the yellow shaded area are considered to be starburst-dominated and have low to no AGN contribution on kpc scales. \textit{(d):} The MIR slope from \cite{stierwalt13}, defined as the logarithmic flux density ratio between 30 and 15\,$\mu$m. The grey shaded area represents the range spanned by the majority of LIRGs in GOALS. In all panels, filled symbols represent ULIRGs, and system with [Ne V] 14.3\,$\mu$m, Fe K 6.4\,keV, and hard ($>10$\,keV) X-ray detections reported in \cite{petric11,iwasawa11,torres-alba18,ricci21} are marked in square symbols in red, black and blue, respectively, with increasing sizes. Nuclei with the highest $L_{\rm 33}$ also have higher observed $L_{\rm 2-10\,keV}$, smallest 6.2\,$\mu$m PAH EW, and steepest MIR, suggesting (dust-obscured) AGN contribution to the 33\,GHz emission in ``AGN" and ``AGN/SBnuc" regions. \label{fig:l33_agn}}
\end{figure*}

\indent Aside from an extreme mode of nuclear starburst, AGN activity likely contributes to the elevated 33\,GHz emission in ``AGN" and ``AGN/SBnuc" regions. TQM05 theorized that efficient AGN fueling on pc scales is accompanied by intense star formation in the nuclear disk over 100\,pc scales above a critical rate. This prediction may explain the relatively low $L_{33}$ and $\Sigma_{L_{33}}$ of the ``SBnuc" regions relative to ``AGN" and ``AGN/SBnuc" regions: star formation in ``SBnuc" do not yet reach the rates required to trigger efficient AGN fueling. In the theoretical context of a merger-quasar evolutionary sequence \citep[e.g.][]{matteo05} where tidal torque of gas-rich galaxy merger drives nuclear fueling, we then would expect the ``SBnuc" regions to reside in systems at earlier interaction stages, and the luminosities of the nuclei to increase towards later interaction stages due to contributions from triggered AGN activity.\\
\indent In the top panel of Figure \ref{fig:merger_hist}, we present a histogram of the region types represented in galaxies from GOALS-ES at different merger stages, normalized by the total number of systems at each stage. We see that the ``SBnuc" regions are indeed preferentially found in early-stage mergers (stage ``b"), which supports the aforementioned evolutionary scenario. Additionally, ``AGN/SBnuc" regions are found at all stages but most frequently in late-stage mergers (stage ``d"). This is consistent with results from MIR analysis of the GOALS systems by \cite{stierwalt13}, who found that among the local U/LIRGs, the fraction of AGN-starburst composite systems increases among late stage mergers. Among the merging systems, ``AGN" regions are also most frequently found in late-stage mergers. As shown in the lower panel of Figure \ref{fig:merger_hist}, it is also at the late-stage that the nuclei have the highest median $L_{33}$. These results are in agreement with the scenario that powerful AGN activity is triggered during gas-rich galaxy-mergers. The marked increase in ``AGN/SBnuc" towards the later stages may reflect increased level of dust obscuration that makes AGN identification more difficult at shorter wavelengths, as suggested in previous GOALS studies in the MIR and X-rays \citep[e.g.][]{stierwalt13, ricci17, ricci21}. We note that nuclei from BM17 were not included in Figure \ref{fig:merger_hist} because the host galaxies do not statistically represent the full GOALS sample. However, including the BM17 nuclei does not alter the overall trend seen in the sky area and flux-limited selection of the GOALS-ES sample.\\
\indent To further investigate whether the elevated 33\,GHz emission is correlated with more powerful AGN activity, we compare AGN diagnostics in the X-rays \citep[hardness ratio, $L_{\rm 2-10 keV}$;][]{iwasawa11,torres-alba18} and MIR \citep[6.2$\mu$m PAH equivalent width, MIR slope;][]{stierwalt13} with $L_{33}$ of ``AGN" and ``AGN/SBnuc" regions in Figure \ref{fig:l33_agn}. We also mark systems with [Ne V] (14.3\,$\mu$m), Fe K (6.4\,keV), and hard X-ray ($> 10$\,keV) detections \citep{petric11, iwasawa11, ricci21}, which are commonly used indicators of AGN activity. The latter two are used to identify heavily-obscured AGN. \\
\indent As shown in Figure \ref{fig:l33_agn}, while $L_{33}$ does not exhibit clear correlation with the X-ray hardness ratio, nuclei with higher $L_{33}$ show higher $L_{\rm 2-10 keV}$, smaller 6.2$\mu$m PAH equivalent width (EW), and steeper MIR slope. The Kendall's Tau correlation coefficients are 0.08, 0.27, -0.29 and 0.29, respectively for comparisons presented in Figure \ref{fig:l33_agn} (a), (b), (c) and (d), indicating stronger (anti)correlations between $L_{33}$ and MIR diagnostics. Nuclei with $L_{33} \gtrsim 10^{29}$\,erg s$^{-1}$ Hz$^{-1}$ mostly reside in ULIRGs, and they also have the smallest 6.2$\mu$m PAH EW and highest $L_{\rm 2-10 keV}$, which suggests that in these nuclei, the 33\,GHz continuum is likely tracing AGN activity that produces strong hard X-ray emission and weak PAH emission. The steeper MIR slope of these nuclei, as shown in \cite{stierwalt13}, suggest the presence of warm dust heated by the accretion disk of the AGN. The dense ISM in such environments can result in the Compton down-scattering of X-ray photons and cause reduced correlations between the radio and observed X-ray luminosity.\\
\indent In Figure \ref{fig:l33_agn}(b) we also show the expected range of X-ray luminosities for radio-quiet AGN \citep[shaded in grey;][]{panessa19} and star-forming galaxies \citep[black dashed line;][]{ranalli03} at the given $L_{\rm 33}$. Many ``AGN/SBnuc" follow the relation established for star-forming galaxies, suggesting that both $L_{\rm 2-10 keV}$ and $L_{\rm 33}$ could be tracing star formation in these nuclei. However, some of them may also host highly-embedded AGN whose X-ray emission is significantly absorbed. Comparison between the observed $L_{\rm 2-10 keV}$ \citep{iwasawa11,torres-alba18} and intrinsic $L_{\rm 2-10 keV}$ derived from spectral model-fitting by \cite{ricci21} for a handful of overlapping systems shows that the latter could be higher by up to two orders of magnitude. Correcting for the effect of host obscuration will allow a more robust comparison between these nuclei to radio-quiet AGN (shaded area), and would require more sensitive X-ray observations and spectral analysis.\\
\indent The inference from the above is that the elevated 33\,GHz emission in ``AGN" and ``AGN/SBnuc" with the highest 33\,GHz-luminosities are likely dominated by contributions from AGN that are obscured in the X-rays. We note that the overall weak correlations between $L_{33}$ and various AGN diagnostics presented above may be driven by the $\sim 5 - 10$ times lower resolutions of the X-ray/MIR observations compared to our 33\,GHz observations. \\

\indent In summary, the elevated 33\,GHz continuum emission of ``AGN" and ``AGN/SBnuc" regions relative to ``SF" regions in local U/LIRGs likely come from a combination of extreme nuclear starburst and AGN activity, with the nuclei with higher 33\,GHz luminosities more dominated by AGN but also experiencing more dust obscuration at shorter wavelengths. This conclusion is in agreement with X-ray studies which show that AGN accretion is accompanied by intense circumnuclear star formation \citep[e.g.][]{lutz18}, and that powerful AGN accretion in mergers are heavily obscured by dust, especially in the final ULIRG stage \citep[e.g.][]{ricci17,ricci21}. However, follow-up observations at higher resolutions are required to fully disentangle the contribution from AGN and starburst.

\subsection{How does star formation in U/LIRGs compare with that in nearby normal galaxies?}
\indent In \S \ref{sec:fth} and Figure \ref{fig:fth_dist} we showed that all regions in the GOALS-ES span a wide range in $f_{\rm th}$, but the median values for ``SF" and ``SBnuc" are significantly higher compared to those for ``AGN" and ``AGN/SBnuc" regions. We note that similarly low $f_\mathrm{th}$ ($< 50\%$) have also been observed by \cite{barcos15} and BM17 in the most luminous local U/LIRGs (including Arp 220), using 6 -- 33 GHz measurements. The authors suggest that in these heavily-obscured systems, thermal emission from the nuclear starburst could be suppressed via dust absorption of ionizing photons, which may be responsible for the apparent dominance of non-thermal emission. Meanwhile, given the discussion in \S \ref{dis:agn},``AGN" and ``AGN/SBnuc" regions may also contain excess non-thermal emission from unresolved jets and/or wind/outflows associated with AGN activity \citep[e.g.][]{panessa19, hayashi21}. Therefore, in this Section we only focus on the comparing 48 ``SF"/``SBnuc" regions in the GOALS-ES with 129 star-forming regions identified in the SFRS (see Section \ref{sec:ancillary} and Appendix \ref{ap:ancillary}).
\subsubsection{Radio spectral indices \& 33\,GHz thermal fraction}\label{dis:fth}
\begin{figure*}[htb]
    \centering
    \includegraphics[scale=0.48]{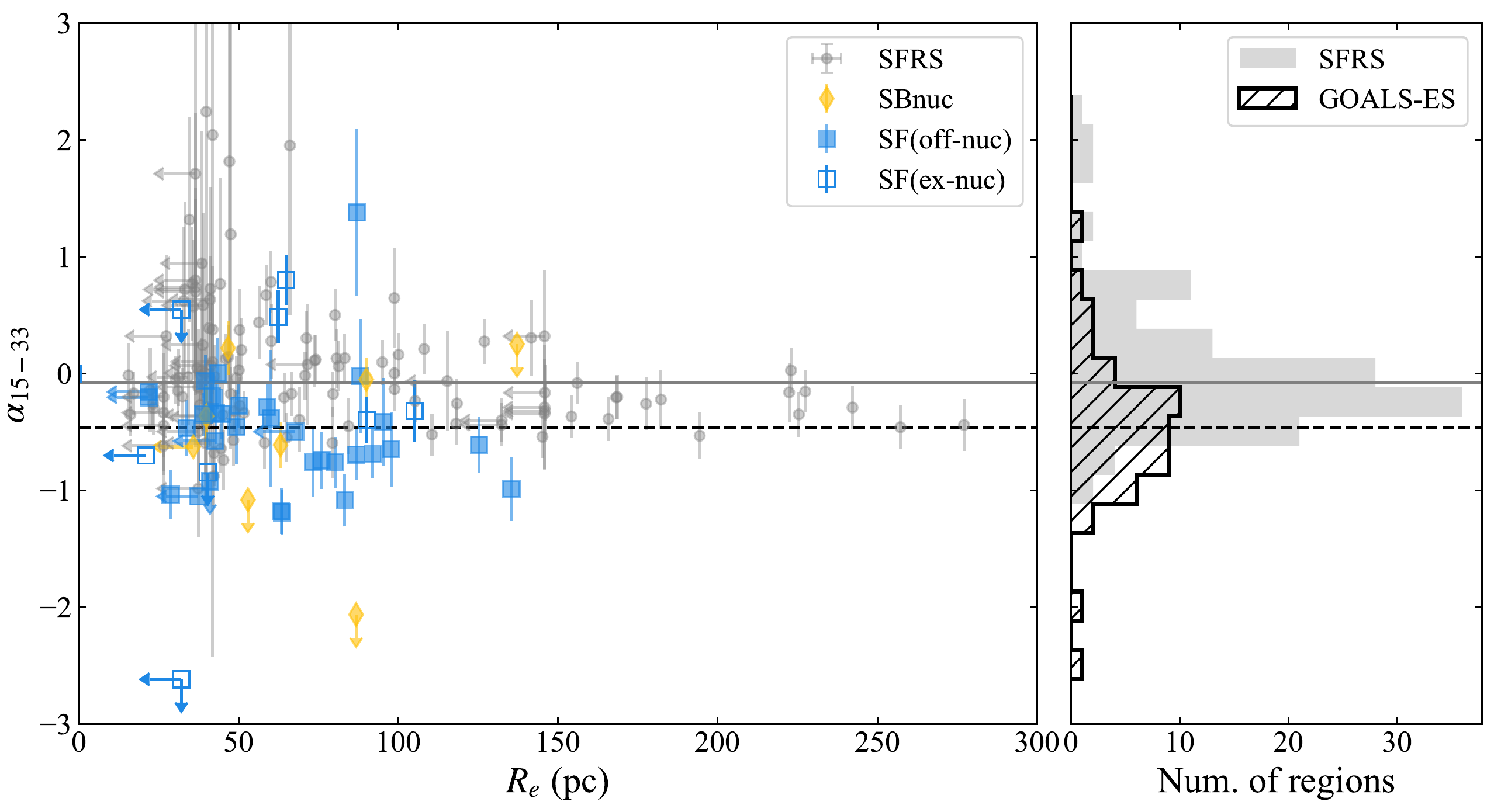}
    \caption{The 15 - 33\,GHz spectral indices ($\alpha_{\rm 15 - 33}$) measured for matched regions identified in nearby star-forming galaxies from the SFRS and local U/LIRGs from GOALS-ES. \textit{Left}: $\alpha_{\rm 15 - 33}$ vs. effective radii, $R_e$, of the region area used to measure $\alpha_{\rm 15 - 33}$. SFRS regions are in grey, GOALS-ES regions are colored in yellow (``SBnuc") and blue (``SF"), with extra-nuclear ``SF" regions in non-filled symbols. Upper-limits in size are indicated with arrows. \textit{Right}: Distribution of $\alpha_{\rm 15 - 33}$ for GOALS-ES regions (hatched black) and for SFRS regions (filled grey). In both panels, the median values for the SFRS ($-0.08\pm0.27$) and GOALS-ES regions ($-0.46\pm$0.29) are shown in solid grey and dashed black lines, respectively. Overall, $\alpha_{\rm 15 - 33}$ spans a wide range for star-forming regions in both local U/LIRGs and normal galaxies, especially at $R_e < 100$\,pc. Regions in U/LIRGs have steeper median $\alpha_{\rm 15 - 33}$ compared with those in nearby normal galaxies, suggesting more dominant non-thermal contribution at 33\,GHz. \label{fig:spx_re}}
\end{figure*}

\indent Studies of nearby normal galaxies with the SFRS have shown that their 33\,GHz continuum emission is largely dominated by thermal free-free emission from HII regions on both kpc and 100\,pc scales, which make 33\,GHz continuum an ideal extinction-free tracer of ongoing massive star formation \citep{murphy11, murphy12, linden20}. \cite{linden19} shows that for extra-nuclear star-forming regions in the GOALS-ES, thermal emission accounts for $\sim$65\% of the 33\,GHz emission on kpc scales, which is similar to values derived for the SFRS galaxies on the same physical scales \citep[$f_{\rm th}\sim$60\% ;][]{murphy12}. To investigate whether this agreement is also seen on 100\,pc-scales, in Figure \ref{fig:spx_re}(right) we compare the distributions of 15 -- 33\,GHz spectral indices ($\alpha_{\rm 15 - 33}$) measured for the GOALS-ES (``SF"/``SBnuc") regions and SFRS star-forming regions. We also show the effective radius of the area we use to measure $\alpha_{\rm 15 - 33}$ for each matched region. We note that $\alpha_{\rm 15 - 33}$ instead of $f_{\rm th}$ is presented because the former can be more straight-forwardly compared without considering any underlying assumptions about the intrinsic non-thermal and thermal spectral shapes.  

\indent Figure \ref{fig:spx_re} shows that star-forming regions in GOALS-ES and SFRS both exhibit a wide range of $\alpha_{\rm 15 - 33}$, from $< -2$ to $1.38\pm0.72$ for GOALS-ES and $-0.98\pm1.20$ to $\gtrsim 2$ for SFRS. The median spectral index of the SFRS regions is $-0.08\pm$0.27, which is consistent with values derived by \cite{linden20} using a different method to measure region photometry. In contrast, the median value for the GOALS-ES regions is $-0.46\pm$0.29, suggesting higher contribution from steep-spectrum non-thermal emission. This value is also steeper than the median value derived on kpc scales for extra-nuclear star-forming regions in the GOALS-ES \citep[$-0.27\pm$0.23;][]{linden19}. A two-sample K-S test on the distributions of $\alpha_{\rm 15 - 33}$ for the GOALS-ES and SFRS regions yields a p-value of $<< 1$, which means that the differences we see between the two sample of regions are likely intrinsic. Several mechanisms may be responsible for the comparatively steep $\alpha_{\rm 15 - 33}$ of the 100\,pc-scale GOALS-ES regions:  \\
\indent First, because U/LIRGs are dusty, thermal free-free emission from HII regions may have been suppressed via dust absorption \citep{barcos15,barcos17}. However, this effect likely only becomes important in the most heavily-obscured systems such as in the ULIRGs, and we also do not find any correlation between $\alpha_{\rm 15 - 33}$ and the MIR $9.7\,\mu$m silicate depths estimated by \cite{stierwalt13}, which measure the level of dust obscuration on kpc scales in these systems. Matched-resolution comparison between the resolved dust and the spectral index distribution will shed light on how much dust absorption affects the 100\,pc scale high-frequency radio properties of local U/LIRGs. \\
\indent Second, the ages of the starbursts also affect the relative contribution of non-thermal and thermal emission \citep[e.g.][]{rabioux14,linden19,linden20}. Using Starburst99 models, \cite{linden20} showed that non-thermal synchrotron emission from supernovae can quickly dominate the radio emission of an instantaneous starburst within 10\,Myr compared with steady continuous star formation that maintains high thermal contribution with relatively flat radio spectrum. Using the same models and NIR hydrogen recombination line observations, \cite{larson20} estimated that star-forming clumps in local U/LIRGs have an age range of 6 - 10\,Myr. Therefore the overall higher non-thermal contribution at 33\,GHz measured in local U/LIRGs could be a reflection of the more recent star formation triggered in local U/LIRGs on 100\,pc scales. \\
\indent Third, the dense ISM in the compact starbursts in local U/LIRGs may produce non-thermal synchrotron spectrum than is intrinsically steeper that those characterized in star-forming regions in nearby normal galaxies \citep[i.e. $\alpha_{\rm NT} \sim -0.85$;][]{murphy11}. Via multi-frequency analysis, \cite{galvin18} measured an average $\alpha_{\rm NT} \sim -1.06$ in a sample of 19 local LIRGs. In the nearby starburst NGC 4945, $\alpha_{\rm NT}$ has been measured to be as steep as $\sim -1.5$ \citep[e.g.][]{bendo16, emig20}. Additionally, spectral steepening of synchrotron emission above 10\,GHz have also been observed in nearby star-forming galaxies \citep[e.g.][]{klein18}, local U/LIRGs \citep[e.g.][]{clemens08,leroy11} as well as high-$z$ star-forming and starburst galaxies \citep[e.g.][]{thomson19, algera21}. As discussed in \cite{klein18}, steep synchrotron spectra either result from energy losses of high-energy electrons due to inverse-Compton scattering and synchrotron radiation in dense ISM environments, or intrinsic lack of high-energy electrons. Therefore, the steeper $\alpha_{\rm 15 - 33}$ measured in GOALS-ES regions may simply reflect intrinsically steep non-thermal spectrum, and does not necessarily require excess non-thermal emission. We note that if we assume a simple two-component power law model without spectral steepening (i.e. Equation \ref{eq_fth}), for $f_{\rm th}$ to be as high as measured in the SFRS regions ($\sim 90\%$) at $\alpha_{\rm 15 - 33} \sim -0.46$, $\alpha_{\rm NT}$ will have to be $\sim -2$, which is also the steepest $\alpha_{\rm 15 - 33}$ measured in the GOALS-ES region. Matched resolution observations at lower radio frequencies are needed to recover the intrinsic non-thermal spectral shape in these extreme systems \citep[e.g.][]{tabatabaei17}. \\
\indent Finally, tidal shocks associated with galaxy mergers may have produced excess non-thermal synchrotron emission in local U/LIRGs \citep{murphy13}. While it is possible that we are detecting traces of shock-driven synchrotron emission, given the high-resolution of our observations, large-scale diffuse emission driven by such dynamical effects are likely to have been resolved out, and would play relatively minimal role in producing the steep $\alpha_{\rm 15 - 33}$ we measure on 100\,pc scales. \\
\indent We emphasize that while the median $\alpha_{\rm 15 - 33}$ of the GOALS-ES regions is significantly steeper than that of the SFRS regions, the wide range of values seen in both samples, especially at $R_e < 100\,$pc, suggests that the balance between thermal and non-thermal emission is more complicated at small scales. Large uncertainties in our measurements due to sparse frequency coverage and short on-source time also limit our ability to draw more definitive conclusions. Matched resolution radio continuum observations at more than two different frequencies are needed to more robustly characterize the radio continuum spectrum of compact star-forming regions in local U/LIRGs. This will also allow us to better understand whether and how the extreme ISM conditions in these dense starbursts may alter the synchrotron production and propagation processes.  
\subsubsection{Star formation rates and surface densities}\label{dis:sfr}
\begin{figure*}[htb]
    \centering
    \includegraphics[scale=0.85]{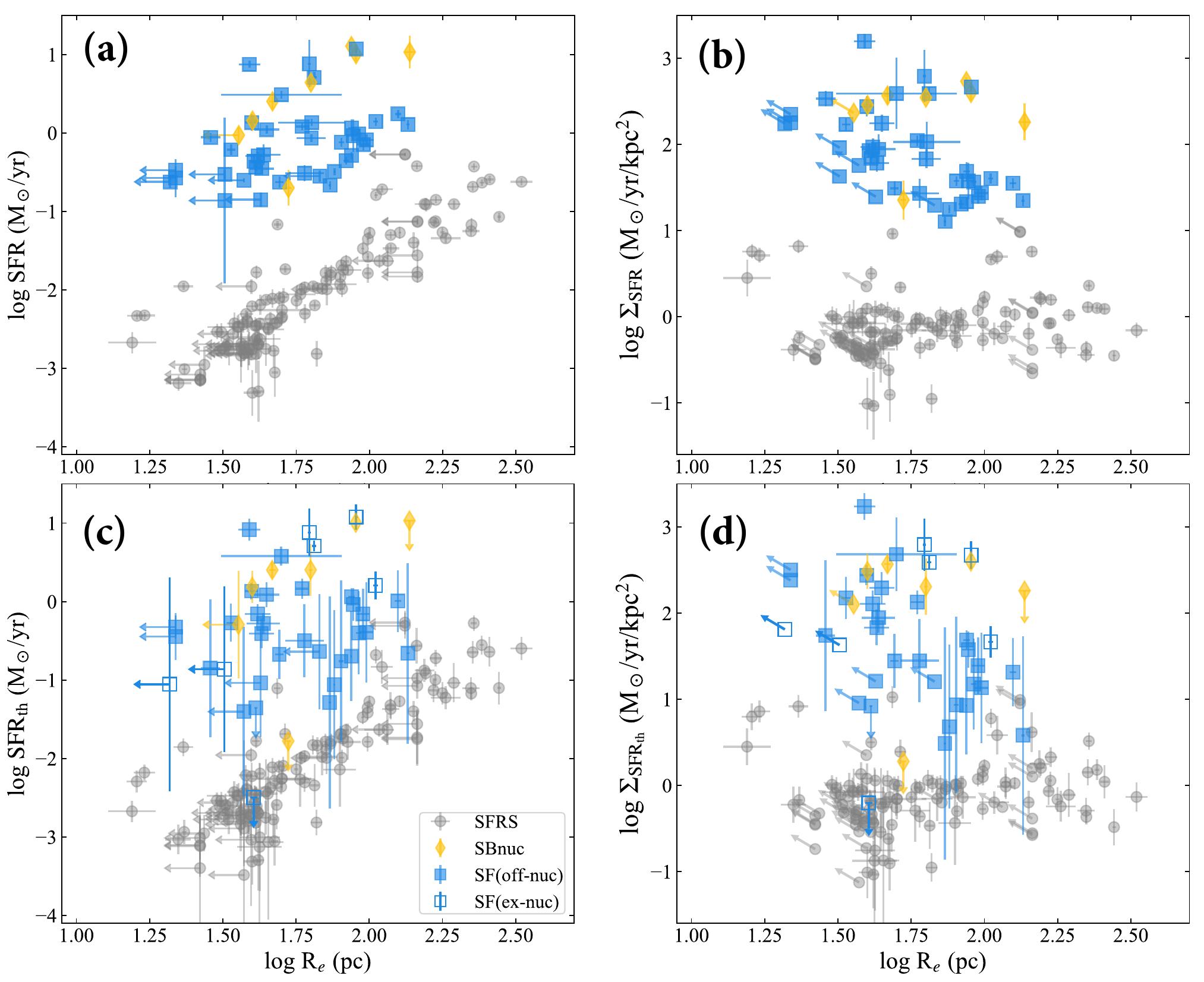}
    \caption{Star formation rates and surface densities vs. effective radii for ``SF" and ``SBnuc” regions characterized in local U/LIRGs in this work, as well as for star-forming regions in nearby normal galaxies from the SFRS characterized using the same methods outlined in \S \ref{sec:analysis}. Values derived for the SFRS sample are in grey circles. We show values derived both from the total 33\,GHz flux (a, b) and thermal free-free only flux based on the measured 15-33\,GHz spectral indices (c, d). Star-forming clumps and starburst nuclei in the GOALS-ES have up to 3\,dex higher star formation rates and surface densities compared with star-forming clumps in the SFRS on $\sim$\,100\,pc scales.\label{fig:sfr_re}}
\end{figure*}
\begin{figure}[bth]
    \centering
    \includegraphics[scale=0.4]{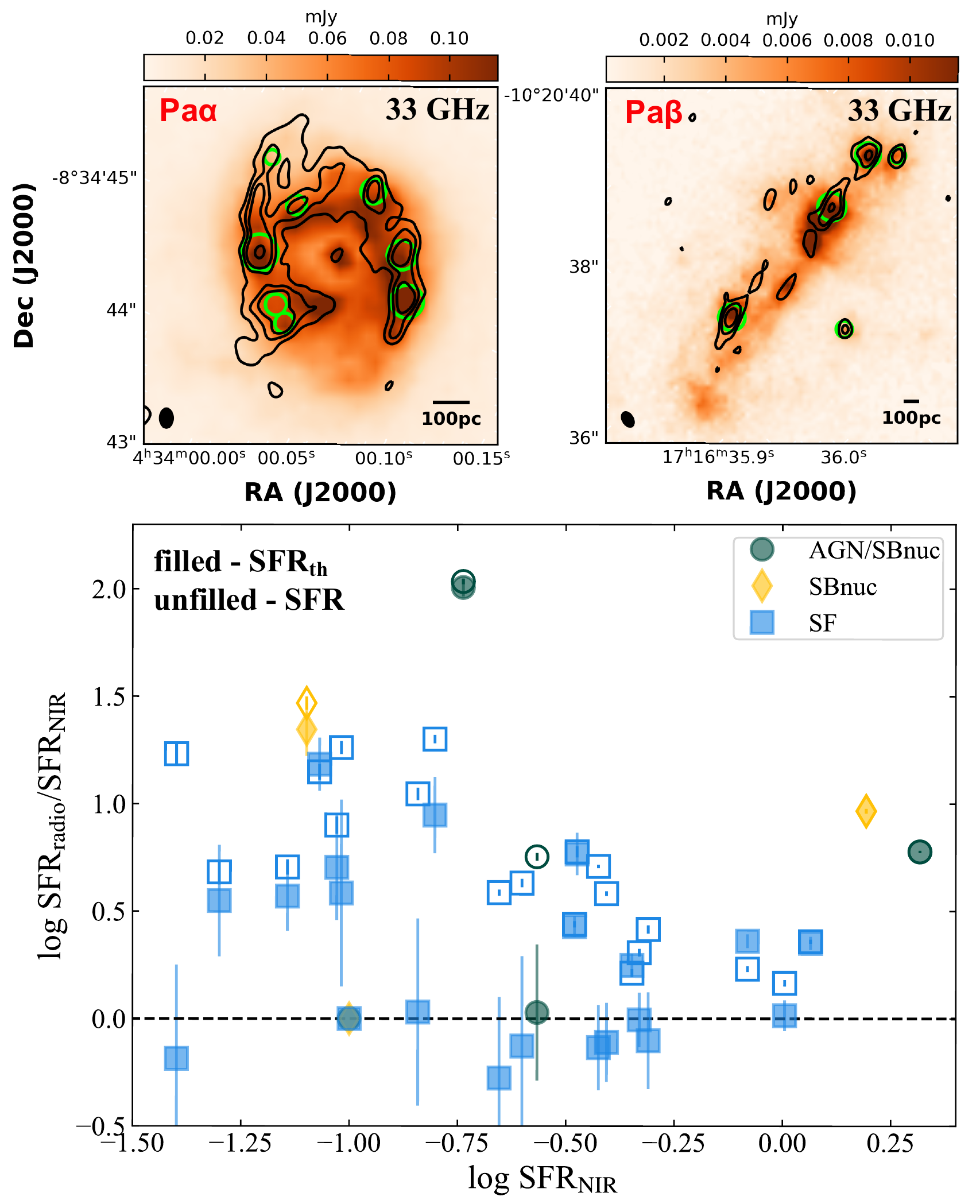}
    \caption{A comparison between radio continuum and Pa$\alpha/\beta$ as SFR tracer. (Upper) Continuum-subtracted \textit{HST} Pa$\alpha$ image of NGC 1614 (\textit{left}) and Pa$\beta$ image of IRAS F17138-1017(\textit{right}) from \citep[][]{larson20}. Black contours show 33\,GHz radio continuum at matched resolutions with 15\,GHz continuum, and the matched beams are shown in the lower left corners in black ellipses. Contour levels are 0.075, 0.15, 0.23, 0.45 mJy/beam for NGC 1614, and 0.032, 0.065, 0.13 mJy/beam for IRAS F17138-1017. Lime circles show the apertures used for measuring and comparing radio- and Pa$\alpha/\beta$-derived SFR. (Lower) The ratio between radio-derived and NIR-derived SFRs for 9 U/LIRGs in the sample. The SFRs derived from thermal free-free radio continuum (filled) show better agreement with NIR-derived SFRs than those derived from total 33\,GHz continuum (unfilled), and deviations from 1:1 relation (dashed line) are likely due to nuclear dust extinction. \label{fig:radio-paschen}}
\end{figure}

\indent In Figure \ref{fig:sfr_re} we show star formation rates and surface densities with respect to effective radii for ``SF" and ``SBnuc" regions in GOALS-ES (see \S \ref{sec:sfr}) and SFRS. We show values derived from both the total 33\,GHz flux (a,b) and the free-free component via 15 -- 33\,GHz spectral decomposition (c,d), as described in Section \ref{sec:fth}. The star-forming regions in local U/LIRGs have 1-3\,dex higher SFR and $\Sigma_{\rm SFR}$ than similarly-sized regions in the nearby normal galaxies, even after accounting for the steeper $\alpha_{\rm 15 - 33}$ measured in the GOALS-ES regions. The median values for SFR and $\Sigma_{\rm SFR}$ for the GOALS-ES regions are 0.8$\pm0.5$\,M$_\odot$yr$^{-1}$ and $86\pm65$\,M$_\odot$yr$^{-1}$kpc$^{-2}$, which is roughly 10 times higher than the median values for the SFRS regions (SFR $\sim 0.1$\,M$_\odot$yr$^{-1}$ and $\Sigma_{\rm SFR}\sim 10$\,M$_\odot$yr$^{-1}$kpc$^{-2}$). As expected, the median values for SFR$_{\rm th}$ and $\Sigma_{\rm SFR_{th}}$  are lower, at $\sim 0.4$\,M$_\odot$yr$^{-1}$ and $\sim 44$\,M$_\odot$yr$^{-1}$kpc$^{-2}$, but still significantly higher than those for the SFRS regions, despite that the latter is more dominated by thermal free-free emission. Given that this comparison is made at the scales of Giant Molecular Clouds (GMC; 10 -- 100\,pc), our result suggests that GMCs in local U/LIRGs are forming more stars compared to those in nearby normal galaxies, at least in these most active star-forming regions detected in these systems that are mostly ``nuclear" and ``off-nuclear".\\
\indent Using HST NIR hydrogen recombination line (i.e. Pa$\alpha$, Pa$\beta$)
observations of 48 local U/LIRGs smoothed to a common resolution of 90\,pc,
\cite{larson20} identified 751 extra-nuclear star-forming clumps in these
systems with median SFR$\sim 0.03$\,M$_\odot$yr$^{-1}$ and $\Sigma_{\rm SFR}
\sim 0.3$\,M$_\odot$yr$^{-1}$kpc$^{-2}$. These values are over 10 times lower
than the values derived for the GOALS-ES regions, which may be due to intrinsic
differences between nuclear and extra-nuclear star formation as suggested in
\cite{linden19}, or systematic offsets introduced by the use of different SFR
tracers. However, due to the 90\,pc resolution limit, the clumps studied in
\cite{larson20} are at least five times larger than the GOALS-ES regions
characterized in this work, which complicates the interpretation. \\
\indent To investigate whether the different SFR tracers used may have introduced a systematic offset, we acquired continuum-subtracted Pa$\alpha$ or Pa$\beta$ images used in \cite{larson20} for 9 non-AGN U/LIRGs also included in the GOALS-ES and directly compare the Pa$\alpha/\beta$ emission with the radio continuum, without smoothing the HST images, as demonstrated in the upper panels of Figure \ref{fig:radio-paschen}. We calculate SFR at each matched region identified in the radio using the the same circular apertures on the NIR, 15 and 33\,GHz maps, following Equation \ref{eq_sfr}, \ref{eq_sfr_th} and the prescription provided in \cite{larson20}. Due to lack of multi-line observations, these NIR images are not corrected for extinction, which has minimal effect on the measurements of extra-nuclear clumps studied in \cite{larson20} but could affect measurements within the central kpc \citep{piqueras13}. \\
\indent As shown in the lower panel of Figure \ref{fig:radio-paschen}, SFR derived from the total 33\,GHz continuum are consistently higher than values derived from the Pa$\alpha/\beta$ emission by up to $\sim 1\,$dex, with the ``AGN/SBnuc" in IRAS F16399-0937 showing the highest discrepancy, possibly due to AGN activity or extreme nuclear obscuration. When only considering the thermal component, the radio-derived values for ``SF" regions show better agreement, with $\rm{SFR_{th}}/\rm{SFR_{NIR}} \sim 0.5 - 15$ and a median of $\sim 2$, which would correspond to $A_{v} \sim 4$ if we assume thermal radio emission is tracing the the same emission. This value is consistent with nuclear extinction estimated from NIR line ratios in previous works \citep[e.g.][]{alonso06,piqueras13}. This suggests that thermal free-free radio continuum is indeed tracing ionized plasma in HII regions that is producing the hydrogen recombination lines, and that radio continuum is more reliably tracing star formation in the dusty nuclear environments of local U/LIRGs.\\
\indent Given the above, while the limited sensitivity of our current radio observations only allow detections of the most energetic regions of nuclear star formation, we expect that radio- and NIR-derived SFR for the mildly obscured extra-nuclear star-forming clumps in local U/LIRGs to be largely consistent with each other. Therefore, nuclear star formation in local U/LIRGs, as probed by the extremely high SFR and $\Sigma_{\rm SFR}$ derived in this work and previous studies \citep[e.g.][]{barcos17,u19}, are likely proceeding at much faster rates at GMC scales than those in the outskirts of local U/LIRGs, as well as those in nearby normal galaxies. Such extreme activity is likely driven by the high molecular gas surface densities in the central kpc of local U/LIRGs, as have been measured with ALMA at $\sim 100\,$pc scales \citep[e.g.][]{wilson19, sanchez-garcia21, sanchez22}. These studies also show that molecular gas forms stars more efficiently in these high density environments, potentially driven by cloud-cloud collisions \citep{jog92} and/or gravitational instability induced by the high stellar mass density \citep[e.g.][]{romeo16}. \\
\indent Meanwhile, it has also been shown that local U/LIRGs host a higher fraction of young ($< 10$\,Myr) and massive ($\gtrsim 10^6$\,M$_\odot$) star clusters compared to normal galaxies \citep[e.g.][]{alonso02,linden17, linden21}. Therefore, the elevated SFR and $\Sigma_{\rm SFR}$ of GOALS-ES regions characterized in this work relative to the SFRS regions may be a reflection of the higher numbers of massive star clusters being produced in the former. These massive clusters better sample the stellar initial mass function (IMF) and thus are more likely to contain a higher number of massive stars that ultimately generate synchrotron emission via supernovae explosions, which possibly contributes to the steep $\alpha_{15 - 33}$ measured in the GOALS-ES regions, as discussed in \S \ref{dis:fth}. \\
\indent Finally, as shown in Figure \ref{fig:sfr_re}, while SFR and SFR$_{\rm
th}$ is clearly correlated with $R_e$ for the SFRS regions, values for the
GOALS-ES regions show relatively weak dependence on the region sizes and larger scatter at a given size. Fitting
the data with a power-law model SFR $\propto L_{\rm radio} \propto r^\eta$ yields
$\eta \sim 2.3$ for the SFRS regions and $\eta \sim 1.1$ for the GOALS-ES
regions, with similar values derived using SFR$_{\rm th}$ which have larger scatter as well as uncertainties. While the limited
sensitivity of the radio observations prevents a direct comparison to
luminosity-size relation established in the optical/NIR
\citep[e.g.][]{piqueras16,cosens18}, the relatively weak size dependence of SFR
and SFR$_{\rm th}$ among the GOALS-ES regions is consistent with a scenario
where the HII region is density-bounded with its luminosity set by the local gas
volume density. In this case, hydrogen atoms in the region recombine faster than
they are ionized and hence a fraction of ionizing photons are not absorbed and
escape the region, resulting in lower luminosity than expected at a given region
size \citep[e.g.][]{beckman2000, wisnioski12}. Hence the relatively constant SFR
and SFR$_{\rm th}$ of the GOALS-ES regions may be reflecting the high density
environments that they reside in. In comparison, the SFRS regions may more
closely resemble photon-bounded HII regions (i.e. Str\"{o}mgren spheres) in
low-density environments, whose luminosities are more or less proportional to
the region volumes as hydrogen recombination balances ionization. \\
\indent A similar dichotomy was also observed by \cite{cosens18} in a large sample of star-forming clumps, and the authors found that clumps with $\Sigma_{\rm SFR} > 1$\,M$_\odot$yr$^{-1}$ kpc$^{-2}$ show weaker size dependence in H$\alpha$ luminosity than clumps with $\Sigma_{\rm SFR} < 1$\,M$_\odot$yr$^{-1}$ kpc$^{-2}$, which are consistent with the ranges of $\Sigma_{\rm SFR}$ and $\Sigma_{\rm SFR_{th}}$ represented by the GOALS-ES and SFRS regions, respectively. Deeper radio observations capable of sampling a wider range of star-forming clumps would allow a more quantitative comparison between the luminosity - size relation observed in the radio and at shorter wavelengths. 
\clearpage
\startlongtable
\begin{longrotatetable}
\begin{deluxetable*}{LrcRRRRRRRR}
\tablecaption{Measured and Derived Quantities for Regions Identified at Native Resolutions \label{tab:native_reg}}
\tablenum{2}
\tablecolumns{11}
\tablehead{
\colhead{ID} &
\colhead{Region} & 
\colhead{Type} & 
\colhead{RA (J2000)} & 
\colhead{Dec (J2000)} & 
\colhead{$S_\nu$} &
\colhead{log R$_e$} &
\colhead{log $L_\nu$} & 
\colhead{log $\Sigma_{L_\nu}$} &
\colhead{log $T_b$} &
\colhead{log $T^{\rm imfit}_b$}\\
\colhead{} &
\colhead{} & 
\colhead{} & 
\colhead{(\degr)} & 
\colhead{(\degr)} & 
\colhead{(mJy)} & 
\colhead{(pc)} & 
\colhead{(\,erg s$^{-1}$ Hz$^{-1}$)} & 
\colhead{(\,erg s$^{-1}$ Hz$^{-1}$ kpc$^{-2}$)} &
\colhead{(K)} &
\colhead{(K)}\\
\colhead{(1)} &
\colhead{(2)} & 
\colhead{(3)} & 
\colhead{(4)} & 
\colhead{(5)} & 
\colhead{(6)} & 
\colhead{(7)} & 
\colhead{(8)} & 
\colhead{(9)} &
\colhead{(10)} &
\colhead{(11)} 
}
\startdata
1    & n1     & AGN       & 2.777289   & -12.107661 & 1.70\pm0.17 & 1.75\pm0.04 & 28.16\pm0.04 & 30.17\pm0.09 & 1.65\pm0.09  & 1.65\pm0.09  \\
2  & n1     & SBnuc     & 4.712000   & -10.376830 & 1.10\pm0.07 & 2.14\pm0.02 & 28.27\pm0.03 & 29.49\pm0.04 & 1.66\pm0.04  & -            \\
3    & n1     & SF        & 16.947427  & -17.507269 & 0.20\pm0.01 & 1.94\pm0.01 & 27.24\pm0.01 & 28.86\pm0.03 & 0.35\pm0.03  & 0.92\pm0.21  \\
     & n2     & SF        & 16.947899  & -17.507178 & 0.13\pm0.02 & 1.62\pm0.04 & 27.07\pm0.08 & 29.34\pm0.12 & 0.82\pm0.12  & 0.93\pm0.05  \\
     & n3     & SF        & 16.947995  & -17.507169 & 0.31\pm0.03 & 1.77\pm0.02 & 27.43\pm0.04 & 29.4\pm0.06  & 0.88\pm0.06  & 1.5\pm0.17   \\
     & n4     & SF        & 16.947943  & -17.507019 & 0.10\pm0.02 & 1.63\pm0.05 & 26.96\pm0.09 & 29.2\pm0.14  & 0.68\pm0.14  & -0.31\pm0.03 \\
     & n5     & SF        & 16.948030  & -17.506986 & 0.11\pm0.02 & 1.64\pm0.04 & 26.98\pm0.07 & 29.21\pm0.11 & 0.69\pm0.11  & 0.68\pm0.04  \\
     & n6     & AGN/SBnuc & 16.948336  & -17.506803 & 0.89\pm0.01 & 2.05\pm0.01 & 27.89\pm0.01 & 29.29\pm0.01 & 0.77\pm0.01  & 2.12\pm0.06  \\
\enddata
\tablecomments{(1) Identifier for the IRAS system, corresponding to column (1)
in Table \ref{tab:sample}. For systems marked with $^*$, 15 GHz measurements are
reported, due to either non- or poor detection (SNR $<$ 5) at 33 GHz. For all other regions, 33\,GHz measurements are used (i.e. $\nu = 33$).
(2) Identifier for regions within each IRAS system identified and characterized
using \textit{Astrodendro}, with``nuclear" and ``off-nuclear" regions labeled with ``n" and
``extra-nuclear" regions labeled with ``e". Regions within each system are numbered
following the order of ascending declination. (3) Region type indicating the
most likely source for the detected radio emission. See Section \ref{sec:reg_type}
and the Appendix notes for details. (4) \& (5): J2000 coordinates of the emission peak
of the region, in degrees. (6): Measured flux density of the region at 33 GHz in mJy. For regions in systems marked with $^*$ in column (1), 15 GHz measurements are
reported instead. We note that in Arp\,240 (NGC\,5257/8), regions in NGC\,5258 were only detected at 15 GHz.} (7) Region effective radius. For unresolved regions smaller than the synthesized beam accounting for uncertainties, the effective radius of the beam is reported instead as an upper-limit for the region size, indicated with "$<$", as described in Section \ref{sec:lum}. (8) Spectral luminosity calculated from column (6) and Table \ref{tab:sample}. (9) Spectral luminosity surface density. For unresolved regions, the reported values are lower-limits and marked with "$>$". (10) \& (11) Brightness temperature of the region in K, derived using \texttt{Astrodendro} measurements, and via Gaussian-fitting using CASA \texttt{imfit}, in $\log_{10}$ units. Values for unresolved regions are marked with "$>$". See Section \ref{sec:T_b}.
\end{deluxetable*}
\end{longrotatetable}

\movetabledown=15mm
\startlongtable
\begin{longrotatetable}
\begin{deluxetable*}{lllRRRRRRRRR}
\tablecaption{Measured and Derived Quantities for Regions Identified at Matched Resolutions \label{tab:kaku_reg}}
\tablenum{3}
\tablecolumns{12}
\tablehead{
\colhead{ID} &
\colhead{Region} & 
\colhead{Type} & 
\colhead{$S_\mathrm{33}$} &
\colhead{$S_\mathrm{15}$} &
\colhead{$\alpha_\mathrm{15 - 33}$} & 
\colhead{$f_\mathrm{th}$} & 
\colhead{log R$_e$} &
\colhead{log SFR} & 
\colhead{log $\Sigma_\mathrm{SFR}$} & 
\colhead{log SFR$_\mathrm{th}$} & 
\colhead{log $\Sigma_\mathrm{SFR_{th}}$} \\
\colhead{} &
\colhead{} & 
\colhead{} & 
\colhead{(mJy)} & 
\colhead{(mJy)} & 
\colhead{} & 
\colhead{(\%)} & 
\colhead{(pc)} & 
\colhead{(M$_\odot$ yr$^{-1}$)} & 
\colhead{(M$_\odot$ yr$^{-1}$ kpc$^{-2}$)} &
\colhead{(M$_\odot$ yr$^{-1}$)} &
\colhead{(M$_\odot$ yr$^{-1}$ kpc$^{-2}$)} \\
\colhead{(1)} &
\colhead{(2)} & 
\colhead{(3)} & 
\colhead{(4)} & 
\colhead{(5)} & 
\colhead{(6)} & 
\colhead{(7)} & 
\colhead{(8)} & 
\colhead{(9)} & 
\colhead{(10)} &
\colhead{(11)} &
\colhead{(12)} 
}
\startdata
1         & n1          & AGN       & 3.54\pm0.37             & 5.01\pm0.51             & -0.44\pm0.19            & 62\pm24                 & 2.08\pm0.01            &              &             &              &                       \\
2$^*$        & n1          & SBnuc     & < 1.34                  & 1.10\pm0.11             & < 0.25                  & < 100                   & 2.14\pm0.02            & 1.03\pm0.21  & 2.26\pm0.21 & < 1.03       & < 2.26                \\
3         & n1          & SF        & 0.20\pm0.02             & 0.07\pm0.04             & 1.38\pm0.72             & 100                     & 1.94\pm0.01            & 0.06\pm0.11  & 1.69\pm0.11 & 0.06\pm0.11  & 1.69\pm0.11           \\
          & n2          & SF        & 0.13\pm0.01             & 0.15\pm0.02             & -0.19\pm0.24            & 91\pm9                  & 1.62\pm0.04            & -0.29\pm0.05 & 1.97\pm0.10 & -0.16\pm0.13 & 2.11\pm0.15           \\
          & n3          & SF        & 0.31\pm0.03             & 0.39\pm0.05             & -0.29\pm0.20            & 80\pm20                 & 1.77\pm0.02            & 0.08\pm0.05  & 2.04\pm0.06 & 0.16\pm0.13  & 2.13\pm0.14           \\
          & n4          & SF        & 0.09\pm0.01             & 0.09\pm0.02             & 0.00\pm0.31             & 100                 & 1.64\pm0.06            & -0.28\pm0.13 & 1.95\pm0.17 & -0.28\pm0.13 & 1.95\pm0.17           \\
          & n5          & SF        & 0.09\pm0.01             & 0.12\pm0.02             & -0.34\pm0.26            & 74\pm26                 & 1.63\pm0.04            & -0.46\pm0.05 & 1.78\pm0.10 & -0.41\pm0.19 & 1.83\pm0.21           \\
          & n6          & AGN/SBnuc & 0.89\pm0.09             & 0.90\pm0.10             & -0.01\pm0.19            & 100                 & 2.05\pm0.01            &              &             &              &                       \\
\enddata
\tablecomments{(1) Identifier for the IRAS system, corresponding to column (1)
in Table \ref{tab:sample}. (2) Identifier for regions corresponding to column
(2) in Table \ref{tab:native_reg}. Due to the lower-resolution of the beam-matched
15 and 33 GHz images, several regions identified at native-resolution are
blended together. In these cases, multiple region identifiers are provided for a
single measurement. (3) Region type indicating the most likely source for the
detected radio emission. See \S \ref{sec:reg_type} and Appendix A for
details. (4): Measured flux of the beam-matched region at 33 GHz. For regions identified at 15\,GHz, upper-limits on the 33\,GHz flux densities are provided. (5): Measured flux density of the beam-matched region at 15 GHz  (6): Measured 15 - 33 GHz spectral index within the beam-matched region. (7) Thermal fraction at 33 GHz derived from (6). (8) Effective radius for the beam-matched region, based on region area ($A$) computed by \textit{Astrodendro}, defined as $R_e=\sqrt{A/\pi}$. For unresolved regions that are smaller than the matched-beam accounting for uncertainties, the effective radius of the matched-beam is
reported instead as an upper-limit for the region size, indicated with "$<$".
(8) Star formation rates of the beam-matched region derived from either 33 GHz or 15 GHz
flux density in column (4) or (5). (10) Star formation rate surface density. For
unresolved regions, the reported values are lower-limits, indicated with "$>$".
(11) Thermal-only star formation rate derived from 33 GHz or 15 GHz flux density in column (4) or (5) and
15 - 33 GHz spectral index from column (6). (12) Thermal-only star formation
rate surface density. For regions identified at 15\,GHz, upper-limits are provided for (6), (7), (11) and (12). For unresolved regions, the reported values are
lower-limits, indicated with "$>$".  For column (9) - (12), we only report values for ``SF" and ``SBnuc".}
\end{deluxetable*}
\end{longrotatetable}
\section{Summary}\label{sec:summary}
\indent Local U/LIRGs provide excellent laboratories for studying the intense dust-obscured phase in the evolution of many massive galaxies across the cosmic time. In this study we have used high-resolution ($\sim 0\farcs1$) 33 and 15\,GHz radio continuum VLA observations of 68 local U/LIRGs from the GOALS ``equatorial" VLA Survey (GOALS-ES) to study the properties of AGN and star formation in these extreme systems at 100\,pc scales. The GOALS-ES sample spans the entire range of IR luminosities, distances and merger stages represented in the local U/LIRG population. Below we provide a summary of our major results and conclusions:\\
\\
\textbullet\ Among the 68 systems in the GOALS-ES sample, compact radio continuum sources was detected in 63 systems with our high-resolution VLA observations at either 33 or 15\,GHz. Using \textit{Astrodendro}, we identified and characterized a total of 133 regions of compact radio continuum emission in these systems at the native resolutions, and found the effective radii ($R_e$) range from 8 to 170\,pc. These regions were further classified as 17 ``AGN" (AGN), 9 ``Jet" (AGN jet), 31 ``AGN/SBnuc" (AGN-starburst composite nucleus), 8 ``SBnuc" (starburst nucleus), 50 ``SF"(star-forming clump) and 17 ``Ud" (unsure) based on their locations in the host galaxies as well as multi-wavelength AGN classifications from the literature. While all regions have low brightness temperatures ($T_b \lesssim 10^{4}\,K$), ``AGN" and ``AGN/SBnuc" regions have consistently higher 33\,GHz luminosities ($L_{\rm 33}$) and surface densities ($\Sigma_{L_{\rm 33}}$) compared with ``SF" and ``SBnuc" regions of similar sizes by up to $\sim$ 3\,dex. Comparisons with analytical models of radiation pressure-supported nuclear starburst and with lower resolution X-ray and IR AGN diagnostics suggest that both extreme mode of nuclear starburst and AGN activity may contribute to the elevated 33\,GHz emission in ``AGN" and ``AGN/SBnuc". \\
\textbullet\ We used resolution-matched 15 and 33\,GHz images to measure the 15 -- 33 spectral indices ($\alpha_{15 - 33}$) of a total of 115 regions, with which we estimated the fractional contribution of thermal free-free emission to the total 33\,GHz continuum (thermal fraction; $f_{\rm th}$) in these regions. The 15 - 33\,GHz spectral indices for these regions span a wide range, from $< -2$ to 1.38$\pm$0.72, corresponding to $f_{\rm} \sim 0 - 100$\% assuming a constant non-thermal spectral index of -0.85. While all region types span a wide range of $\alpha_{15 - 33}$, ``SF" and ``SBnuc" have flatter median spectral indices compared with ``AGN" and ``AGN/SBnuc" regions. However, the median spectral index for ``SF" and ``SBnuc" ($\alpha_{15 - 33} \sim -0.46\pm$0.29) are significantly steeper than star-forming regions in nearby normal galaxies measured at similar physical scales, suggesting higher contribution of non-thermal synchrotron emission at 33\,GHz in local U/LIRGs. \\
\textbullet\ For the 48 ``SF" and ``SBnuc" regions measured at matched resolution, we estimated their star formation rates and surface densities from both total 33\,GHz (or 15 GHz) flux densities as well as thermal free-free emission extracted using the estimated $f_{\rm th}$ for each region.  We found that with effective radii of 20 - 140\,pc, these regions have star formation rates and surface densities of 0.14 - 13\,M$_\odot$/yr and 13 - 1600\,M$_\odot$/yr/kpc$^2$, respectively, which are consistently higher than similarly-sized star-forming regions in nearby normal galaxies. Even after accounting for the relatively low estimated 33\,GHz thermal fractions, the estimated thermal-only star formation rates and surface densities still have median values of 0.4\,M$_\odot$/yr and 44\,M$_\odot$/yr/kpc$^2$, respectively, and are at least 2\,dex higher than star-forming regions in normal galaxies. \\

Throughout this study we have demonstrated the elevated star-forming activities in local U/LIRGs relative to nearby normal galaxies at the scales of giant molecular clouds, which motivates comprehensive investigation of the cold molecular gas properties at high resolution in these extreme environments. We have also shown the ubiquity of compact and powerful nuclear activity in local U/LIRGs with a wide range of host properties, despite the fact that the origin for these luminous high-frequency radio emission remains highly debatable. Future multi-frequency high-resolution observations with wider frequency coverage will allow more accurate characterization of the radio SED of these compact radio sources to future investigate their nature, and VLBI observations will help determine the prevalence and contribution from AGN activity. Meanwhile, {\it JWST} will provide crucial information of dust and multi-phase ISM at matched resolutions.\\

\noindent
Y.S. would like to thank J. Molden and M. Perrez-Torres for sharing preliminary results on AGN identification using e-MERLIN,  T. Thompson for providing helpful insights on comparisons with the TQM05 models, J. Hibbard for helpful discussions on locating galactic nuclei using ALMA datasets, and J. Rich for additional information on optical AGN classifications. Support for this work was provided by the NSF through the Grote Reber Fellowship Program administered by Associated Universities, Inc./National Radio Astronomy Observatory. A.S.E.
and Y.S. were supported by NSF grant AST 1816838. A.S.E. was also supported by the Taiwan, ROC,
Ministry of Science and Technology grant MoST 102-2119-M001-MY3.V. U acknowledges funding support from NASA Astrophysics Data Analysis Program (ADAP) Grant 80NSSC20K0450. H. I. acknowledges support from JSPS KAKENHI Grant Number JP19K23462. The National Radio Astronomy Observatory is a facility of the National Science Foundation operated under cooperative agreement by Associated Universities, Inc. We acknowledge the usage of the HyperLeda database (http://leda.univ-lyon1.fr), and the NASA/IPAC Infrared Science Archive, which is funded by the National Aeronautics and Space Administration and operated by the California Institute of Technology. 
This research made use of APLpy, an open-source plotting package for Python \citep[][]{aplpy2012,aplpy2019}.

\facility{VLA, Spitzer, HST, IRSA, NASA/ADS, IPAC/NED}
\software{Ned Wright’s Cosmology Calculator \citep[][]{wright06},
Astropy \citep[][]{astropy:2018}, Astrodendro \citep[http://www.dendrograms.org;][]{astrodendro}, CASA \citep[][]{mcmullin07}, APLpy \citep[][]{aplpy2012,aplpy2019}.}
\bibliography{reference.bib}

\appendix

\section{Notes on Individual Systems}\label{ap:source_notes}
Here we provide details on the regions identified in each system, along with
their classifications via comparisons with archival optical and IR datasets as
well as information from the literature, when available. Unless otherwise
specified, merger stage classifications are from \cite{stierwalt13},
descriptions of the optical and IR comparisons are based on \textit{y-}band images from
the Pan-STARRS1 database \citep{flewelling20}, and channel maps from
\textit{Spitzer} IRAC \citep[Mazzarella in prep;][]{goalsdoi}. When describing 6.2$\mu$m PAH equivalent width (EW) as an AGN diagnostic, we follow \cite{stierwalt13} and \cite{vardoulaki15} and consider sources with  6.2$\mu$m PAH EW $< 0.27\mu$m to be AGN-dominated, and those with $ 0.27 <$ 6.2$\mu$m PAH EW $ < 0.54\mu$m to have mixed contribution from AGN and starburst, and those with 6.2$\mu$m PAH EW $ > 0.54\mu$m to be starburst-dominated. For galaxies with regions with undetermined nature due to lack of high-resolution ancillary data, we also show comparison between the ancillary data and our radio images for clarity.
\begin{enumerate}[label={[\textit{\arabic*}]}]
    \item \textit{NGC 0034} : This system is a late-stage
    merger. We identified one bright nuclear region in this system at both 15
    and 33 GHz. This region aligns with the galaxy center in the optical and IR.
    X-ray study suggests that this galaxy hosts a buried AGN
    \citep[$N_\mathrm{H} \sim 10^{23}\rm cm^{-2}$;][]{torres-alba18}, and this
    galaxy is classified as a type-2 Seyfert in the optical
    \citep{veilleux95,yuan10}. Therefore we classify the nuclear region
    identified in the radio as ``AGN". 
    \item \textit{MCG -02-01-051} : This galaxy is the
    southern component of the early stage interacting pair Arp 256. Two nuclear
    regions were identified at 33 GHz at native resolution, but the image
    quality is poor (SNR $\lesssim$ 5), therefore in Table \ref{tab:native_reg} we
    report region properties characterized using the native resolution 15 GHz
    image where the two regions blend together into one larger region. Note that
    the 15 -- 33\,GHz spectral index of this region, reported in Table \ref{tab:kaku_reg}, is an upper-limit given the low SNR of the detection at 33 GHz. This region aligns with the optical and IR peak, and no evidence of AGN activity has been reported, therefore here we classify this region as ``SBnuc".
    \item \textit{IC 1623 (VV 114)} : We detect the eastern component
    of this mid-stage merger at both 15 and 33 GHz. In total, 6
    nuclear regions are identified with \textit{Astrodendro}. The brightest
    region aligns with the optical center and is identified as an AGN by
    \cite{iono13} based on elevated HCN/HCO$^+$ ratio. However, analysis of JWST/MIRI imaging data shows that this region has mid-IR colors (e.g. F770W/F560W) consistent with pure star formation, while a much fainter radio region to the southwest shows mid-IR colors more consistent with AGN activity \citep[][]{evans22}. No clear
    signatures of AGN have been found in the X-rays, optical or MIR on global scales, and high excitation MIR coronal lines indicative of AGN activity were not detected on 100-pc scales using JWST/MIRI-MRS spectroscopic datasets in any of the radio-selected regions (Rich et al. in prep). Given the uncertainties, we tentatively classify the brightest region as ``AGN/SBnuc", and
    the rest as off-nuclear ``SF".
    \item \textit{MCG -03-04-014} : Two nuclear regions are
    identified at both 15 and 33 GHz in this non-interacting galaxy. The
    brighter region aligns with the optical and IR peak, as well as the
    dynamical center of the warm molecular gas as revealed in ALMA CO(J=3-2)
    dataset (2013.1.01165.S, PI: S. Haan). The fainter region lies
    on a nuclear spiral arm that connects to the dynamical center. No clear
    detection of AGN has been reported for this galaxy but it has been
    classified as an AGN/SB composite system in the optical by \cite{yuan10}.
    Therefore we classify the brighter region as ``AGN/SBnuc" and the
    fainter region as off-nuclear ``SF".
    \item \textit{CGCG 436-030} : In the native resolution 33
    GHz image, we detect one bright and two faint knots at the optical and IR
    peak of this western component of an early-stage merger. At matched
    resolution at 15 and 33 GHz, these three knots are blended together and were
    identified as one larger extended nuclear region with \textit{Astrodendro}.
    This nuclear region is detected in soft and hard X-ray with \textit{Chandra}
    \citep{torres-alba18}, and the nuclear Mid-IR spectra (slit width $\sim
    4''$) indicate clear dominance of emission from star formation
    \citep{diaz-santos17, inami18}. However, this galaxy is classified as an
    AGN/SB composite system in the optical \citep{vega2008} and radio
    \citep{vardoulaki15}. VLBI observations of this galaxies revealed a high brightness temperature ($T_b > 10^{7}$\,K) component that can be explained with a clustered radio supernovae model \citep{smith98}. Given these uncertainties, here we classify this nuclear region as an ``AGN/SBnuc".
    \item \textit{IRAS F01364-1042} : In this late-stage
    merger, we detect one bright nuclear region close to the optical and IR peak
    of the galaxy. \textit{Chandra} detected both soft and hard X-ray at the
    center of this galaxy, and \citep{iwasawa11} attributed their origin to
    highly-obscured high mass X-ray binaries or AGN. This system is classified
    as an LINER in the optical \citep{veilleux95, yuan10}, and as a likely AGN
    candidate in the radio due to the compactness of its 33 GHz emission
    \citep{barcos17}. Given the above, here we classify this nuclear region as
    ``AGN/SB nuc".
    \item \textit{III Zw 035} : We detect the northern
    component of this pre-merger at both 15 and 33 GHz. At native resolution,
    one bright knot and a much fainter knot are identified at 33 GHz. Analysis of 0\farcs03 resolution ALMA Band 6 continuum (2018.1.01123.S, PI: A. Medling) suggests that the brighter northern region likely hosts the AGN, while the fainter region in the south is part of a clumpy dust ring-like structure that is also producing strong OH maser emission \citep{philstrom01}. At matched
    resolution, these two knots are blended and were identified as one extended
    nuclear region. This region aligns with the optical and IR peak. While no
    direct evidence for AGN currently exists, \cite{gonzalez09} found indirect
    X-ray signatures for a Compton-thick AGN, which is supported by an extremely
    high gas surface density estimated by \cite{barcos17}, who also reported
    that this galaxy has the most compact 33 GHz emission among U/LIRGs in the
    GOALS sample. Given the above, here we classify this extended nuclear region
    as ``AGN/SBnuc". More precisely, the nucleus is located at the brighter
    knot detected at native resolution, which coincides with the dynamical
    center of the molecular gas as revealed by ALMA (2018.1.01123.S, PI: A. Medling). Observation with e-MERLIN at 5\,GHz reveals compact continuum emission at the location of the ``AGN/SBnuc" with peak brightness temperature of 10$^{4.8}$\,K (J. Molden, private communication).
    \item \textit{NGC 0838} : This galaxy is the north-east component of a
    complex pre-merging system with three components, one of which is a closely
    interacting galaxy pair formed by NGC 0833 and NGC 0835. A bright
    region lying within the MIR galaxy ``core" is visible at both 15 and 33 GHz in NGC 0838. This region is
    also detected in the hard X-ray with \textit{Chandra} \citep{torres-alba18},
    and lies to the south of the optical and IR peak, with a faint
    counterpart in the NIR as revealed by archival \textit{HST} NICMOS images
    (11080, PI: D. Calzetti). The soft X-ray emission of this system is very
    extended, likely associated with wind from a starburst\citep{turner01, torres-alba18}, which also shows up in our low resolution $C-$configuration
    image. Given the above, here we tentatively classify this region as off-nuclear ``SF". At matched resolution, detection at 33 GHz is poor (SNR $<$ 5),
    therefore spectral index reported in Table \ref{tab:kaku_reg}
    is an upper-limit. 
    
    \item \textit{IC 0214} : Two regions are identified in this late-stage merger
    at both 15 and 33 GHz. The fainter region aligns with the optical and IR
    peak of the system, and the brighter region lies outside of the galaxy, with no visible IR or optical counterpart. No evidence for AGN has
    been reported in the literature and the high PAH 6.2$\mu$m equivalent width
    measured with \textit{Spitzer} indicates that this system is dominated by
    star formation, which is consistent with optical BPT diagnostics using
    VLT/MUSE (ID: 097.B-0427, PI: G. Privon). Therefore here we
    classify the fainter region as the starburst-dominated nucleus. The bright
    extra-nuclear radio source does not have bright counterparts in optical or
    IR, therefore we tentatively classify this region as ``Bg".
    We report its measured quantities in Table \ref{tab:native_reg} and
    \ref{tab:kaku_reg} for completeness but exclude it from further analysis. 
    \item \textit{NGC 0877} : Due to the limited sensitivity of the
    \textit{A-}configuration observations, we did not detect any 15 or 33 GHz
    emission in this pre-merging system. 
    \item \textit{NGC 0958} : Due to the limited sensitivity of the
    \textit{A-}configuration observations, we did not detect any 15 or 33 GHz
    emission in this isolated galaxy. 
    \item \textit{NGC 1068} : This well-studied isolated Seyfert 2 galaxy
    \citep[e.g.]{yuan10} is the nearest LIRG in our sample. At both 15 and 33
    GHz, three luminous nuclear regions are identified with
    \textit{Astrodendro}. These regions are aligned almost linearly along the
    N-S direction, with the central region being the brightest at both 15 and 33
    GHz. All three regions have been previously identified as radio jets
    associated with a highly obscured AGN, which is likely located within the
    southern region \citep[e.g.][]{gallimore96, gallimore04}. Therefore we have
    classified the southern region as ``AGN", and the other two as ``Jet".
    \item \textit{UGC 02238} : We detect one nuclear region
    in this late-stage merger at both 15 and 33 GHz. This bright region aligns
    with the optical and IR peak of the galaxy. This region is also detected in
    the soft and hard X-ray with\textit{Chandra}, but clear X-ray AGN signature was
    not found \citep{torres-alba18}. The galaxy also does not show excess radio emission relative to FIR emission, as expected from radio AGN \citep[][]{condon95}. However, optical observations have classified this system as a LINER
    \citep{veilleux95} or AGN/SB composite galaxy \citep{yuan10}. Given the
    above, we classify this nuclear region as ``AGN/SBnuc".  
    \item \textit{UGC 02369} : In this pair of early-stage merger, we detect two
    nuclear regions that coincide with the optical and IR peaks of the two galaxies
    at both 15 and 33 GHz. The nucleus of the southern galaxy is detected in
    both soft and hard X-ray with \textit{Chandra}, and is classified as an AGN
    by \cite{vardoulaki15} based on steep radio spectral index between 1.4 and
    8.4 GHz. The northern nucleus is very faint in the X-ray and IR, with no
    reported signatures of AGN. The entire interacting system has been
    classified as HII \citep{veilleux95} or AGN/SB composite system
    \citep{yuan10} in the optical, and MIR diagnostics indicate that the system
    is dominated by star formation \citep{stierwalt13,inami18}. Given the above,
    we classify the southern nucleus as ``AGN/SBnuc", and the northern nucleus
    as ``SBnuc". 
    \item \textit{IRAS F03359+1523} : We detect one bright
    nuclear region in this edge-on eastern component of a late-stage merger at
    both 15 and 33 GHz. This radio source is classified as AGN/SB composite
    based on 1.4 - 8.4 GHz spectral index profile in \cite{vardoulaki15}, but
    the galaxy is classified as starburst-dominated based on optical
    \citep{veilleux95,yuan10} and MIR diagnostics \citep{inami18}. Observation with e-MERLIN at 5\,GHz reveals compact continuum emission at the nucleus with a peak brightness temperature of 10$^{4.5}$\,K (J. Molden, private communication). Given the
    above, here we tentatively classify this region as ``AGN/SBnuc".
    \item \textit{CGCG 465-012} : In this mid-stage merger,
    one bright extra-nuclear region is detected and identified at 15 GHz. No
    emission is detected at 33 GHz, therefore we do not report measurements of
    this region in Table \ref{tab:kaku_reg}. This extra-nuclear region lies in
    the tidal tail of the merger and has a bright counterpart in the X-ray
    \citep{torres-alba18}. We classify this region as extra-nuclear ``SF" following \cite{torres-alba18}. 
    \item \textit{UGC 02982} : Due to the limited sensitivity of the
    \textit{A-}configuration observations, we did not detect any 15 or 33 GHz
    emission in this late-stage merger.
    \item \textit{ESO 550-IG025} : We detect the two nuclei
    of this pair of pre-merger at both 15 and 33 GHz. However, nuclear emission
    from the southern galaxy is much fainter and more diffuse and does not have
    good detection, therefore we only report measurements for the nucleus of the
    northern galaxy in Table \ref{tab:native_reg} and \ref{tab:kaku_reg}. Both
    nuclei are detected in the X-rays with \textit{Chandra} in
    \cite{torres-alba18}. While no clear signatures of AGN have been reported,
    both galaxies have been separately classified as LINER/composite system by \cite{veilleux95} and
    \cite{yuan10} in the optical. Given the above, here we tentatively classify
    the identified nucleus of the northern component as ``AGN/SBnuc".
    \item \textit{NGC 1614} : We detect 13 individual
    star-forming regions along the nuclear star-forming ring of this late-stage
    minor merger at native resolution 33 GHz. At matched resolution, several
    smaller regions are blended together, resulting a total of 8 regions
    identified at both 15 and 33 GHz. A faint nucleus is visible at 33 GHz, but
    the detection is poor, therefore not characterized in this work. The
    property of the nuclear star-forming ring has been studied at various
    wavelengths \citep[e.g.]{alonso-herrero01,konig13,xu15}, and analysis of the
    ring based on radio datasets used in this work is presented in 
    \cite{song21}. Although this galaxy has been classified as an AGN/SB
    composite system in the optical \citep{yuan10}, deep VLBI observation has
    found no evidence for AGN \citep{herrero17}, and the lack of molecular gas
    at the nucleus also excludes the possibility of a Compton-thick AGN
    \citep{xu15}. Following these studies, we classify all these nuclear ring regions as
    off-nuclear ``SF".
    \item \textit{UGC 03094} : We detect one nuclear region
    in this isolated spiral galaxy at both 15 and 33 GHz. This region aligns
    with the optical and IR peak of the galaxy center, and was detected in the
    ultra-hard X-ray with \textit{SWIFT}/BAT \citep{koss13}. While this galaxy does not show excess radio emission relative to FIR emission as expected for radio AGN \citep{condon95}, fine-structure gas
    emission line [Ne V] 14.3\,$\mu$m is clearly detected with \textit{Spitzer} in this galaxy,
    which is a strong signature of AGN presence \citep{petric11}. Given the above, here we classify the
    identified radio nucleus as ``AGN".
    \item \textit{NGC 1797} : This galaxy is in a pre-merging
    system IRAS F05053-0805, together with NGC 1799. We detected four nuclear
    regions at both 15 and 33 GHz, which are star-forming regions along a
    nuclear star-forming ring, whose diffuse emission is detected in
    $C-$configuration 33 and 15 GHz observations \citep{song21}. This galaxy is
    classified as a starburst galaxy in the optical \citep{veilleux95,yuan10}
    and based on PAH 6.2$\mu$m equivalent width \citep{stierwalt13}, with no
    emission detected at the center of the ring in our radio observations. We
    classify these four nuclear regions as off-nuclear ``SF". 
    \item \textit{CGCG 468-002} : In this mid-merging galaxy pair, we detect the
    nuclei of both galaxies at 15 and 33 GHz. Detection of hard X-ray emission
    with \textit{NuSTAR} \citep{ricci17} and [Ne V] 14.3$\mu$m line emission with
    \textit{Spitzer} \citep{petric11} in the southwestern galaxy strongly suggests
    AGN presence, which is not detected in the eastern galaxy. Based on the PAH
    6.2$\mu$m equivalent width, the northeastern galaxy is dominated by star
    formation. \cite{ps15} classified the northeastern galaxy as an AGN/SB composite
    galaxy based on optical diagnostics. Because only the northeastern component is
    a LIRG, here we only report measurements of the northeastern
    nucleus, which we tentatively classify as ``AGN/SBnuc". 
    \item \textit{IRAS F05187-1017} : We detect one nuclear
    region in this isolated galaxy at both 15 and 33 GHz. This region coincides with
    with the optical and IR peak of the galaxy. This galaxy is classified as a
    LINER in the optical \citep{veilleux95, yuan10} and the PAH 6.2\,$\mu$m
    equivalent width indicates that both AGN and star formation could be
    contributing to the emission in this galaxy. Given the above, we classify
    the detected radio nucleus as ``AGN/SBnuc".
    \item \textit{IRAS 05442+1732} : We detect 5 nuclear
    regions in this east-most component of a pre-merging system at both 15 and
    33 GHz. This galaxy is likely dominated by star formation given the
    relatively high PAH 6.2\,$\mu$m\ equivalent width \citep{stierwalt13}, and
    its IR SED agrees well with a pure starburst model \citep{dopita11}. These
    regions have counterparts in the NIR of various brightness based on
    comparison with archival high-resolution WFC3 F110W \textit{HST} image
    (15241, PI: K. Larson). Given the lack of evidence for AGN, these regions
    are likely ``SBnuc" and off-nuclear ``SF". However,
    currently available ancillary information does not allow us to determine the
    precise nature of radio emission in these regions.
    \begin{figure*}[h!]
    \centering
    \includegraphics[scale=0.6]{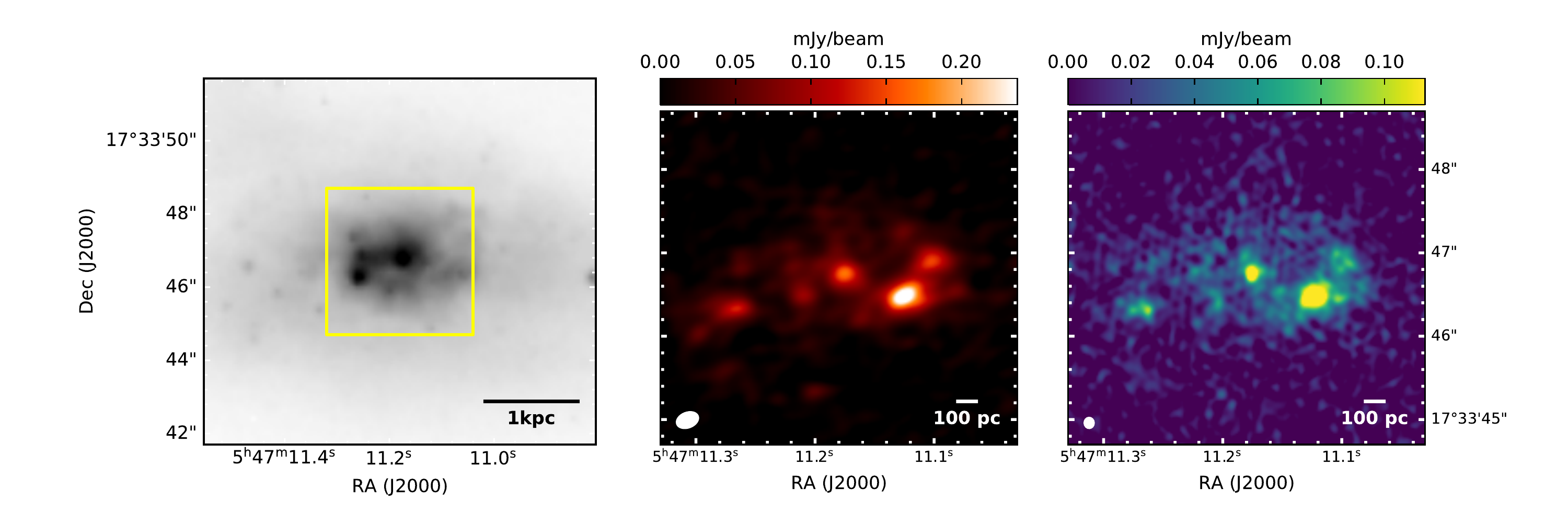}
    \caption{IRAS 05442+1732: \textit{left:} Archival {\it HST} 110W image, displayed in arcsinh stretch with bilinear interpolation. Yellow square outlines the field-of-view of \textit{middle:} 33 GHz VLA continuum observed with \textit{B-}configuration. \textit{right:} 15 GHz VLA continuum observed with \textit{A-}configuration. \label{fig:iras05442}}
    \end{figure*}
    \item \textit{ESO 557-G002} : We detect the nucleus of
    this northern component of a pre-merging galaxy pair at both 15 and 33 GHz.
    Detection at 33 GHz is poor (S/N $<$ 5), therefore we use the 15 GHz image
    to measure the properties of the detected nucleus, and the
    reported 15 - 33 GHz spectral index is an upper-limit. Emission in this
    galaxy is dominated by star formation based on optical \citep{corbett03} and
    MIR PAH 6.2\,$\mu$m equivalent width \citep{stierwalt13}, and no clear
    signatures for AGN have been reported. Therefore here we tentatively
    classify this radio nucleus as ``SBnuc".
    \item \textit{IRAS F07251-0248} : In this late-stage
    merger we detect one nuclear region at 33 GHz. Observation at 15 GHz was
    severely affected by malfunction of the re-quantizer therefore we do not
    report matched resolution measurements of this region in Table
    \ref{tab:kaku_reg}. This region lies within the optical and IR peak of the
    galaxy, and is detected in the soft X-ray with \textit{Chandra}. This galaxy
    is classified as a ``hard X-ray quiet (HXQ)" source by \cite{iwasawa09}, and
    the X-ray emission may come from obscured high-mass X-ray binary or AGN. The
    low PAH 6.2\,$\mu$m equivalent width detected in this galaxy \citep
    {stierwalt13} also indicates potential AGN presence. Given the above, we
    classify the radio nucleus as ``AGN/SB nucleus". 
    \item \textit{MCG +02-20-003} : In this northern
    component of a pre-merging system, we identify 3 radio regions at 33 GHz.
    Observation at 15 GHz was severely affected by malfunction of the
    re-quantizer therefore we do not report matched resolution measurements of
    these regions in Table \ref{tab:kaku_reg}. These regions all coincide with
    the optical and IR peak of the galaxy, and lie within the elongated galaxy
    nucleus detected in Pa$\alpha$ with \textit{HST} NICMOS \citep[(10169, PI: A.
    Alonso-Herrero);][]{alonso09, larson20} (see Figure \ref{fig:mcg+02-20-003}). This galaxy is classified as an AGN/SB composite galaxy in
    the optical by \cite{alonso09}. While a low PAH 6.2\,$\mu$m is detected
    \citep{stierwalt13}, \cite{alonso12} did not find evidence for AGN via IR
    spectral decomposition. Currently available ancillary information does not allow us to determine the precise nature of radio emission in these
    regions.
    \begin{figure*}[h!]
    \centering
    \includegraphics[scale=0.6]{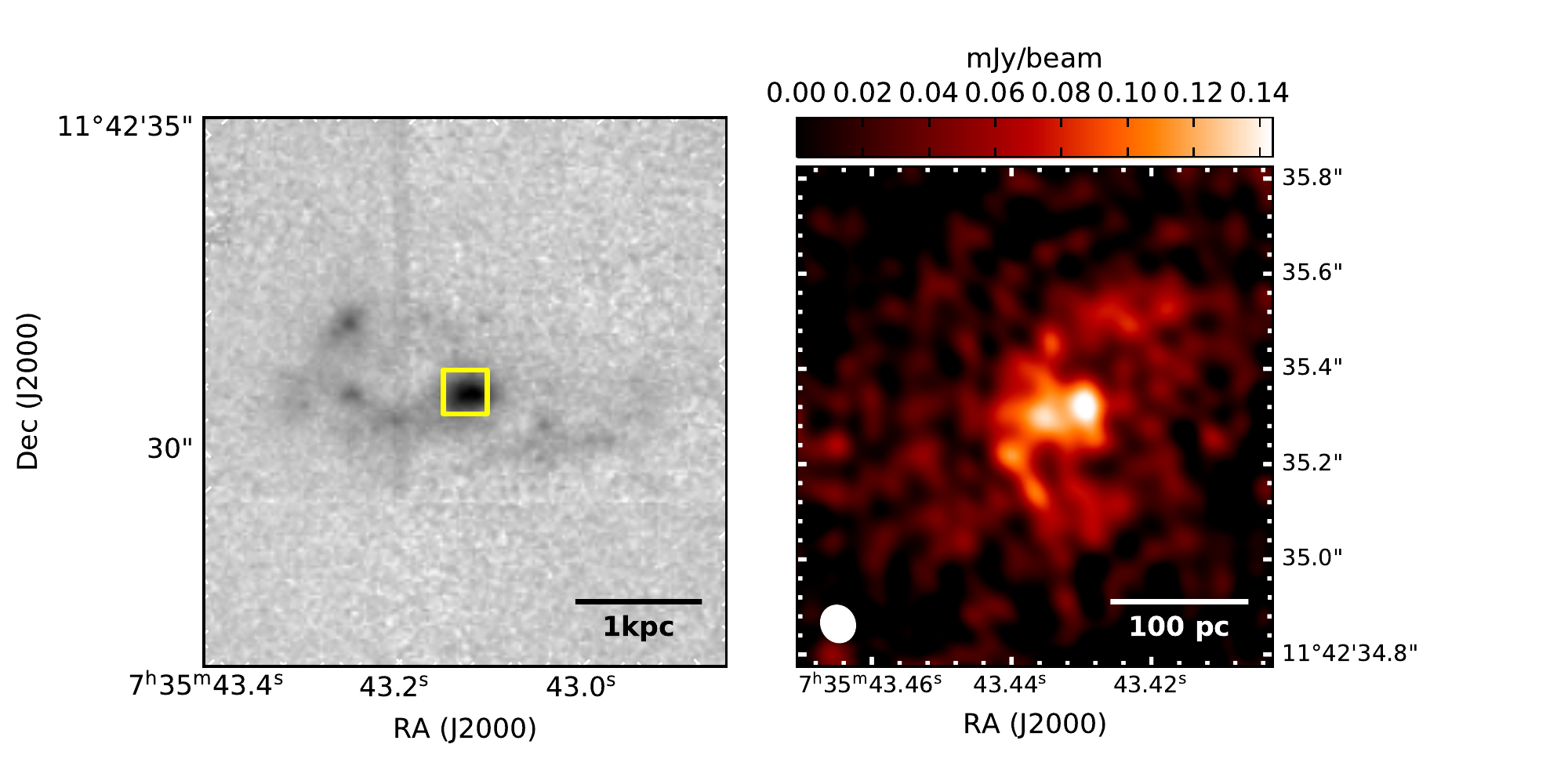}
    \caption{MCG +02-20-003: {\it HST} Pa$\alpha$ image (\textit{left}), displayed in arcsinh stretch with bilinear interpolation. The yellow square outlines the field-of-view of the 33 GHz VLA continuum observed with \textit{B-}configuration (\textit{right}), showing bright clumpy nuclear star-forming structures. \label{fig:mcg+02-20-003}}
    \end{figure*}
    \item \textit{IRAS 09111-1007} : Due to the limited sensitivity of the
    \textit{A-}configuration observation, we did not detect 33 GHz emission in
    this mid-stage merging galaxy pair. At 15\,GHz, we detect three regions within the MIR peak of the eastern component, with the central region aligned with optical peak. This galaxy was not classified as X-ray AGN, but strong SI XIII line was detected with Chandra \citep[][]{iwasawa11}, which may come from a buried AGN. [Ne V] 14.3$\mu$m line is detected in the MIR on kpc scales \citep{petric11}, and the system is classified as an Seyfert 2 or LINER in the optical \citep{duc97}. Therefore we classify the central region as ``AGN", and the other two regions as ``Ud" as we do not have sufficient information from other wavelengths to identify the nature of their radio emission. 
    \begin{figure*}[h!]
    \centering
    \includegraphics[scale=0.6]{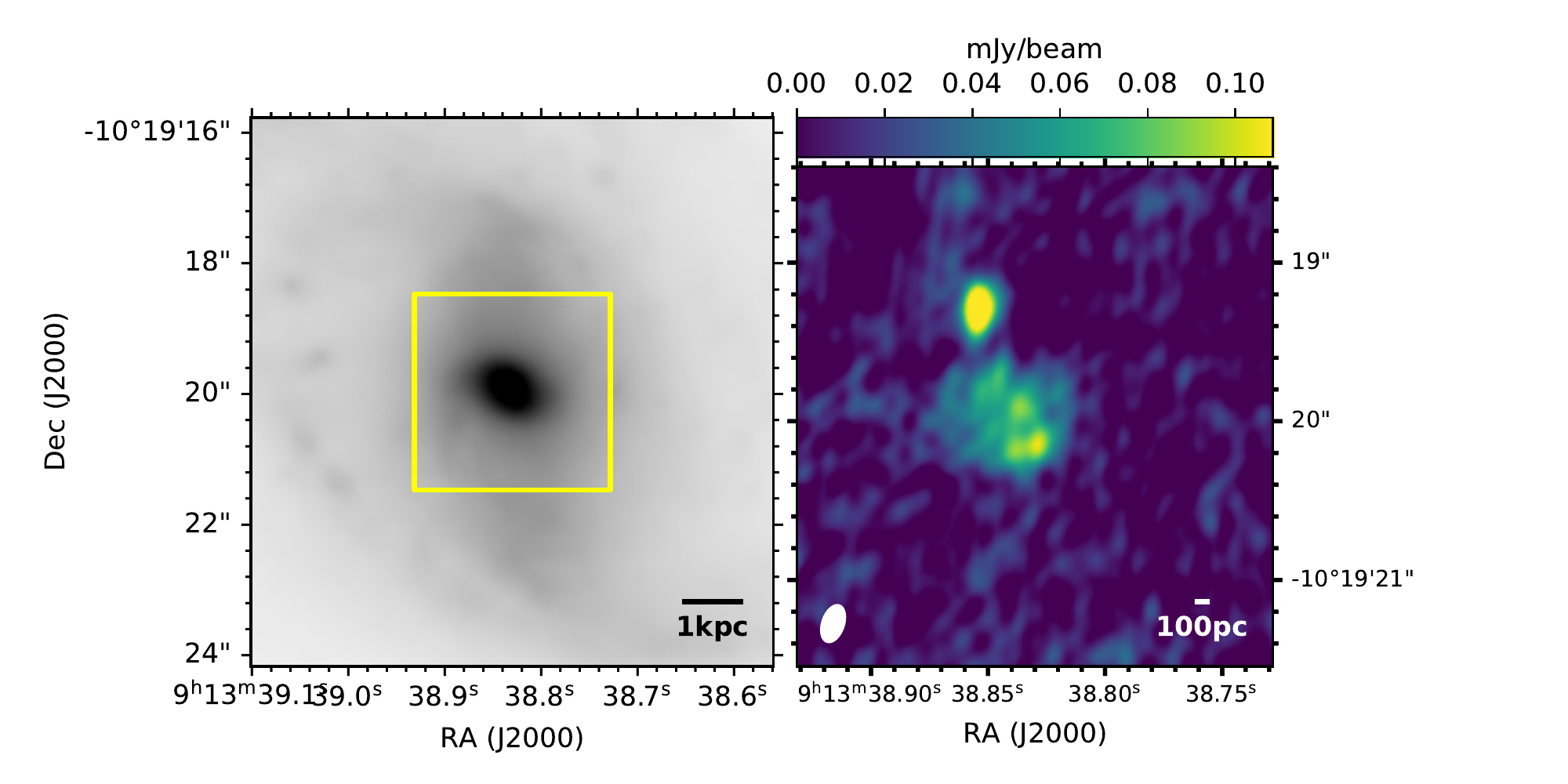}
    \caption{IRAS 09111-10: \textit{left:} Archival {\it HST} 160W image, displayed in arcsinh stretch with bilinear interpolation. \textit{right:} 15 GHz VLA continuum observed with \textit{A-}configuration. \label{fig:iras09111}}
    \end{figure*}
    \item \textit{Arp 303 (IC 0563/4)} : Due to the limited sensitivity of the
    \textit{A-}configuration observation, we did not detect 33 GHz emission in
    this pre-merging galaxy pair. Only the nucleus of the southern component (IC 0563) is detected at 15 GHz. The southern galaxy show excess hard X-ray coming from an off-nuclear ULX region, which is possibly a background AGN. Given that no AGN evidence has been reported in this system, we classify the detected nucleus as ``SBnuc".
    \item \textit{NGC 3110} : Due to the limited sensitivity of the
    \textit{A-}configuration observations, we did not detect emission in
    this late-stage merger at 33 or 15\,GHz. 
    \item \textit{IRAS F10173+0828} : We detect the nucleus
    of this galaxy in pre-merging stage at both 15 and 33 GHz. This nucleus is faint in the soft X-ray and is
    classified as an HXQ source by \cite{iwasawa09}, potentially containing a
    Compton-thick AGN. Optical \citep{vardoulaki15} and MIR diagnostics
    \citep{stierwalt13} indicate a mixture of AGN- and SB-driven emission in
    this galaxy, which also hosts 26 OH mega-masers \citep{yu05}.
    \cite{vardoulaki15} classified this galaxy as a radio AGN based on its
    negative 1.4 - 8.4 GHz spectral index, a signature for face-on AGN with
    jets. Given the above, here we classify the nucleus as ``AGN/SBnuc". 
    \item \textit{CGCG 011-076} : We detect one nuclear
    region in this pre-merging galaxy at both 15 and 33 GHz. This region coincides
    with the optical and IR peak of the galaxy. While no clear evidence for AGN
    presence has been reported, its intermediate MIR PAH equivalent widths
    \citep{stierwalt13, yamada13} indicate potential contribution from an AGN.
    Given the above, we classify the radio nucleus as ``AGN/SBnuc".
    \item \textit{IC 2810} : We detect one nuclear and one extra-nuclear region
    in this northwestern component of a per-merging galaxy pair at both 15 and 33 GHz.
    The fainter region aligns with the optical
    and IR peak of the galaxy, while the brighter region lies in the galaxy disk
    about 5\arcsec south to the nucleus. Optical, MIR and radio diagnostics all
    indicate a mixture of AGN- and SB-driven emission in the galaxy
    \citep{imanishi10,stierwalt13,vardoulaki15}, therefore we classify the
    fainter region as ``AGN/SBnuc", and the brighter region as
    extra-nuclear ``SF". 
    \item \textit{IRAS F12112+0305} : In this late-stage
    merger ULIRG, we detect the two nuclei of the north-south aligned galaxy
    pair at both 15 and 33 GHz, with the southern nucleus about 5 times fainter than the
    northern nucleus. Both nuclei are detected with \textit{Chandra} in the hard
    X-ray but the northern nucleus is slightly fainter and has softer spectrum
    than the southern nucleus \citep{iwasawa11}. Neither of the two nuclei was
    detected with \textit{NuSTAR} and analysis of the
    \textit{Chandra/XMM-Newton} spectrum indicates no presence of AGN
    \citep{ricci21}. While no clear AGN signatures are present, this system has
    been classified as a type-2 Seyfert in the optical \citep{yuan10}, and the
    southern nucleus is classified as AGN/SB in the radio by
    \cite{vardoulaki15}. Given the above, here we tentatively classify both
    nuclei as ``AGN/SBnuc".
    \item \textit{IRAS F12224-0624} : In this isolated
    galaxy, we detect one nuclear region, at both 15 and 33 GHz, that coincides
    with the galaxy's optical and IR peak. Small equivalent width of PAH
    emission at both 6.2 and 3.3 $\mu$m strongly suggests AGN presence
    \citep{stierwalt13,yamada13}, and this galaxy has also been classified as a
    Seyfert 2 galaxy in the optical \citep{yuan10}. Therefore here we classify
    the nucleus as ``AGN".
    \item \textit{NGC 4418} : We detect one nuclear region in
    this isolated galaxy at both 15 and 33 GHz. This region coincides with the
    optical and IR peak of the galaxy, and the nature of this nucleus has been
    under active debate. While small equivalent widths of PAH emission at both
    6.2 and 3.3 $\mu$m indicate AGN-dominated emission
    \citep{stierwalt13,yamada13}, this galaxy is not detected in the ultra-hard
    X-rays \citep{koss13, ricci21}. The flat hard X-ray spectrum
    potentially points to an Compton-thick AGN \citep{maiolino03}, and VLBI
    observation at 5 GHz with \textit{EVN} indicates that the nuclear radio
    emission in this galaxy comes from a mixture of AGN and star formation
    \citep{varenius14}. Given the above, we classify this nucleus as
    ``AGN/SBnuc".
    \item \textit{CGCG 043-099} : We detect 3 nuclear regions
    in this late-stage merger at 33 GHz at native resolution, two of which blend
    into one at 15 - 33 GHz matched resolution. All these regions coincides with
    the IR and optical peak of the system, whose emission is dominated by star
    formation given the relatively large PAH 6.2\,$\mu$m equivalent width. Optical studies have classified this system as an type-2 Seyfert
    \citep{toba13}, or a shock-dominated starburst \citep{rich15}. Archival \textit{HST}/WFC3 F160W image (11235, PI: J.
    Surace) shows a unresolved galaxy nucleus that encompass all radio regions (Figure \ref{fig:cgcg043-099}).
    Given the late merger stage, we are likely detecting the obscured double/triple nuclei of this system. However, currently available ancillary information does not allow us to determine the precise nature of radio emission in these regions.
    \begin{figure*}[h!]
    \centering
    \includegraphics[scale=0.6]{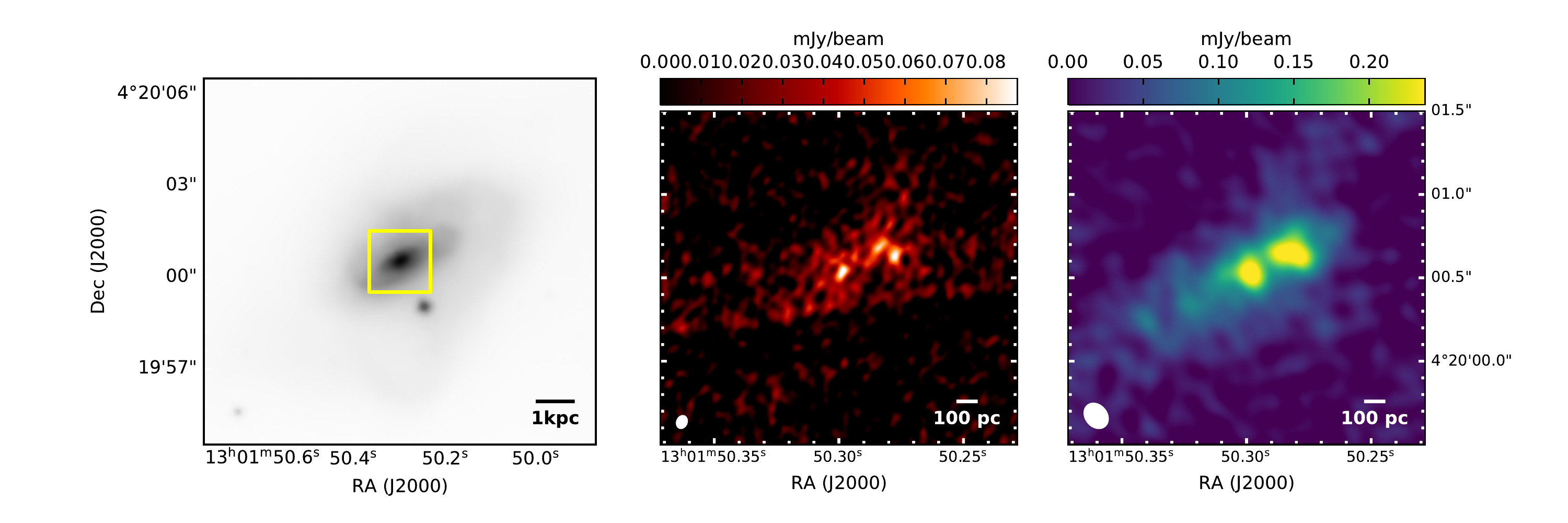}
    \caption{CGCG 043-099: \textit{left:} Archival {\it HST} 160W image, displayed in arcsinh stretch with bilinear interpolation. \textit{middle:} 33 GHz VLA continuum observed with \textit{A-}configuration. \textit{right:} 15 GHz VLA continuum observed with \textit{A-}configuration. \label{fig:cgcg043-099}}
    \end{figure*}
    \item \textit{MCG-02-33-098} : We detect the nucleus of
    this western component of the early-stage merger IRAS F12596-1529 at both 15
    and 33 GHz. Based on optical and MIR diagnostics
    \citep{veilleux95,yuan10,stierwalt13}, emission in this galaxy is dominated
    by star formation, and no AGN is detected in the ultra-hard X-rays with
    \textit{Swift}/BAT \citep{koss13}. Therefore we classify this nucleus as an
    SB-dominated nucleus.
    \item \textit{NGC5104} : In this isolated galaxy, we
    detect one nuclear region at the location of the optical and IR peak of the
    galaxy at both 15 and 33 GHz. While no signatures of AGN are present in the
    X-rays \citep{koss13,privon20}, this galaxy is classified as a LINER
    \citep{veilleux95} or AGN/SB composite system \citep{yuan10} in the optical,
    which is supported by an intermediate PAH 6.2\,$\mu$m equivalent width
    \citep{stierwalt13}. Give the above, we classify the radio nucleus as    ``AGN/SBnuc".
    \item \textit{MCG-03-34-064} : In this southern component
    of a pre-merging galaxy pair, we detect four nuclear regions aligned
    linearly on NE-SW direction in native resolution 33 GHz image. At 15 and 33
    GHz with matched resolutions, these four regions are blended together into
    one elongated region that lies within the IR and optical peak of the galaxy.
    Clear signatures of AGN are present in the X-rays
    \citep{torres-alba18,ricci17}, optical \citep{corbett02} and MIR
    \citep{petric11}. This galaxy also show significant radio excess relative to
    the radio-FIR correlation \citep{condon95,corbett03}, therefore the four linearly
    aligned regions we observe at native resolution are very likely AGN and
    radio jets, which is confirmed by steep 15 - 33 GHz spectral index ($< -0.9$)
    measured at matched resolution. Although the precise location of the AGN is
    unclear from the currently available observations, the morphology of radio
    emission resembles commonly observed one-sided radio jets, with the brighter
    and more compact knots resulting from Doppler boosting effect from
    relativistic jets traveling close to the line-of-sight
    \citep[e.g.][]{bridle84,singal16}. Given the above, we tentatively attribute
    the nuclear radio emission detected in this galaxy to radio ``jets".
    \item \textit{Arp 240 (NGC 5257/8)} : Both the eastern
    and western components in this early-stage merger have been observed. We
    detect two regions in the western component NGC 5257, both at native
    and at matched resolutions. These two regions
    are about 12\arcsec apart along the N-S direction, and the northern region
    coincides with the optical and IR peak of the galaxy while the southern
    region lies at the tip of a spiral arm. This galaxy is classified as a
    starburst galaxy in the optical \citep{veilleux95}, MIR \citep{stierwalt13}
    and radio \citep{vardoulaki15}, therefore we classify the northern region as
    ``SBnuc" and the southern region as extra-nuclear ``SF". For the eastern component NGC 5258, we detect two extra-nuclear regions in one of the spiral arms, but only at 15\,GHz due to limited sensitivity at 33 GHz. These two regions are classified as ``SF". 
    \item \textit{NGC 5331} : In this mid-merging galaxy
    pair, we only detect radio emission in the southern galaxy NGC 5331S given
    the sensitivity limits. At both 15 and 33 GHz, we identify 7 regions in
    total in NGC 5331S, with 6 regions residing within the bulk of MIR emission
    at the galaxy center, and 1 region on the northern spiral arm. Comparison
    with high-resolution ALMA CO(J=2-1) data (2017.1.00395.S, PI: T.
    D\'{i}az-Santos) indicates that the nuclear radio knots lie
    along the edge of a inclined rotating nuclear disk. This galaxy is
    classified as a starburst galaxy in the optical \citep{wu98}, in agreement
    with a high PAH 6.2\,$\mu$m equivalent width\citep{stierwalt13}, and no AGN
    is detected in the X-rays \citep{torres-alba18}. Therefore we classify the
    nuclear regions as off-nuclear ``SF", and the region on the spiral
    arm as extra-nuclear ``SF". 
    \item \textit{IRAS 14348-1447} : In this late-stage
    merger, we detect the two nuclei of the SW-NE aligned merging galaxy pair at
    both 15 and 33 GHz. Both galaxies have been separately classified as LINERs
    \citep{veilleux95} or AGN/SB composite galaxies \citep{yuan10} in the
    optical, as well as in the radio \citep{vardoulaki15}. Both nuclei are
    detected in the X-rays with \textit{Chandra}, but only the southwestern
    nucleus is bright in the hard X-ray with a spectrum consist with the model
    for buried AGN \citep{iwasawa11}. Therefore we classify the southern nucleus as
    ``AGN", and the northern nucleus as ``AGN/SBnuc".
    \item \textit{CGCG 049-057} : In this isolated galaxy we
    detect one nuclear region at both 15 and 33 GHz. This region coincides with
    the optical and IR peak of the galaxy. While no AGN signatures are present
    based on IR and optical studies \citep[e.g.][]{stierwalt13, imanishi10,
    alonso12}, this galaxy likely contains an buried AGN based on X-ray
    \citep{torres-alba18} and radio \citep{baan06} analysis, which is supported
    by the high gas column density measured in the nucleus \citep{falstad15}.
    Therefore we classify this nucleus as ``AGN".
    \item \textit{NGC 5936} : In this isolated galaxy we detect the nucleus at
    both 15 and 33 GHz. However, detection at 33 GHz is poor (S/N $<$ 5), so we
    only report measurements at 15 GHz for this nucleus in Table
    \ref{tab:native_reg} and its spectral index reported in \ref{tab:kaku_reg}
    is an upper-limit. This galaxy has relatively large PAH 6.2\,$\mu$m
    equivalent width indicating SB-dominated emission, but it has been
    classified as an AGN/SB composite galaxy in the optical \citep{yuan10,
    alonso12}. Given the above, we tentatively classify this nucleus as ``AGN/SBnuc". 
    \item \textit{NGC 5990} : We detect the nucleus of the bright southern
    component of this pre-merger at both 15 and 33 GHz. However, detection at 33
    GHz is poor (S/N $<$ 5), so we only report measurements at 15 GHz for this
    nucleus in Table \ref{tab:native_reg} and its spectral index reported in
    \ref{tab:kaku_reg} is an upper-limit. While this galaxy does not show excess radio emission relative to FIR emission as expected for radio AGN \citep[][]{condon95}, it has been classified as a
    type-2 Seyfert in the optical \citep{yuan10} and [Ne V] 14.3\,$\mu$m line is detected at the
    nucleus on kpc scale \citep{petric11}, which is a clear signature of AGN.
    Therefore we classify this nucleus as ``AGN".
    \item \textit{IRAS F16164-0746} : In this late-stage
    merger, at both 15 and 33 GHz we detect a compact luminous nuclear region
    with faint elongated emission on its sides along the E-W direction and
    perpendicular to the galaxy's optical disk. Only emission on the west side
    of this region has strong enough detection to be characterized with
    \textit{Astrodendro}, along with the bright region itself. Despite having a
    relatively large PAH 6.2\,$\mu$m equivalent width \cite{stierwalt13}, this
    galaxy is classified as an AGN in the X-rays \citep{torres-alba18} and [Ne V] 14.3\,$\mu$m 
    line is detected at the nucleus on kpc scales \citep{petric11}, which is a
    clear signature for AGN. It has also been classified as a LINER \citep{veilleux95}, or Seyfert 2 \citep{yuan10} in the optical. The bright region coincides with the dynamical
    center of the molecular gas as revealed by ALMA (2017.1.00395.S, PI: T.
    D\'{i}az-Santos), therefore here we assume it to be the location of the AGN.
    The ALMA data also shows an edge-on rotating nuclear molecular disk, with
    the west side of the disk coinciding with the fainter elongated radio
    region, which has a relatively flat 15 - 33 GHz spectral index ($\sim
    -0.3$). Therefore here we tentatively classify this fainter region as
    off-nuclear ``SF".
    \item \textit{CGCG 052-037} : In this isolated galaxy,
    multiple nuclear regions were detected at native resolutions, but only one
    region has high enough S/N to be consistently identified with
    \textit{Astrodendro} at both 15 and 33 GHz. No clear signatures of AGN have
    been reported in the literature, and existing MIR studies classify this
    galaxy as a starburst galaxy \citep{imanishi10,yamada13,stierwalt13}.
    Therefore this region is likely ``SBnuc", or off-nuclear ``SF". However, currently available information does not allow us to determine the nature of the radio emission.
    \item \textit{IRAS F16399-0937N} : In this late-stage
    interacting pair we detect the northern nucleus at both 33 and 15 GHz. While
    no clear evidence for AGN has been presented, a buried AGN possibly exists
    and is producing the OH mega-maser observed in this system
    \cite{sales15,torres-alba18}. Given the above, we classify this nucleus as ``AGN/SBnuc".
    \item \textit{NGC 6240} : In this well-studied late-stage merger, we detect
    the two nuclei of the merging galaxy pair at both 15 and 33 GHz. This system has been classified as a LINER in the optical \citep{veilleux95,yuan10}. Despite
    having a relatively large PAH 6.2\,$\mu$m equivalent width, MIR line
    diagnostics on kpc scales indicate a strong presence of one or two buried
    AGN \cite{armus06}. Analysis in the X-rays show that both nuclei have clear
    AGN signatures \citep{iwasawa11}. These two active nuclei have also been
    resolved with radio VLBI \citep{gallimore04}. Therefore here we classify
    these two nuclei as ``AGN".
    \item \textit{IRAS F16516-0948} : In this late-stage
    merger, we detect two regions that lie outside of the MIR galaxy ``core" at both 15 and 33 GHz. These two extra-nuclear radio-emitting regions were also identified by \cite{herrero17} at 8\,GHz. However, only the region to the east has good enough detection to be identified by \textit{Astrodendro} and only at 15 GHz, hence we only report the 15 GHz measurements in Table \ref{tab:native_reg}
    and the spectral index reported in Table \ref{tab:kaku_reg} are upper-limits. This region lies outside the
    bulk of optical and 3.5$\mu$m IR emission (\textit{Spitzer} IRAS Channel 1),
    yet coincides with the peak of 8$\mu$m IR emission (\textit{Spitzer} IRAS
    Channel 4), as shown in Figure \ref{fig: irasf16516}. While relatively large PAH 6.2\,$\mu$m equivalent width was
    measured in this system \citep{stierwalt13}, the measurement was performed
    centering on the optical peak of the system and misses the region we
    identify at 15 GHz, which may be a highly-obscured nucleus or off-nuclear ``SF". Currently available ancillary information does not
    allow us to determine the precise nature of radio emission in this region.
    \begin{figure*}[h!]
    \centering
    \includegraphics[scale=0.5]{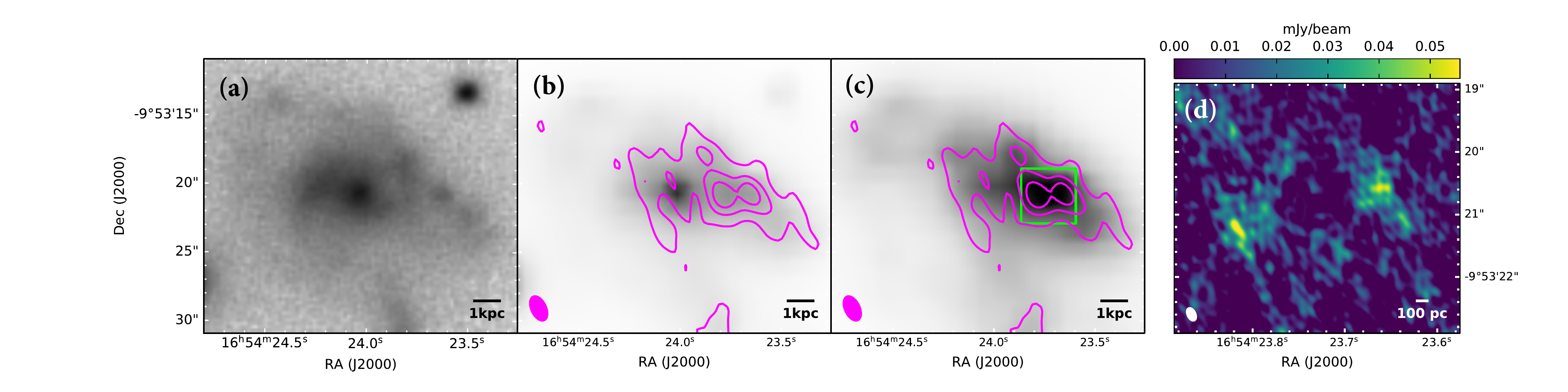}
    \caption{IRAS F16516-0948: \textit{(a):} PanSTARRS1 y band image, displayed in arcsinh stretch with bilinear interpolation. \textit{(b):} Spitzer channel 1 image, overlaid with 33 GHz VLA continuum (magenta contours) observed with \textit{C-}configuration (beam shown on the lower left in magenta). \textit{(c):} Same as \textit{(b)} but on Spitzer channel 4 image. Lime square outlines the field-of-view of \textit{(d):}  15 GHz VLA continuum observed with \textit{A-}configuration. \label{fig: irasf16516}}
    \end{figure*}
    \item \textit{IRAS F17138-1017} : In this late-stage
    merger, we detect 5 regions at both 15 and 33 GHz, 4 of which are aligned
    along N-S direction following bulk of the nuclear optical and IR emission in
    the galaxy. The north-most region is at the location of the nucleus
    \citep{colina15}, and is also the brightest among all regions. Although this
    galaxy is detected in hard X-ray with \textit{NuSTAR} and \textit{Chandra},
    it is inconclusive from analysis of the X-ray spectra whether a buried AGN
    is present \citep{ricci17,torres-alba18}. In the optical this galaxy is
    classified as an AGN/SB composite galaxy \citep{yuan10}, while the PAH 6.2
    $\mu$m equivalent width and NIR line diagnostics indicate that its nuclear
    emission is dominated by SF \citep{stierwalt13,colina15}. Give the above, we
    classify the north-most region as ``AGN/SBnuc", the rest 4 nuclear
    regions as off-nuclear ``SF".
    \item \textit{IRAS 17208-0014} : In this late-stage
    merger, we detect three nuclear regions at native resolutions at both 15 and
    33 GHz. These regions are aligned linearly along the E-W direction, and the
    west-most region appears much more luminous than the rest. At matched
    resolution, these regions are blended together into a larger elongated
    region at 15 and 33 GHz. Prominent shock features have been observed in this
    galaxy \citep{medling15,u19}, which has been attributed to star formation \citep[][]{rich15}. While VLBI observations do not find compact radio cores indicative of AGN activity \citep{momjian03iras, momjian06}, the system has been classified as a AGN-starburst composite systems in the optical \citep{yuan10} and based on an intermediate 6.2\,$\mu$m PAH equivalent width \citep{stierwalt13}. It has been argued in several studies that the system likely hosts buried AGN \citep[e.g.][]{iwasawa11, falstad21, baba22}. The brightest region on the west
    is located at the dynamical center of the molecular gas as revealed by ALMA
    (2018.1.00486.S, PI: M. Pereira-Santaella). This region and the faint radio region on the east coincide with the locations of the merging dual nuclear disks revealed in milli-arcsecond resolution Keck observations \citep{medling15}. Therefore we classify these two regions as ``AGN/SBnuc". The other faint region in between may be associated with shock or clumpy star formation in one of the nuclear disk, but currently available information is not sufficient to clearly identify its nature. Figure \ref{fig: irasf17207} shows the HST image of the galaxy along with VLA radio continuum images of the nuclear regions studied in this work.
    \begin{figure*}[h!]
    \centering
    \includegraphics[scale=0.5]{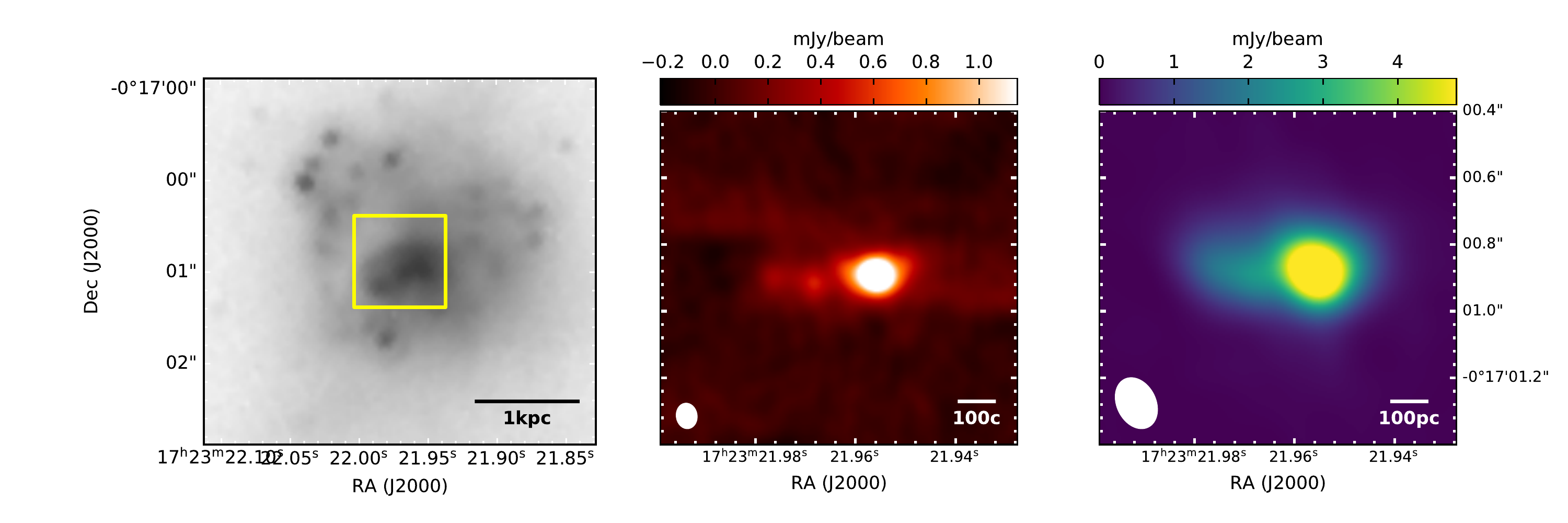}
    \caption{IRAS 17208-0014: \textit{(a):} {\it HST}/NICMOS F160W image, displayed in arcsinh stretch with bilinear interpolation. Yellow square outlines the field-of-view of \textit{(middle):} 33 GHz VLA continuum observed with \textit{A-}configuration (beam shown on the lower left in white). \textit{(right):} 15 GHz VLA continuum observed with \textit{A-}configuration. \label{fig: irasf17207}}
    \end{figure*}

    \item \textit{IRAS 17578-0400} : In the northern
    component of this early-stage merging galaxy triplet, we detect one nuclear
    region at both 15 and 33 GHz. This region aligns with the optical and IR
    peak of this galaxy. This galaxy is not detected in the ultra-hard X-rays
    with \textit{SWIFT}/BAT \citep{koss13}, and it has a large PAH 6.2\,$\mu$m
    equivalent width \citep{stierwalt13}, indicating SB-dominated emission, which is also consistent with the optical classification \citep{rich15}.
    However, this galaxy may host a highly embedded AGN \citep{falstad21}. Therefore we tentatively
    classify this nucleus as ``AGN/SBnuc".
    \item \textit{IRAS 18090+0130W} : We detect two nuclear
    regions and one extra-nuclear region in this western component of a
    early-stage merger at both 15 and 33 GHz. However, the detection at 33 GHz
    is poor (S/N $<$ 5), therefore we only report measurements made at 15 GHz in Table \ref{tab:native_reg}
    and the derived spectral indices reported in Table \ref{tab:kaku_reg} are upper-limits. The two nuclear
    regions lie closely besides each other, and both are located at the IR and optical peak of the galaxy. Archival HST 132N image (ID: 14095, PI: G. Brammer) shows an unresolved galaxy nucleus encompassing the radio regions we detect, as shown in Figure \ref{fig:iras18090}. This galaxy is not detected in the X-rays with \textit{Chandra} or \textit{SWIFT}/BAT
    \citep{koss13,torres-alba18}, but an intermediate PAH 6.2\,$\mu$m equivalent
    width \citep{stierwalt13} indicates that the emission in this galaxy may be
    partially driven by AGN. Therefore the two nuclear regions that we detect
    are likely the AGN/SB nucleus and a off-nuclear SF region or a radio jet.
    Archival However, currently available ancillary information does not allow us to
    determine the precise nature of radio emission in these two nuclear regions.
    \begin{figure*}[h!]
    \centering
    \includegraphics[scale=0.6]{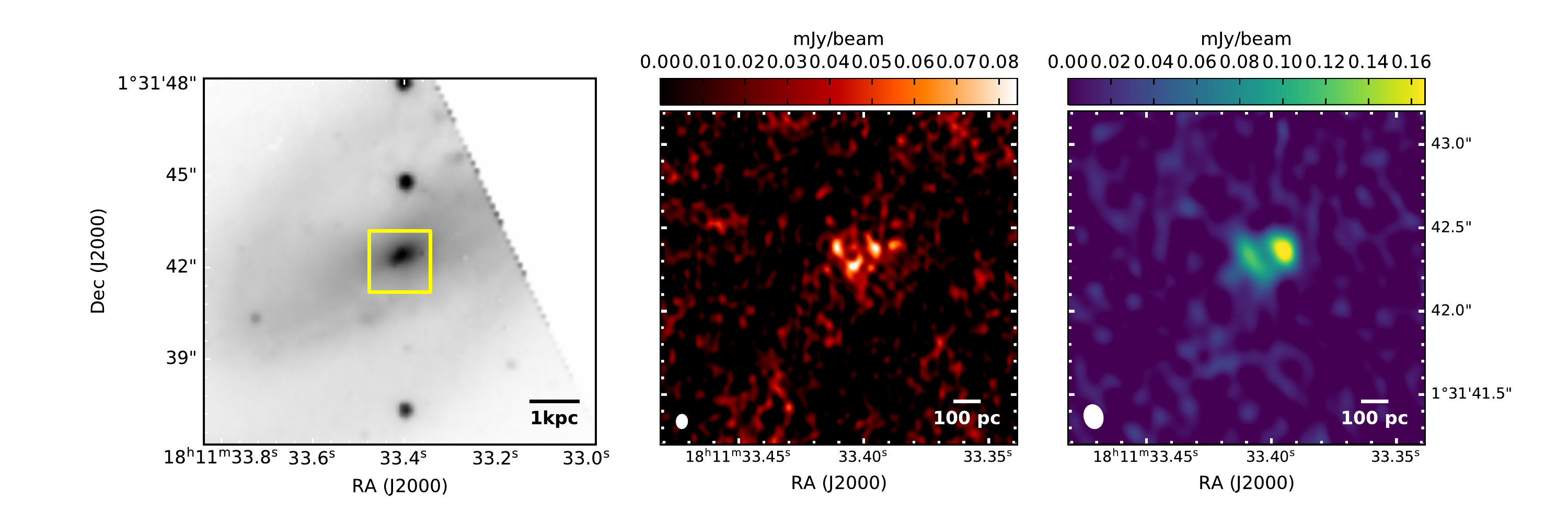}
    \caption{IRAS 18090+0130: {\it HST} F132N image (\textit{left}), displayed in arcsinh stretch with bilinear interpolation. The yellow square outlines the field-of-view of the 33 GHz (\textit{middle}) and 15 GHz (\textit{right}) VLA continuum observed with \textit{A-}configuration, showing at least two radio components of undetermined nature. \label{fig:iras18090}}
    \end{figure*}
    \item \textit{IRAS F19297-0406} : We detect the nucleus
    of this late-stage merger at both 15 and 33 GHz. This galaxy is detected in
    both soft and hard X-rays with \textit{Chandra}, and \cite{iwasawa09}
    classified it as an HXQ, and may contain an buried AGN. Optical and MIR
    diagnostics all point to an AGN/SB composite system
    \citep{yuan10,stierwalt13}. Therefore we classify this nucleus as ``AGN/SBnuc".
    \item \textit{IRAS 19542+1110} : In this isolated galaxy
    we detect its nucleus at both 15 and 33 GHz. While MIR diagnostics do not
    clearly identify AGN signatures \citep{imanishi10, stierwalt13}, this galaxy
    is compact and bright in the hard X-ray and has a very steep spectrum and
    hence was classified as an AGN by \cite{iwasawa11}. AGN signatures were
    not detected in the optical \citep{fluetsch20}. Therefore here we classify this nucleus as ``AGN/SBnuc".
    \item \textit{NGC 6926} : We detect the nucleus of this
    late-stage merger at both 15 and 33 GHz. This galaxy is an optical type-2
    Seyfert \citep{veilleux95,yuan10}, and has clear detection of [Ne V] 14.3\,$\mu$m line at
    the nucleus on kpc scale, which is a clear signature of AGN presence.
    Therefore we classify this nucleus as AGN.
    \item \textit{II Zw 096} : In this mid-stage merger, we
    detect one region at both 15 and 33 GHz. This region coincides with the site
    of an extremely obscured off-nuclear starburst that is responsible for 70\%
    of the IR luminosity of this system \citep{inami10}, which also has the
    hardest X-ray spectrum among all X-rays sources detected in this system due
    to the obscuration \citep{iwasawa11}. Here we tentatively classify this region as
    an extra-nuclear ``SF" following \citep{inami10}. We note that while VLBI observations of the OH megamaser in this region have suggested that it may host an obscured AGN \citep{migenes11}, recent multi-frequency analysis of the radio spectrum shows that it is well described by pure star formation \citep{wu22}.
    \item \textit{ESO 602-G025} : In this isolated galaxy we detect its nucleus
    at both 15 and 33 GHz. However, detection at 33 GHz is poor therefore we
    report the 15 GHz measurements instead in Table \ref{tab:native_reg} and its
    spectral index reported in \ref{tab:kaku_reg} is highly uncertain. Both MIR
    and optical diagnostics point to an AGN/SB composite scenario \citep{yuan10,
    stierwalt13,yamada13}, therefore we classify this nucleus as ``AGN/SBnuc"
    \item \textit{IRAS F22491-1808} : We detect the eastern
    nucleus of this late-stage merger at both 33 and 15 GHz. This nucleus is
    detected in both soft and hard X-rays with \textit{Chandra}, and is
    classified as an HXQ by \cite{iwasawa11} which may contain a
    buried AGN. This system is classified as SB-dominated in the optical
    \citep{veilleux95, yuan10}, but MIR diagnostics indicate potential AGN
    contribution to the emission \citep{stierwalt13}. Given the above, we
    classify this nucleus as ``AGN/SBnuc".
    \item \textit{NGC 7469} : In this southern component of a
    pre-merging galaxy pair we detect its nucleus at both 15 and 33 GHz. This
    galaxy does not show excess radio emission relative to FIR emission, as expected for radio AGN\citep[][]{condon95}, but it is a well-studied optical Seyfert 1 galaxy
    \citep[e.g.][]{veilleux95,yuan10}, and [Ne V] 14.3\,$\mu$m line is clearly detected at its
    nucleus on kpc scales \citep{petric11}. Therefore we classify the detected
    radio nucleus as ``AGN".
    \item \textit{CGCG 453-062} : In this isolated galaxy we
    detect the nucleus at both 33 and 15 GHz. While MIR diagnostics point
    to a SB-dominated scenario \citep{imanishi10,stierwalt13,yamada13}, this
    galaxy has been classified as a LINER \citep{veilleux95} or Seyfert 2
    \citep{yuan10} in the optical and strong [Ne V] 14.3\,$\mu$m line emission has been detected at the nuclear position at kpc scales \citep{petric11}. Therefore we classify this nucleus as an
    ``AGN".
    \item \textit{NGC 7591} : In this isolated galaxy, we
    detect a nuclear star-forming ring at both 33 and 15 GHz. The nucleus and 6
    individual star-forming regions along the ring are resolved only at 15 GHz
    at native resolution, and analysis of these regions is presented in
    \cite{song21}. This ring has also been detected and studied in NIR hydrogen
    recombination lines by \cite{larson20}. For consistency, here we mainly use
    the 33 GHz measurements for our analysis. At 33 GHz, the circum-nuclear ring
    was observed at lower resolution (i.e. \textit{B-}configuration),
    and 3 distinct regions were identified by \textit{Astrodendro}, which
    we classify as off-nuclear ``SF".
    \item \textit{NGC 7592} : In this early-stage merging
    galaxy triplet, we detect the nuclei of the eastern and western components
    at both 15 and 33 GHz. While [Ne V] 14.3\,$\mu$m line is clearly detected for the entire
    unresolved system with \textit{Spitzer} \citep{petric11}, only the western
    galaxy is classified as a Seyfert 2 in the optical
    \citep{veilleux95,yuan10}, which also shows compact emission in the X-rays
    with hard X-ray excess \cite{torres-alba18} and MIR AGN signatures
    \citep{imanishi10}. The eastern component is classified as a starburst
    galaxy in the optical \citep{veilleux95,yuan10}, and no AGN signatures have
    been identified in the X-ray or MIR
    \citep{torres-alba18,imanishi10,stierwalt13}. Therefore here we classify the
    eastern nucleus as ``SBnuc", and the western nucleus as an
    ``AGN".
    \item \textit{NGC 7674} : In this western component of a
    pre-merging galaxy pair, we detect four nuclear regions at both 33 and 15
    GHz. These regions are aligned almost linearly along the E-W direction. This
    galaxy is detected with \textit{NuSTAR} \citep{gandhi17} and classified as a Seyfert 2 in the optical \citep{veilleux95,yuan10}, with a strong
    detection of [Ne V] 14.3\,$\mu$m line \citep{petric11} and small PAH equivalent widths
    \citep{imanishi10,stierwalt13}. This galaxy also shows excess radio emission relative to FIR emission, as expected from radio AGN \citep{condon95}. Using VLBI, \cite{momjian03} also concluded that
    the nuclear radio continuum emission are mostly likely all associated with
    AGN activity and that the AGN itself is located in between the
    brightest two radio components. At 15 and 33 GHz, we detect a faint region
    in between the two brightest radio continuum sources that were observed by
    \cite{momjian03}, and this is the only region that shows a flat 15 - 33 GHz
    spectral index ($\alpha \sim -0.35$) among all four regions, which likely
    marks the location of the AGN. Therefore we classify this faint
    region as ``AGN", and the others as ``Jet". 
    \item \textit{NGC 7679} : In this pre-merging galaxy, we
    detect one nuclear region and one extra-nuclear region at 15 and 33 GHz. The
    nuclear region coincides with the optical and IR peak of the galaxy.
    Observations in the X-ray revealed that this galaxy hosts an unobscured AGN
    \citep{della01}, while it is classified as a Seyfert 2 in the optical
    \citep{veilleux95, yuan10}. Although the large PAH 6.2\,$\mu$m equivalent
    width \citep{stierwalt13} indicates that this galaxy is dominated by star
    formation, [Ne V] 14.3\,$\mu$m  line is clearly detected at the nucleus on kpc scales
    \citep{petric11}, supporting the optical and X-ray classification. Therefore we classify the nuclear region as AGN, and the extra-nuclear region as
    extra-nuclear ``SF".
    \item \textit{MCG -01-60-022}: We did not detect any radio continuum
    emission in this pre-merging galaxy due to incorrect pointing setup.
\end{enumerate}

\startlongtable
\begin{deluxetable*}{lcrrrrrr}
\tablecaption{VLA 15 and 33 GHz observations \label{tab:vlaimaging}}
\tablenum{A1}
\tablecolumns{8}
\tablehead{
\colhead{ID} &
\colhead{Project Code} &
\colhead{$\theta^{33 \mathrm{GHz}}_{M\times m}$} & 
\colhead{P.A.$^{33 \mathrm{GHz}}$} & 
\colhead{$\sigma^{33 \mathrm{GHz}}_\mathrm{rms}$} & 
\colhead{$\theta^{15 \mathrm{GHz}}_{M\times m}$} & 
\colhead{P.A.$^{15 \mathrm{GHz}}$} & 
\colhead{$\sigma^{15 \mathrm{GHz}}_\mathrm{rms}$}\\
\colhead{} &
\colhead{} &
\colhead{} & 
\colhead{(deg)} & 
\colhead{($\mu$Jy\,bm$^{-1}$)} & 
\colhead{} & 
\colhead{(deg)} & 
\colhead{($\mu$Jy\,bm$^{-1}$)}
}
\startdata
1  & 14A$-$471          & 0\farcs14$\times$0\farcs06 & $-$41    & 21.5   & 0\farcs17$\times$0\farcs11 & $-$6     & 21.3   \\
2  & 14A$-$471          & 0\farcs14$\times$0\farcs06 & $-$40    & 20.9   & 0\farcs19$\times$0\farcs12 & $-$7     & 11.6   \\
3  & 20A$-$401,14A$-$471  & 0\farcs28$\times$0\farcs18 & 16     & 8.0    & 0\farcs19$\times$0\farcs10  & $-$15    & 13.6   \\
4  & 20A$-$401,14A$-$471  & 0\farcs28$\times$0\farcs19 & $-$15    & 6.0    & 0\farcs22$\times$0\farcs13 & $-$15    & 11.7   \\
5  & 14A$-$471          & 0\farcs09$\times$0\farcs06 & $-$41    & 16.0   & 0\farcs14$\times$0\farcs12 & 42     & 15.6   \\
6  & 14A$-$471          & 0\farcs16$\times$0\farcs05 & $-$43    & 21.3   & 0\farcs16$\times$0\farcs13 & $-$7     & 12.9   \\
7  & 14A$-$471          & 0\farcs08$\times$0\farcs05 & $-$44    & 18.4   & 0\farcs11$\times$0\farcs10  & $-$18    & 13.2   \\
8  & 14A$-$471          & 0\farcs20$\times$0\farcs06 & $-$43    & 18.9   & 0\farcs20$\times$0\farcs12  & $-$26    & 12.3   \\
9  & 14A$-$471          & 0\farcs11$\times$0\farcs05 & $-$48    & 16.6   & 0\farcs14$\times$0\farcs11 & $-$39    & 13.6   \\
10 & 14A$-$471          & 0\farcs11$\times$0\farcs06 & $-$42    & 14.5   & 0\farcs13$\times$0\farcs12 & $-$28    & 11.4   \\
11 & 14A$-$471, 16A$-$204 & 0\farcs12$\times$0\farcs06 & 1      & 11.5   & 0\farcs18$\times$0\farcs14 & 14     & 8.6    \\
12 & 14A$-$471          & 0\farcs10$\times$0\farcs06  & 1      & 35.3   & 0\farcs14$\times$0\farcs11 & 12     & 120.0  \\
13 & 16A$-$204, 20A$-$401 & 0\farcs24$\times$0\farcs18 & $-$65    & 5.2    & 0\farcs16$\times$0\farcs14 & 15     & 10.4   \\
14 & 14A$-$471, 16A$-$204 & 0\farcs09$\times$0\farcs06 & $-$6     & 11.7   & 0\farcs13$\times$0\farcs11 & 8      & 9.9    \\
15 & 14A$-$471          & 0\farcs09$\times$0\farcs06 & 12     & 15.8   & 0\farcs13$\times$0\farcs11 & $-$174   & 9.4    \\
16 & 14A$-$471, 16A$-$204 & 0\farcs10$\times$0\farcs07  & 7      & 15.3   & 0\farcs13$\times$0\farcs11 & 0      & 9.8    \\
17 & 14A$-$471, 16A$-$204 & 0\farcs10$\times$0\farcs06  & 12     & 12.2   & 0\farcs13$\times$0\farcs11 & 5      & 9.7    \\
18 & 14A$-$471, 16A$-$204 & 0\farcs15$\times$0\farcs07 & 12     & 13.7   & 0\farcs22$\times$0\farcs12 & $-$5     & 11.2   \\
19 & 14A$-$471, 16A$-$204 & 0\farcs12$\times$0\farcs06 & 16     & 13.7   & 0\farcs17$\times$0\farcs11 & 0      & 15.1   \\
20 & 14A$-$471          & 0\farcs07$\times$0\farcs07 & 58     & 15.0   & 0\farcs13$\times$0\farcs11 & 8      & 10.4   \\
21 & 14A$-$471, 20A$-$401 & 0\farcs27$\times$0\farcs19 & $-$34    & 4.0    & 0\farcs16$\times$0\farcs11 & 13     & 10.4   \\
22 & 14A$-$471          & 0\farcs08$\times$0\farcs06 & 63     & 15.3   & 0\farcs14$\times$0\farcs13 & $-$3     & 13.0   \\
23 & 14A$-$471          & 0\farcs10$\times$0\farcs06  & 33     & 19.3   & 0\farcs18$\times$0\farcs13 & 17     & 11.5   \\
24 & 14A$-$471, 20A$-$401 & 0\farcs27$\times$0\farcs18 & $-$69    & 7.2    & 0\farcs13$\times$0\farcs11 & $-$178   & 10.6   \\
25 & 14A$-$471          & 0\farcs15$\times$0\farcs06 & 29     & 23.9   & 0\farcs19$\times$0\farcs10  & $-$2     & 12.7   \\
26 & 14A$-$471, 20B$-$313          & 0\farcs08$\times$0\farcs06 & 6      & 17.8   & 0\farcs15$\times$0\farcs11 &  $-$5    &  13.9    \\
27 & 14A$-$471, 20B$-$313          & 0\farcs08$\times$0\farcs07 & 24     & 17.8   & 0\farcs14$\times$0\farcs11 & $-$10    &  9.8    \\
28 & 14A$-$471, 20B$-$313          & 0\farcs10$\times$0\farcs07  & $-$12    & 16.4   &
0\farcs24$\times$0\farcs13 &  $-$19   & 11.2    \\
29 & 14A$-$471, 20B$-$313          & 0\farcs08$\times$0\farcs07 & $-$17    & 15.5   & 0\farcs19$\times$0\farcs13 &  $-$22   & 11.4    \\
30 & 14A$-$471, 20B$-$313          & 0\farcs09$\times$0\farcs06 & $-$13    & 17.5   & 0\farcs23$\times$0\farcs13 & $-$25    & 11.8    \\
31 & 14A$-$471, 20B$-$313          & 0\farcs07$\times$0\farcs06 & $-$15    & 16.5   & 0\farcs14$\times$0\farcs11 & $-$24    & 15.9    \\
32 & 14A$-$471          & 0\farcs08$\times$0\farcs06 & $-$11    & 12.1   & 0\farcs28$\times$0\farcs13 & 47     & 19.3   \\
33 & 14A$-$471, 20B$-$313          & 0\farcs07$\times$0\farcs06 & $-$37    & 17.0   & 0\farcs15$\times$0\farcs12 &   $-$41  & 13.5    \\
34 & 14A$-$471          & 0\farcs07$\times$0\farcs06 & $-$20    & 12.0   & 0\farcs18$\times$0\farcs13 & 48     & 11.8   \\
35 & 14A$-$471          & 0\farcs09$\times$0\farcs06 & $-$28    & 15.6   & 0\farcs20$\times$0\farcs12  & 33     & 11.4   \\
36 & 14A$-$471          & 0\farcs08$\times$0\farcs06 & $-$19    & 20.6   & 0\farcs19$\times$0\farcs13 & 41     & 27.4   \\
37 & 14A$-$471          & 0\farcs08$\times$0\farcs06 & $-$22    & 12.8   & 0\farcs16$\times$0\farcs13 & 40     & 11.3   \\
38 & 14A$-$471          & 0\farcs10$\times$0\farcs06  & $-$16    & 14.9   & 0\farcs25$\times$0\farcs12 & 31     & 14.5   \\
39 & 14A$-$471, 20A$-$401 & 0\farcs25$\times$0\farcs17 & $-$42    & 7.3    & 0\farcs14$\times$0\farcs12 & 11     & 11.9   \\
40 & 14A$-$471          & 0\farcs10$\times$0\farcs06  & $-$14    & 12.9   & 0\farcs26$\times$0\farcs12 & 31     & 11.1   \\
41 & 14A$-$471          & 0\farcs08$\times$0\farcs06 & 18     & 13.4   & 0\farcs14$\times$0\farcs11 & 14     & 9.7    \\
41 & 14A$-$471          & 0\farcs08$\times$0\farcs06 & 18     & 13.3   & 0\farcs14$\times$0\farcs11 & 14     & 9.7    \\
42 & 14A$-$471, 20A$-$401 & 0\farcs21$\times$0\farcs18 & 63     & 6.2    & 0\farcs14$\times$0\farcs11 & 11     & 9.8    \\
43 & 14A-471          & 0\farcs11$\times$0\farcs06 & $-$26    & 18.4   & 0\farcs21$\times$0\farcs12 & 16     & 13.5   \\
44 & 14A$-$471          & 0\farcs07$\times$0\farcs06 & 0      & 31.1   & 0\farcs13$\times$0\farcs11 & $-$7     & 55.2   \\
45 & 14A$-$471          & 0\farcs08$\times$0\farcs06 & $-$7     & 15.0   & 0\farcs13$\times$0\farcs11 & $-$7     & 10.2   \\
46 & 14A$-$471          & 0\farcs08$\times$0\farcs06 & $-$1     & 13.5   & 0\farcs14$\times$0\farcs11 & -8     & 10.0   \\
47 & 14A$-$471          & 0\farcs08$\times$0\farcs06 & $-$1     & 15.4   & 0\farcs16$\times$0\farcs11 & $-$12    & 19.3   \\
48 & 14A$-$471, 20A$-$401 & 0\farcs20$\times$0\farcs18  & 51     & 9.1    & 0\farcs14$\times$0\farcs11 & $-$10    & 9.8    \\
49 & 14A$-$471          & 0\farcs08$\times$0\farcs06 & $-$1     & 16.1   & 0\farcs17$\times$0\farcs11 & -14    & 14.0   \\
50 & 14A$-$471          & 0\farcs07$\times$0\farcs06 & 0      & 28.7   & 0\farcs14$\times$0\farcs11 & $-$13    & 54.7   \\
51 & 14A$-$471          & 0\farcs09$\times$0\farcs06 & 11     & 16.6   & 0\farcs23$\times$0\farcs14 & 26     & 10.0   \\
52 & 14A$-$471, 20A$-$401 & 0\farcs27$\times$0\farcs18 & 31     & 9.8    & 0\farcs19$\times$0\farcs11 & 23     & 10.3   \\
53 & 14A$-$471          & 0\farcs07$\times$0\farcs06 & 9      & 24.9   & 0\farcs16$\times$0\farcs11 & 26     & 11.7   \\
54 & 14A$-$471          & 0\farcs08$\times$0\farcs07 & $-$48    & 24.6   & 0\farcs15$\times$0\farcs11 & 16     & 11.4   \\
55 & 14A$-$471, 16A$-$204 & 0\farcs08$\times$0\farcs06 & $-$2     & 16.7   & 0\farcs14$\times$0\farcs11 & 15     & 10.9   \\
56 & 14A$-$471          & 0\farcs08$\times$0\farcs06 & $-$11    & 16.4   & 0\farcs15$\times$0\farcs11 & $-$1     & 10.1   \\
57 & 14A$-$471          & 0\farcs07$\times$0\farcs06 & $-$9     & 16.5   & 0\farcs15$\times$0\farcs13 & $-$11    & 11.4   \\
58 & 14A$-$471          & 0\farcs08$\times$0\farcs06 & $-$19    & 16.1   & 0\farcs18$\times$0\farcs11 & $-$7     & 11.5   \\
59 & 14A$-$471          & 0\farcs07$\times$0\farcs06 & $-$37    & 17.1   & 0\farcs16$\times$0\farcs13 & 60     & 10.2   \\
60 & 14A$-$471          & 0\farcs12$\times$0\farcs05 & $-$29    & 20.4   & 0\farcs21$\times$0\farcs11 & $-$12    & 9.8    \\
61 & 14A$-$471          & 0\farcs12$\times$0\farcs05 & $-$31    & 20.9   & 0\farcs21$\times$0\farcs11 & $-$15    & 9.2    \\
62 & 14A$-$471          & 0\farcs09$\times$0\farcs06 & $-$27    & 20.7   & 0\farcs13$\times$0\farcs12 & $-$5     & 9.3    \\
63 & 14A$-$471          & 0\farcs07$\times$0\farcs06 & $-$37    & 17.9   & 0\farcs12$\times$0\farcs11 & $-$31    & 10.8   \\
64 & 14A$-$471, 20A$-$401 & 0\farcs30$\times$0\farcs17  & 55     & 7.9    & 0\farcs13$\times$0\farcs11 & $-$18    & 9.3    \\
65 & 14A$-$471          & 0\farcs09$\times$0\farcs06 & $-$23    & 19.0   & 0\farcs15$\times$0\farcs11 & $-$19    & 9.9    \\
66 & 14A$-$471          & 0\farcs08$\times$0\farcs05 & $-$34    & 20.8   & 0\farcs13$\times$0\farcs11 & $-$19    & 12.2   \\
67 & 14A$-$471          & 0\farcs09$\times$0\farcs06 & $-$26    & 17.7   & 0\farcs13$\times$0\farcs11 & $-$13    & 9.3    \\
68 & 14A$-$471          & 0\farcs09$\times$0\farcs06 & $-$23    & 15.7   & 0\farcs15$\times$0\farcs11 & $-$12    & 9.9   
\enddata
\tablecomments{All observations were taken with the VLA in $A-$configuration
except for $Ka-$Band observations carried out in project 20A-401, which used
$B-$configuration instead. For each image, the $\sigma_\mathrm{rms}$ is
measured in an emission-free region before primary-beam correction. $\theta_{M\times m}$ are the synthesized beam FWHM (major $\times$ minor) for the native resolution images produced following procedures described in \S \ref{sec:sampledata}.}
\end{deluxetable*}

\startlongtable
\begin{deluxetable*}{lcccclc}
    \centering
\tablecaption{Sample AGN Classification \label{tab:AGN}}
\tablecolumns{11}
\tablenum{A2}
\tablehead{
\colhead{ID} &
\colhead{X-Ray} & 
\colhead{Optical} & 
\colhead{MIR} & 
\colhead{Radio} &
\colhead{Reference(X-ray/Optical/MIR/Radio)} &
\colhead{Adopted} \\
\colhead{(1)} &
\colhead{(2)} & 
\colhead{(3)} & 
\colhead{(4)} & 
\colhead{(5)} & 
\colhead{(6)} & 
\colhead{(7)} 
}
\startdata
1  & Y            & Y          & N, M, M    & Y           & T18/V95,Y10/I10, S13, I18/V15                                                                      & Y            \\
2  & N            & N          & N*         & -           & R17/V95,Y10/I10, S13, I18/-                                                                          & N            \\
3  & M            & N          & N, M       & N, Y, Y           & G06, G20/V95,Y10/I10, S13/C95, V15, I13                                                                 & M            \\
4  & N            & N,M        & N          & -           & K13/V95,Y10/I10, S13, I18/-                                                                          & M            \\
5  & N            & M          & N, M       & M           & T18/V08/I10, S13/V15                                                                               & M            \\
6  & M            & M          & M, Y       & M           & I11/V95,Y10/S13, I18/V15                                                                           & M            \\
7  & N,M          & N          & N*         & M           & G09, TA18/V95,Y10/I18/PC                                                                            & M            \\
8  & N            & N          & N          & -           & T18,T01/T01/S13/-                                                                                  & N            \\
9  & N            & N          & N          & -           & K13/PC/S13/-                                                                                          & N            \\
12 & Y            & Y          & Y          & Y           & T18/V95,Y10/P11/G96                                                                                & Y            \\
13 & N            & M          & N          & N           & T18/V95,Y10/I10, S13, I18/C95                                                                        & M            \\
14 & N            & N,M        & N*         & Y(S)        & T18/V95,Y10/S13, I18/V15                                                                           & M(S), N(N)   \\
15 & N            & N          & N          & M           & T18/V95,Y10/I10, I18/V15, PC                                                                           & M            \\
18 & N            & M          & -          & Y(S), M(N)  & T18/V95,Y10/-/V15                                                                                  & M(S), M(N)   \\
19 & N            & M          & N          & N           & T18/V95,Y10/I10, S13, I18/H17                                                                      & N            \\
20 & Y            & -          & Y, M       & -           & K13/-/P11, S13/C95                                                                                    & Y            \\
21 & N            & N          & N          & -           & K13/V95,Y10/S13/-                                                                                    & N            \\
22 & N            & M          & N          & -           & R17/P15/S13/-                                                                                      & M            \\
23 & N            & M          & M          & -           & K13/V95,Y10/S13/-                                                                                    & M            \\
24 & -            & -          & N          & -           & -/-/S13/-                                                                                          & N            \\
25 & N            & N          & N          & -           & K13/C03/S13/-                                                                                      & N            \\
26 & M            & -          & Y          & -           & I09, I11/-/S13/-                                                                                   & M            \\
27 & N            & M          & Y, N       & -           & K13/A09/S13, A12/-                                                                                   & M            \\
31 & M            & M          & N, M, N    & Y           & I09, I11/V15/I10, S13, I18/V15                                                                     & M            \\
32 & -            & -          & M          & -           & -/-/S13/-                                                                                          & M            \\
33 & N            & M          & Y, N       & M           & T18/V08,V15/I10, S13/V15                                                                           & M            \\
34 & M            & Y*         & M*         & N(N), M(S)  & I09, I11/V95,Y10/S13/V15                                                                           & M            \\
35 & N            & Y          & Y          & -           & K13/V95,Y10/I10, S13, Y13/-                                                                          & Y            \\
36 & N            & -          & Y          & M           & K13, R21/-/I10, S13, Y13/V13                                                                       & M            \\
37 & N            & Y, N          & N          & -           & T18/T13, R15/S13/-                                                                                      & M            \\
38 & N            & N          & N          & -           & K13/V95,Y10/S13/-                                                                                  & N            \\
39 & N            & M       & M          & -           & K13,P20/V95,Y10/S13/-                                                                              & M            \\
40 & Y            & Y          & Y          & Y           & R17, T18/R93/P11, S13/C95, C03                                                                            & Y            \\
41 & N            & N          & N          & N           & K13/V95/S13/V15                                                                                      & N            \\
42 & N            & N          & N          & -           & T18/W98/S13/-                                                                                      & N            \\
43 & Y(SW), N(NE) & M       & Y*         & M           & I11/V95,Y10/I10, S13/V15                                                                           & Y(SW), M(NE) \\
44 & Y            & N, M       & N, M       & Y           & T18/V95,A12/I10, S13/BK06, F15                                                                     & Y            \\
45 & N            & N, M       & N          & -           & K13/V95,Y10/S13/-                                                                                    & M            \\
46 & N            & Y          & Y          & N           & K13/Y10/P11, S13/C95                                                                                   & Y            \\
47 & Y            & M, Y          & Y, N       & -           & T18/V95,Y10/P11, S13/-                                                                             & Y            \\
48 & -            & -          & N          & -           & -/-/I10, S13, Y13/-                                                                                & M            \\
49 & M            & M          & M          & N           & T18/Y10, R15/S13/BK06                                                                                   & M            \\
50 & Y     & M          & Y*, M*     & Y           & I11/V95,Y10/A06, S13/G04                                                                           & Y            \\
51 & -            & -          & N          & N           & -/-/S13/C02                                                                                          & N            \\
52 & M            & M          & N          & -           & R17, T18/Y10/S13, C15/-                                                                            & M            \\
53 & M            & M, N          & M          & -           & I11/Y10, R15/S13, M15/-                                                                                 & M            \\
54 & N            & N          & N          & M           & K13/R15/S13/F21                                                                                      & M            \\
55 & N            & -          & M          & -           & T18/-/S13/-                                                                                        & M            \\
56 & M            & N, M       & M          & -           & I11/V95,Y10/S13/-                                                                                  & M            \\
57 & Y            & N          & N,M        & -           & I11/F20/I10, S13/-                                                                                 & M            \\
58 & Y            & Y          & Y, M       & -           & K13/V95,Y10/P11, S13/-                                                                               & Y            \\
60 & N            & M       & M          & -           & K13/V95,Y10/S13, Y13/-                                                                               & M            \\
61 & N            & N          & N, M       & -           & I11/V95,Y10/I10, S13/-                                                                             & M            \\
62 & Y            & Y          & Y          & N           & B86/V95,Y10/I10, S13, Y13/C95                                                                        & Y            \\
63 & -            & M, Y          & Y          & -           & -/V95,Y10/P11/-                                                                          & Y            \\
65 & Y(W), N(E)   & Y(W), N(E) & Y(W), N(E) & -           & T18/V95,Y10/I10, P11/-                                                                             & Y(W), N(E)   \\
66 & Y            & Y          & Y          & Y           & K13,G17/V95,Y10/P11, S13, I13/M03, K17                                                                      & Y            \\
67 & Y            & Y          & Y, N       & -           & D01,K13/V95,Y10/P11, S13/-                                                                             & Y
\enddata
\tablecomments{(1): Identifier for each IRAS system matched with Table
\ref{tab:sample}. (2): Whether any AGN has been detected in the X-ray (i.e. ultra-hard X-ray detection, hardness ratio, Fe line detection) where
Y=Yes, N=No, M=Maybe (i.e. analysis is unable to identify origin of X-ray emission). (3) Whether AGN has been detected in the system at optical
wavelengths via optical line ratios (i.e. BPT diagram). We consider ``LINER" to be potentially hosting AGN (i.e. classified as ``M).  (4): Whether any AGN has been detected via Mid-IR diagnostics (i.e.
6.2\,$\mu$m PAH equivalent width, 15 - 30$\mu$m spectral slope, [Ne V] 14.3\,$\mu$m line detection). (5) Whether AGN
has been detected in the radio. For (2) - (5): If AGN classifications from multiple studies conducted at the same wavelength range disagree, the classification from individual studies is presented, separated by comma.  (6): Coded references for AGN identification in
the X-ray/Optical/Mid-IR/Radio. See end of the table caption for full
references.  (7): Adopted AGN classification in this study. If an AGN has been
identified at least two different wavelength ranges, then we consider the system as an
AGN host in this work (i.e. labeled as Y=Yes); if evidence for AGN is identified
at only one wavelength range (Y or M), or if evidence is ambiguous at all wavelengths, then we
consider the system as an potential AGN host (i.e. classified as M=Maybe); If no
AGN evidence is currently identified at any wavelength ranges (i.e. classified as N=No). $^{*}$ indicates classification for the
entire IRAS system instead of individual galaxy components.}
\tablerefs{PC: private communication, A06 \citep{armus06},
A09 \citep{alonso09}, A12 \citep{alonso12}, B86 \citep{barr1986},
BK06 \citep{baan06}, C02 \citep{corbett02},C03 \citep{corbett03}, C15 \citep{colina15}, C95 \citep{condon95}
D01 \citep{della01}, F15 \citep{falstad15}, F20 \citep{fluetsch20},
F21 \citep{falstad21}, G04 \citep{gallimore04}, G06 \citep{grimes06}, G09
\citep{gonzalez09}, G17	\citep{gandhi17}, G20 \citep{garofali20}, G96
\citep{gallimore96}, H17 \citep{herrero17}, I09 \citep{iwasawa09}, I10
\citep{imanishi10}, I11	\citep{iwasawa11}, I13 \citep{iono13}, I18
\citep{inami18}, K13	\citep{koss13}, K17 \citep{kharb17}, M03	\citep{momjian03}, M15
\citep{medling15}, P10	\citep{petric11}, P15	\citep{ps15}, P20 \citep{privon20}, R15 \citep{rich15},
R17	\citep{ricci17}, R21 \citep{ricci21}, R93 \citep{rush93}, S13
\citep{stierwalt13}, T01	\citep{turner01}, T13 \citep{toba13}, T18
\citep{torres-alba18}, V08	\citep{vega2008}, V14 \citep{varenius14}, V15
\citep{vardoulaki15}, V95	\citep{veilleux95}, W98 \citep{wu98}, Y10
\citep{yuan10}, Y13	\citep{yamada13}.}
\end{deluxetable*}

\section{Analysis result for ancillary VLA Data}\label{ap:ancillary}
\indent Images from BM17 and SFRS (see \S \ref{sec:ancillary}) were re-analyzed following methods described in \S \ref{sec:analysis}. Five of the 22 systems in BM17 are also in the GOALS-ES sample, and emission in two other systems are resolved out in the \textit{A-} or \textit{B-}configuration), therefore we only include results derived for 26 regions in 15 systems from BM17 to complement our discussions in \S \ref{sec:discuss}. These regions are classified into seven ``AGN", ten ``AGN/SBnuc", six ``SF", two ``Ud" and one ``Bg", using procedures described in \S \ref{sec:reg_type}. Regions classified as ``Ud" and ``Bg" were further removed from comparison. One galaxy in the SFRS, NGC 2146, is a LIRG included in the full GOALS sample \citep{armus09}, and hence was also excluded from re-analysis and comparison to the GOALS-ES results. A total of 187 regions are identified at characterized at both 15 and 33\,GHz in 35 SFRS galaxies, but we remove 58 regions identified as AGN, background galaxies and AME candidates by \cite{linden20} from comparison to the GOALS-ES ``SF" and ``SBnuc" regions in \S \ref{dis:fth} and \S \ref{dis:sfr}. Below we present measured and derived quantities for a total of 152 regions identified in the BM17 and SFRS sample that were included in Figure \ref{fig:lum-size}, \ref{fig:l33_agn}, \ref{fig:spx_re} and \ref{fig:sfr_re} in \S \ref{sec:discuss}.

\startlongtable
\begin{deluxetable*}{lcRRRRRR}
\tablecaption{Measured and Derived Quantities for BM17 Regions \label{tab:native_reg_bm17}}
\tablenum{A3}
\tablecolumns{8}
\tablehead{
\colhead{Region} &
\colhead{Type} & 
\colhead{RA} & 
\colhead{Dec} & 
\colhead{log R$_e$} &
\colhead{$S_{\rm 33}$} &
\colhead{log $L_{\rm 33}$} & 
\colhead{log $\Sigma_{L_{\rm 33}}$} \\
\colhead{} &
\colhead{} & 
\colhead{(\degr)} & 
\colhead{(\degr)} & 
\colhead{(pc)} & 
\colhead{(mJy)} & 
\colhead{(\,erg s$^{-1}$ Hz$^{-1}$)} & 
\colhead{(\,erg s$^{-1}$ Hz$^{-1}$ kpc$^{-2}$)} \\
\colhead{(1)} &
\colhead{(2)} & 
\colhead{(3)} & 
\colhead{(4)} & 
\colhead{(5)} & 
\colhead{(6)} & 
\colhead{(7)} & 
\colhead{(8)} 
}
\startdata
IRASF08572+3915\_n1 & AGN    & 135.105726             & 39.065042               & 2.40\pm0.01            & 2.08\pm0.04             & 29.24\pm0.01            & 29.94\pm0.01 \\
UGC04881\_n1        & AGN/SB & 138.981308             & 44.332764               & 2.28\pm0.01            & 1.39\pm0.03             & 28.72\pm0.01            & 29.65\pm0.02 \\
UGC05101\_n1        & AGN    & 143.964993             & 61.35325                & 2.59\pm0.01            & 10.4\pm0.12             & 29.59\pm0.01            & 29.92\pm0.01 \\
MCG+07-23-019\_n1   & AGN/SB & 165.974877             & 40.849986               & 2.29\pm0.01            & 3.95\pm0.09             & 29.07\pm0.01            & 29.99\pm0.03 \\
NGC3690\_e1         & SF     & 172.127739             & 58.563694               & 1.72\pm0.03            & 0.72\pm0.08             & 27.35\pm0.05            & 29.42\pm0.06 \\
NGC3690\_n1         & AGN    & 172.129125             & 58.561333               & 1.91\pm0.01            & 6.82\pm0.1              & 28.33\pm0.01            & 30.01\pm0.02 \\
NGC3690\_e2         & SF     & 172.130562             & 58.563889               & 1.88\pm0.01            & 2.41\pm0.04             & 27.88\pm0.01            & 29.62\pm0.01 \\
NGC3690\_n2         & AGN    & 172.140122             & 58.562958               & 2.13\pm0.01            & 36.54\pm0.32            & 29.06\pm0.01            & 30.31\pm0.01 \\
\enddata
\tablecomments{(1) Region identifier including galaxy name, with ``n" and ``e" indicating region location: ``n" - within the MIR galaxy core measured by \cite{diaz-santos10}, ``e" - extra-nuclear (outside MIR galaxy core). 
(2) Region type indicating the
most likely source for the detected radio emission, classified following \S \ref{sec:reg_type}. (3) \& (4) J2000 coordinates of the emission peak
of the region. (5) Region effective radius. (6) 33\,GHz flux density. (7) 33\,GHz spectral luminosity calculated from column (6). (8) 33\,GHz spectral luminosity surface density, calculated from (5) and (7).}
\end{deluxetable*}

\movetabledown=18mm
\startlongtable
\begin{longrotatetable}
\begin{deluxetable*}{lllRRRRRRRRR}
\tablecaption{Measured and Derived Quantities for SFRS Star-Forming Regions \label{tab:kaku_reg_sfrs}}
\tablenum{A4}
\tablecolumns{12}
\tablehead{
\colhead{Region} &
\colhead{RA} &
\colhead{Dec} &
\colhead{log R$_e$} &
\colhead{$S_\mathrm{15}$} &
\colhead{$S_\mathrm{33}$} &
\colhead{$\alpha_\mathrm{15 - 33}$} & 
\colhead{$f_\mathrm{th}$} & 
\colhead{log SFR} & 
\colhead{log $\Sigma_\mathrm{SFR}$} & 
\colhead{log SFR$_\mathrm{th}$} & 
\colhead{log $\Sigma_\mathrm{SFR_{th}}$} \\
\colhead{} &
\colhead{(\degr)} & 
\colhead{(\degr)} & 
\colhead{(pc)} &
\colhead{(mJy)} & 
\colhead{(mJy)} & 
\colhead{} & 
\colhead{(\%)} & 
\colhead{(M$_\odot$ yr$^{-1}$)} & 
\colhead{(M$_\odot$ yr$^{-1}$ kpc$^{-2}$)} &
\colhead{(M$_\odot$ yr$^{-1}$)} &
\colhead{(M$_\odot$ yr$^{-1}$ kpc$^{-2}$)} \\
\colhead{(1)} &
\colhead{(2)} & 
\colhead{(3)} & 
\colhead{(4)} & 
\colhead{(5)} & 
\colhead{(6)} & 
\colhead{(7)} & 
\colhead{(8)} & 
\colhead{(9)} & 
\colhead{(10)} &
\colhead{(11)} &
\colhead{(12)} 
}
\startdata
NGC0337\_e4  & 14.95841               & -7.57622                & 2.25\pm0.01                & 0.34\pm0.04             & 0.28\pm0.04             & -0.26\pm0.23            & 83\pm17                 & -1.27\pm0.06               & -0.26\pm0.06                  & -1.17\pm0.14                  & -0.16\pm0.14 \\
NGC0337\_e1  & 14.96135               & -7.58297                & 2.52\pm0.04                & 1.66\pm0.17             & 1.24\pm0.13             & -0.37\pm0.19            & 70\pm23                 & -0.62\pm0.05               & -0.16\pm0.09                  & -0.6\pm0.15                   & -0.13\pm0.17 \\
NGC0337\_e3  & 14.96622               & -7.57672                & < 2.06                     & 0.09\pm0.01             & 0.08\pm0.02             & -0.07\pm0.43            & 100                     & -1.63\pm0.20               & > -0.25                       & -1.63\pm0.2                   & > -0.25      \\
NGC0337\_e2  & 14.96757               & -7.57772                & 2.15\pm0.02                & 0.11\pm0.02             & 0.14\pm0.03             & 0.30\pm0.32             & 100                     & -1.40\pm0.13               & -0.20\pm0.13                  & -1.4\pm0.13                   & -0.20\pm0.13 \\
NGC0628\_e1  & 24.14878               & 15.83535                & < 1.53                     & 0.07\pm0.01             & 0.07\pm0.02             & -0.03\pm0.41            & 100                     & -2.56\pm0.19               & > -0.13                       & -2.56\pm0.19                  & > -0.13      \\
NGC0628\_e1  & 24.15686               & 15.75200                & 1.63\pm0.01                & 0.16\pm0.02             & 0.16\pm0.03             & -0.02\pm0.27            & 100                     & -2.19\pm0.13               & 0.06\pm0.13                   & -2.19\pm0.13                  & 0.06\pm0.13  \\
\enddata
\tablecomments{(1) Region identifier, with ``n" and ``e" indicating the location of the region within the galaxy: ``n" -- nuclear (within 250\,pc of the galaxy center), ``e" -- extra-nuclear. (2) \& (3) region coordinates. (4) Region effective radius. For regions with area smaller than the matched 15 -- 33\,GHz synthesized beam (i.e. unresolved), the effective radius of the synthesized beam is reported and marked as upper-limits (``$<$"). (5) 15\,GHz flux density. (6) 33\,GHz flux density. (7) 15 -- 33\,GHz spectral index. (8) Thermal fraction at 33\,GHz, estimated from (6) using Equation \ref{eq_fth}. (9) Star formation rates derived from total 33\,GHz continuum emission from (6) using Equation \ref{eq_sfr}. (10) Star formation rate surface density. For
unresolved regions, the reported values are lower-limits, indicated with ``$>$". (11) Thermal-only star formation rate derived from 33 GHz flux in column (6) and
15 - 33 GHz spectral index from column (7), using Equation \ref{eq_sfr_th}. (12) Thermal-only star formation
rate surface density. For unresolved regions, the reported values are
lower-limits, indicated with ``$>$". }
\end{deluxetable*}
\end{longrotatetable}

\end{document}